\documentclass[aps,prd,twocolumn,superscriptaddress,nofootinbib,eqsecnum,secnumarabic]{revtex4-1}
\usepackage[utf8]{inputenc}
\usepackage{amsmath,amsfonts,amssymb,graphicx,subfigure,xcolor}
\usepackage[colorlinks=true, linkcolor = blue]{hyperref} 
\hypersetup{bookmarks=true,unicode=true,pdftoolbar=true,pdfmenubar=true,
  pdffitwindow=false,pdfstartview={FitH},
  pdftitle={Doubly Charged Higgs Boson Production at Hadron Colliders - Fuks, Nemevsek, Ruiz - Phys. Rev. D101, 075022D 101, 075022 (2020) [arXiv:1912.08975]},
  pdfauthor={Fuks, Nemevsek, Ruiz},
  pdfsubject={Subject},pdfcreator={Creator},pdfproducer={Producer},
  pdfkeywords={Type II Seesaw}{Colliders}{Monte Carlo Tools},
  pdfnewwindow=true,colorlinks=true,
  linkcolor=blue,citecolor=purple,filecolor=magenta,urlcolor=cyan}
\providecommand{\tabularnewline}{\\}
\newcommand{\be}{\begin{equation}}
\newcommand{\ee}{\end{equation}}
\def\bsp#1\esp{\begin{split}#1\end{split}}
\def\bpm{\begin{pmatrix}}
\def\epm{\end{pmatrix}}
\def\lag{{\cal L}}
\def\dpp{\Delta^{++}}
\def\dpx{\Delta^{+}}
\def\dmm{\Delta^{--}}
\def\dmx{\Delta^{-}}
\def\dxx{\Delta^0}
\def\dpmpm{\Delta^{\pm\pm}}
\def\dpm{\Delta^{\pm}}
\def\dmp{\Delta^{\mp}}

\def\DYCC{{\rm CC~DY}}
\def\DYNC{{\rm NC~DY}}

\def\VBF{{\rm VBF}}
\def\GF{{\rm GF}}
\def\AF{{\rm AF}}
\def\zb{{\rm ~zb}}
\def\ab{{\rm ~ab}}
\def\fb{{\rm ~fb}}
\def\pb{{\rm ~pb}}
\def\invab{{\rm~ab}^{-1}}

\def\GeV{{\rm ~GeV}}
\def\TeV{{\rm ~TeV}}

\newcommand{\tr}[1]{{{\rm Tr}\big[#1\big]}}
\newcommand{\libName}{\textsc{TypeIISeesaw}}
\newcommand{\mgFull}{\textsc{MadGraph5\_aMC@NLO}}
\newcommand{\mgamc}{\textsc{MG5aMC}}

\newcommand{\mogre}{\textsc{MoGRe}}
\newcommand{\nloct}{\textsc{NloCT}}
\newcommand{\fr}{\textsc{FeynRules}}
\newcommand{\fa}{\textsc{FeynArts}}
\def\ie{{\it i.e.}}
\def\eg{{\it e.g.}}
\definecolor{darkgreen}{rgb}{0.0, 0.5, 0.13}
\definecolor{asparagus}{rgb}{0.53, 0.66, 0.42}
\definecolor{darkmagenta}{rgb}{0.55, 0.0, 0.55}
\definecolor{darkblue}{rgb}{0,0,.5}
\newcommand{\confirm}[1]{{\color{black}#1}}
\begin{document}
\leftline{}
\rightline{CP3-19-61, MCnet-19-29, VBSCAN-PUB-11-19}

\title{Doubly Charged Higgs Boson Production at Hadron Colliders}

\author{Benjamin Fuks}
\email{fuks@lpthe.jussieu.fr}
\affiliation{Laboratoire de Physique Th\'eorique et Hautes Energies (LPTHE),
  UMR 7589, Sorbonne Universit\'e et CNRS, 4 place Jussieu,
  75252 Paris Cedex 05, France}
\affiliation{Institut Universitaire de France, 103 boulevard Saint-Michel,
   75005 Paris, France}

\author{Miha Nemev\v{s}ek}
\email{miha.nemevsek@ijs.si}
\affiliation{Jo\v{z}ef Stefan Institute, Jamova cesta 39, 1000 Ljubljana, Slovenia}

\author{Richard Ruiz}
\email{richard.ruiz@uclouvain.be}
\affiliation{Centre for Cosmology, Particle Physics and Phenomenology {\rm (CP3)},\\
Universit\'e Catholique de Louvain, Chemin du Cyclotron, Louvain la Neuve, B-1348, Belgium}

\date{\today}

\begin{abstract}
We present a systematic comparison of doubly charged Higgs boson production
mechanisms at hadron colliders in the context of the Type II Seesaw model,
emphasizing the importance of higher-order corrections and subdominant
channels. We consider the Drell-Yan channel at next-to-leading order in QCD,
photon fusion at leading order, 
gluon fusion with resummation of threshold logarithms up to
next-to-next-to-next-to-leading logarithmic accuracy, 
and same-sign weak boson fusion at next-to-leading order in QCD.
For Drell-Yan processes, we study the impact of a static jet veto
at next-to-leading order matched to the resummation of jet veto scale logarithms at next-to-next-to-leading logarithmic accuracy.
For the photon fusion channel, the dependence on modeling photon
parton distribution functions is definitively assessed. To model vector boson
fusion at next-to-leading order, we include all interfering topologies at
$\mathcal{O}(\alpha^4)$  and propose a method for introducing generator-level
cuts within the MC@NLO formalism. Our results are obtained using a Monte Carlo
tool chain linking the \textsc{FeynRules}, \textsc{NloCT} and
\textsc{MadGraph5\_aMC@NLO} programs and have necessitated the development of
new, publicly available, Universal FeynRules Output libraries that encode the
interactions between the Type II Seesaw scalars and Standard Model
particles. Libraries are compatible with both the normal and inverted
ordering of Majorana neutrino masses.
\end{abstract}

\maketitle

\section{Introduction} \label{sec:intro}
The Type II Seesaw mechanism hypothesizes extending the Standard Model (SM) of particle physics
 by a single scalar multiplet $\hat\Delta$ in the  $({\bf 1}, {\bf 3})_1$ representation its
$SU(3)_c\times SU(2)_L\times U(1)_Y$ gauge symmetry~\cite{Magg:1980ut,
Schechter:1980gr, Cheng:1980qt, Mohapatra:1980yp, Lazarides:1980nt}.
In doing so, it is arguably the simplest known way to account for the smallness
of the neutrino masses and all neutrino oscillation data
in a renormalizable and gauge invariant manner,
without hypothesizing the existence of right-handed neutrinos as in the 
Type I Seesaw paradigm~\cite{Minkowski:1977sc,Yanagida:1979as,GellMann:1980vs,Glashow:1979nm,Mohapatra:1979ia,
Shrock:1980ct,Schechter:1980gr},
$SU(2)_L$ triplet fermions as in the Type III Seesaw case~\cite{Foot:1988aq},
or new symmetries as in, for example, loop-level and gauge-extended neutrino mass models~\cite{Cai:2017jrq,Cai:2017mow}.

In contrast to Type I or III scenarios which explain neutrino masses through an
admixture of Dirac and right-handed Majorana masses, the Type~II Seesaw
mechanism  dynamically  generates left-handed Majorana neutrino
masses through Yukawa couplings between the SM leptonic doublet and the scalar
triplet $\hat\Delta$. After the breaking of the electroweak (EW) symmetry,
mixing with the SM Higgs sector arises and the degrees of freedom are organized
into two electrically neutral $CP$-even ($h$, $\Delta^0$), 
one electrically neutral $CP$-odd state ($\chi$), 
one singly charged state ($\Delta^\pm$), and
one doubly charged state ($\Delta^{\pm\pm}$), 
with the $\Delta^0$ and $\chi$ fields being dominated by their triplet component.
The strength of the interactions
of the triplet scalars with the SM charged leptons is then proportional to
the neutrino Yukawa coupling, and hence to the neutrino oscillation parameters.
This connection is perhaps the most appealing aspect of the Type II Seesaw model,
as it directly relates the neutrino oscillation data with possible collider signals~\cite{Chun:2003ej, Garayoa:2007fw, Kadastik:2007yd,Perez:2008zc,Perez:2008ha}.

The production of triplet Higgs bosons at hadron colliders has been studied in numerous renown works
including, for example, Refs.~\cite{Rizzo:1981xx,Gunion:1996pq,Huitu:1996su,Muhlleitner:2003me,Akeroyd:2005gt,Han:2007bk,Akeroyd:2007zv,
Perez:2008zc,Perez:2008ha,Akeroyd:2011zza,Melfo:2011nx,Alloul:2013raa},
as well as in more recent investigations~\cite{Li:2018jns,Dev:2018sel,Crivellin:2018ahj,Du:2018eaw,Antusch:2018svb,Primulando:2019evb,deMelo:2019asm,Padhan:2019jlc}.
For comprehensive reviews, we refer to Refs.~\cite{Deppisch:2015qwa,Cai:2017mow}.
Current experimental searches by both the ATLAS and CMS collaborations report 95\% confidence level (CL) exclusion limits on triplet scalar masses up to
about $400 \GeV$ in multileptonic channels~\cite{Chatrchyan:2012ya,ATLAS:2012hi} and
about $200\GeV$ in bosonic channels~\cite{Aaboud:2018qcu},  assuming a degenerate mass spectrum.

\begin{figure*}
\includegraphics[width=0.90\textwidth]{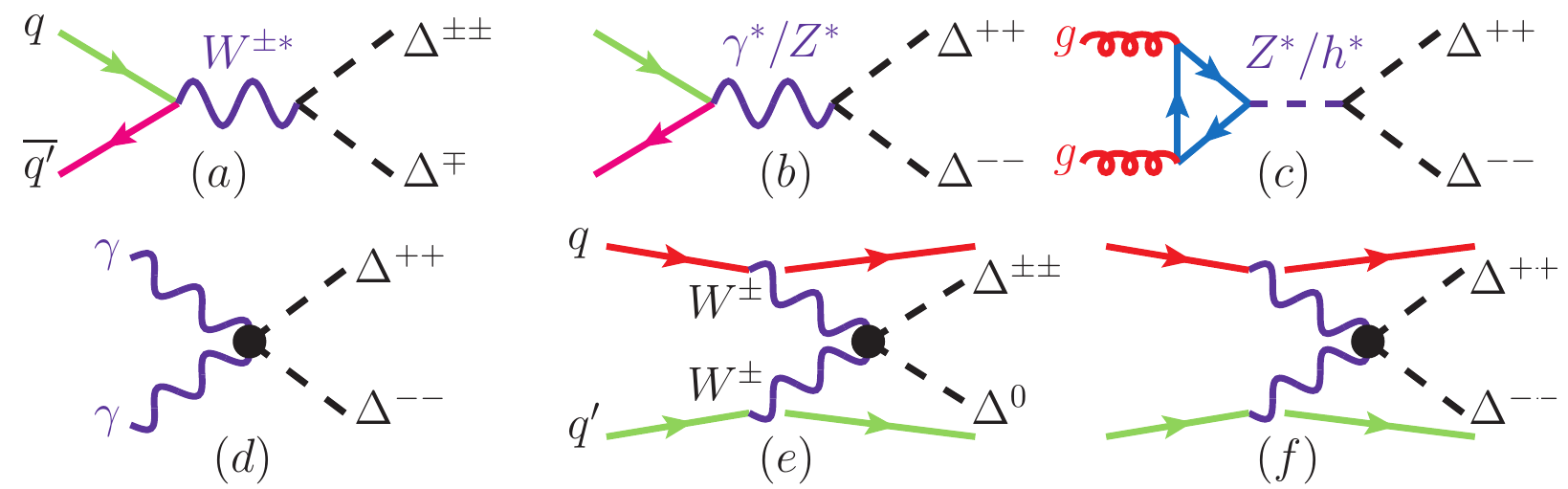}
\caption{
Born-level, diagrammatic representation of doubly charged $(\dpmpm)$, singly charged $(\dpm)$, and neutral $(\dxx)$ scalar production 
in the Type II Seesaw, through the
(a) charged current Drell-Yan,
(b) neutral current Drell-Yan,
(c) gluon fusion,
(d) photon fusion,
(e) $W$ fusion, and
(f) electroweak boson fusion mechanisms.
Diagrams are drawn with \textsc{JaxoDraw}~\cite{Binosi:2008ig}.
}
\label{fig:diagrams}
\end{figure*}

Notably, most of these works extract their sensitivity or exclusion limits from
leading-order (LO) descriptions matched with parton showers, and consider
$\dpmpm\dpm$ associated production or $\dpp\dmm$ pair production 
via the charged current (CC) and neutral current (NC) Drell-Yan (DY) mechanisms \ie,
quark-antiquark annihilation. The reported results however generally
include a normalization of the total production cross section
at next-to-leading order (NLO) in QCD.
As shown in Figs.~\ref{fig:diagrams}(a) and (b), these processes respectively
read at the Born level
\begin{eqnarray}
\text{CC~DY:}\quad q ~\overline{q'} \rightarrow &W^{(*)} 			& \rightarrow \dpmpm \dmp,  
\label{eq:dyAssociated}
\\
\text{NC~DY:}\quad q ~\overline{q} \rightarrow &\gamma^*/Z^{(*)} 	& \rightarrow \dpp \dmm,		 
\label{eq:dyPair}
\end{eqnarray}
for quark $q \in \{u,c,d,s\}$.
While NLO normalizations for Eq.~(\ref{eq:dyPair}) at the {$14\TeV$} LHC have been known for some time~\cite{Muhlleitner:2003me},
fully differential predictions are only publicly available from the LO Monte Carlo (MC) 
event generators\footnote{Doubly charged scalars in various $SU(2)_L$ representations can also be studied in a model-independent way following Ref.~\cite{delAguila:2013mia}.}
\textsc{Pythia}~\cite{Huitu:1996su,Sjostrand:2006za,Sjostrand:2014zea} and \textsc{CalcHEP}~\cite{Pukhov:2004ca,Belyaev:2012qa}.
Even after matching events to parton showers, such descriptions model the associated hadronic activity to at most the leading logarithmic (LL) accuracy~\cite{Dasgupta:2018nvj}.
Consequently, studies and searches for Type II scalars using these tools are blind to
significant, qualitative differences in jet activity between triplet scalar signal events and the leading background processes,
and therefore cannot reliably exploit selection cuts that discriminate accordingly, \eg, jet vetoes.
This is noteworthy as recent investigations show that event-based jet vetoes
can substantially increase the discovery prospects of anomalous multileptonic events~\cite{Pascoli:2018rsg,Pascoli:2018heg,Fuks:2019iaj}.

Beyond this limitation, no publicly available event generator describes loop-induced production and decay modes of triplet scalars.
This includes $\dpp\dmm$ pair production through gluon fusion (GF),
\begin{equation}
\text{GF}:\quad g ~g \rightarrow h^*/Z^* \rightarrow \dpp \dmm,
 \label{eq:gfPair}
\end{equation}
as shown in Fig.~\ref{fig:diagrams}(c). 
Assuming that the EW quantum numbers of the doubly charged scalars $\dpmpm$
(and singly charged scalar $\dpm$) can
be determined through their potential discovery in the DY production mode,
Eq.~\ref{eq:gfPair} and the $gg\to\dpx\dmx$ analog process offer direct probes of their couplings to the SM Higgs~\cite{Hessler:2014ssa}.

The situation is only slightly better for modeling triplet Higgs production through electroweak boson fusion processes at $pp$ colliders using public MC tools.
Computations relevant for $2\to2$ photon fusion (AF) processes, such as the one
illustrated at the Born level in Fig.~\ref{fig:diagrams}(d),
\begin{equation}
\text{AF}:\quad \gamma \gamma \to \dpp \dmm \ ,
 \label{eq:afPair}
\end{equation}
can be achieved with \textsc{CalcHEP}. However, extensions to more exclusive
(in the jet multiplicity) processes are computationally prohibitive. For
\textsc{Pythia} the situation is also bleak. In particular, $\dpmpm\dxx jj$
associated
production and $\dpp\dmm jj$ pair production through EW vector boson fusion
(VBF), as shown representatively in Figs.~\ref{fig:diagrams}(e) and (f),
\begin{eqnarray}
\text{VBF [associated]:}\quad		q_1 q_2 &\to& \dpmpm \dxx q_1' q_2'		 \label{eq:vbfAss}
\\
\text{VBF [pair]:}\quad			q_1 q_2 &\to& \dpp \dmm q_1' q_2'		 \label{eq:vbfPair}
\end{eqnarray}
cannot be simulated as their matrix elements are not hard-coded into the event generator.
More specifically, only $\dpmpm jj$ single production in the context of the Left-Right Symmetric Model is possible.

In light of these impediments to studying the canonical Type II Seesaw model, we revisit the modeling of triplet Higgs boson production at current and 
proposed~\cite{Arkani-Hamed:2015vfh,CEPC-SPPCStudyGroup:2015csa,Golling:2016gvc,Benedikt:2018csr,Abada:2019ono} hadron collider facilities.
Though our work holds for all mass eigenstates arising from 
the scalar triplet field $\hat \Delta$, for conciseness, 
we limit our investigation to the pair and associated production of the doubly charged scalar $\dpmpm$. 
In particular, we consider for the first time a systematic comparison of all the
production mechanisms described in Eqs.~\ref{eq:dyPair}-\ref{eq:vbfAss}.
For quark-initiated processes, we include QCD corrections up to NLO;
for gluon-initiated processes, soft gluon threshold logarithms are resummed up to next-to-next-to-next-leading logarithmic accuracy (N$^3$LL).
For the VBF-associated process in Eq.~\ref{eq:vbfAss},
we propose a way to introduce  generator-level cuts within the MC@NLO~\cite{Frixione:2002ik} formalism to enrich the VBF contribution over all $\mathcal{O}(\alpha^4)$ contributions.
For the AF channel, we investigate the dependence on modeling photon parton distribution functions (PDFs),
which we assert are responsible for renewed claims~\cite{Babu:2016rcr,Ghosh:2017jbw,Ghosh:2018drw} 
and (correct) counterclaims~\cite{Cai:2017mow,Crivellin:2018ahj} of the mechanism dominance.
We also present jet veto predictions for the DY modes, up to NLO and matched to jet veto resummation at next-to-next-to-leading logarithm (NNLL).

To facilitate this work, we report the development of the \libName~UFO libraries, 
a new, publicly\footnote{Available from \href{https://feynrules.irmp.ucl.ac.be/wiki/TypeIISeesaw}{feynrules.irmp.ucl.ac.be/wiki/TypeIISeesaw}.} 
available UFO~\cite{Degrande:2011ua} containing relevant QCD ultraviolet (UV) counterterms and
$R_2$ rational terms that enable 
automated computations of tree-induced processes up to NLO plus parton showers
(NLO+PS) in QCD and
loop-induced processes up to LO+PS using the precision MC event generator \mgFull~(\mgamc)~\cite{Alwall:2014hca}.

The remainder of this work is outlined in the following manner:
In Sec.~\ref{sec:theory} we describe the main features of canonical Type II Seesaw model, present experimental constraints,
and introduce the \libName~UFO libraries.
In Sec.~\ref{sec:mcSetup}, we describe our computational setup and benchmark input parameters.
Our main results are reported in Sec.~\ref{sec:results},
and we conclude in Sec.~\ref{sec:Conclusions}.
Additional technical details are reported in the appendices.

\section{The Canonical Type II Seesaw}\label{sec:theory}
In this section, we introduce the canonical Type II Seesaw model and 
the development of the publicly available \libName~UFO
libraries.
In conjunction with state-of-the-art MC event generators, these UFO libraries
allow for the simulation of tree-induced collider processes up to NLO in QCD with parton shower (PS)
and loop-induced processes at the LO+PS.

\subsection{The Type II Seesaw Model}\label{sec:TH_lag}
To generate neutrino masses, the Type II Seesaw postulates extending the 
SM field content by one complex scalar {multiplet} $\hat\Delta$, which lies in the 
adjoint representation of the weak group, and has hypercharge
$Y_{\hat\Delta}=1$. The scalar sector of the theory is thus defined by two
gauge eigenstates,  $\hat\Delta$ and the SM Higgs doublet $\varphi$,
\be
  \hat\Delta =   \bpm
    \frac{1}{\sqrt{2}} \hat\Delta^+ & \hat\Delta^{++}\\
    \hat\Delta^0 & -\frac{1}{\sqrt{2}} \hat\Delta^+
  \epm \ ,\qquad
  \varphi = \bpm \varphi^+\\ \varphi^0 \epm \ .
\label{eq:scalars}\ee
The corresponding Lagrangian includes, in addition to the SM Lagrangian
$(\lag_{\rm SM})$, gauge-invariant kinetic terms, Yukawa couplings
$(\lag_{Y_\Delta})$ for the scalar multiplet $\hat\Delta$, and extra
contributions to the scalar potential ($V_\Delta$),
\be
  \lag_{\rm Type II} = \lag_{\rm SM} +
    {\rm Tr}\big[D_\mu \hat\Delta^\dag D^\mu\hat\Delta\big]
   - V_\Delta + \lag_{{\rm Y}_\Delta} \ .
\label{eq:LagTypeII}\ee
The electroweak covariant derivative acting on the triplet
field $\hat\Delta$ (for which we use a matrix representation) reads,
\be
  D_\mu \hat\Delta =  \partial_\mu\hat\Delta
   - \frac{i}{2} g W_\mu^k \big[\sigma_k\hat\Delta-\hat\Delta\sigma_k\big]
   - i g' B_\mu \hat\Delta \ .
\ee
In our notation, $g$ ($g'$) and $W_\mu$ ($B_\mu$) consist of the usual weak
(hypercharge) gauge coupling and boson respectively, and the $\sigma$ matrices
denote the Pauli matrices. The full scalar potential $V=V_H+V_\Delta$
(where $V_H$ stands for the SM Higgs potential) is
 written as
\be\bsp
  V = &\
    - \mu_h^2 \varphi^\dagger \varphi
    + \lambda_h \big(\varphi^\dagger \varphi \big)^2
    + m_{\hat{\Delta}}^2 \tr{\hat\Delta^\dagger \hat\Delta}
  \\&\
    + \lambda_{\Delta 1} \big(\tr{\hat\Delta^\dagger \hat\Delta}\big)^2
    + \lambda_{\Delta 2} \tr{(\hat\Delta^\dagger \hat\Delta)^2} \\
  &\ + \lambda_{h \Delta 1}\ \varphi^\dagger \varphi\
          \tr{\hat\Delta^\dagger\hat\Delta}
     + \lambda_{h \Delta 2}\ \varphi^\dagger \hat\Delta\hat\Delta^\dagger\varphi
  \\&\
     + \mu_{h\Delta} \Big(\varphi^\dagger \hat\Delta \cdot \varphi^\dagger +
        \text{H.c.} \big)\ .
\esp\label{eq:vpot}\ee
Here the dot denotes the $SU(2)$-invariant product of two objects lying
in its (anti)fundamental representation. Finally, the triplet Yukawa Lagrangian
is given by
\be
  \lag_{{\rm Y}_\Delta} = -{\bf Y}_\Delta \bar L\cdot\hat\Delta L+\text{H.c.}\ ,
\ee
where flavor indices are omitted for clarity and $L$ denotes the 
{SM left-handed weak lepton doublet.}

\begin{figure*}[t]
  \includegraphics[width=\textwidth]{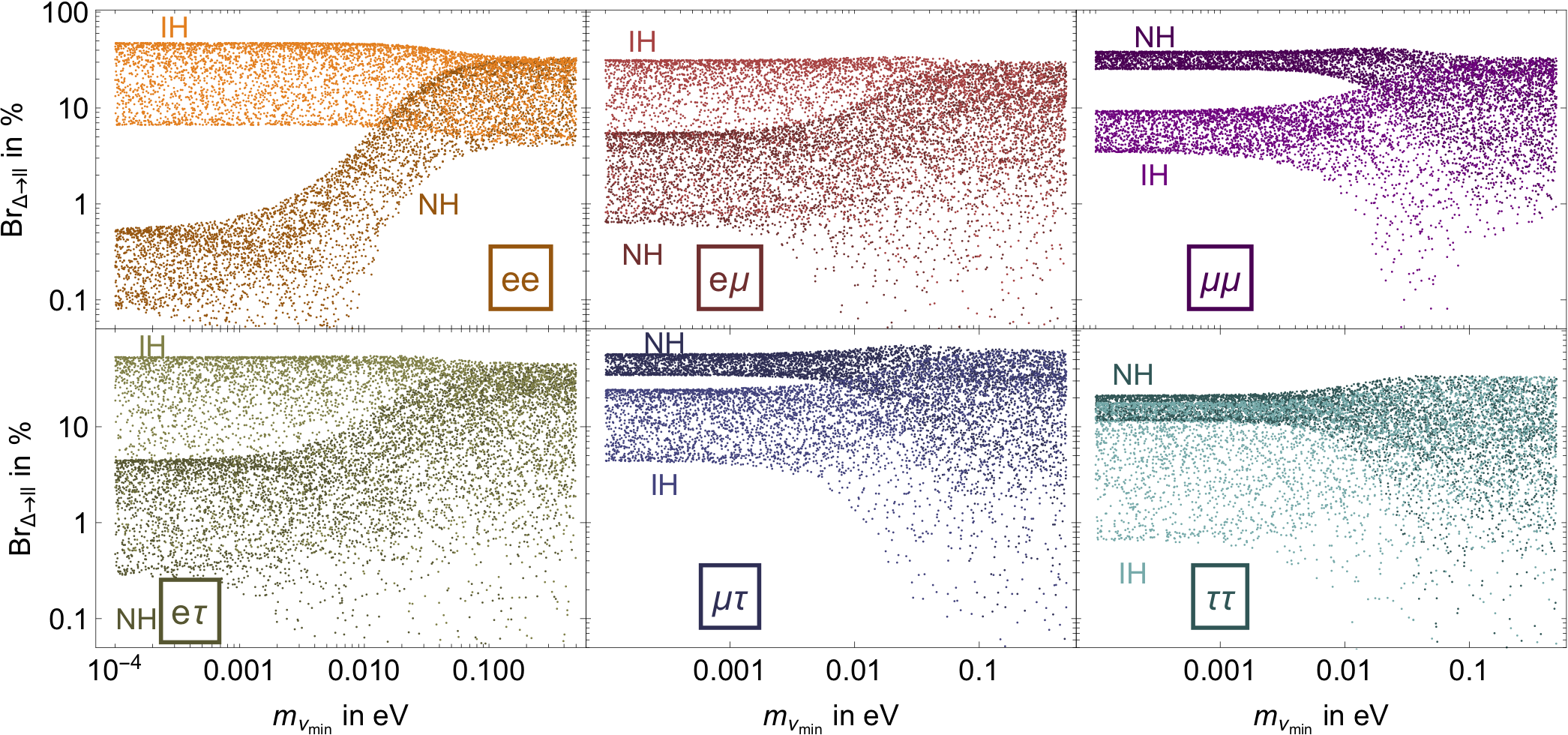}
  \caption{
  Flavor composition of the $\Delta^{++} \to \ell_i \ell_j$ decay channel branching ratios (Br)
  with $v_\Delta = 10^{-7}\GeV$, as a function of the lightest neutrino mass $m_{\nu_{\min}}$, for both 
  normal (NH, darker) and inverted (IH, lighter) hierarchy. The mixing angles in the PMNS matrix and $\Delta m^2$ are fixed
  by oscillation data~\cite{Esteban:2018azc} and Majorana phases are varied over the range $[ 0, 2 \pi)$.
  }
  \label{fig:typeIInlo_BrFlavor}
\end{figure*}

After breaking EW symmetry to electromagnetism, the neutral components of the two
scalar fields get nonvanishing vacuum expectation values (vevs) $v$ and $v_\Delta$,
\be
  \langle\hat\Delta\rangle =  \frac{1}{\sqrt{2}} \bpm 0&0\\ v_\Delta & 0\epm\ ,
  \qquad
  \langle\varphi\rangle =  \frac{1}{\sqrt{2}} \bpm 0\\v \epm \ ,
\ee
and the degrees of freedom carrying the same electric charges mix.
The two, complex neutral fields $\varphi^0$ and $\hat\Delta^0$ mix into two massive $CP$-even states $h$ and $\Delta^0$, 
one massive $CP$-odd state $\chi$, and one Goldstone boson $G^0$ that is absorbed by the $Z$ boson. 
Expressing $\varphi^0$ and $\hat\Delta^0$ in terms of their real degrees of freedom,
\be\bsp
  \varphi^0 =& \frac{1}{\sqrt{2}} \Big[v + \Re[\varphi^0] + i\Im[\varphi^0]\Big]
  \ , \\
  \hat\Delta^0 =& \frac{1}{\sqrt{2}} \Big[v_\Delta + \Re[\hat\Delta^0] +
    i \Im[\hat\Delta^0]\Big] \ ,
\esp\ee
the mixing relations read
\be\renewcommand{\arraystretch}{1.2}
\bsp
  \bpm G^0 \\ \chi\epm  = &\ \frac{1}{\sqrt{v^2+4 v_\Delta^2}}
      \bpm v & 2v_\Delta\\ - 2v_\Delta & v \epm
      \bpm  \Im[\varphi^0] \\  \Im[\hat\Delta^0]\epm \ , \\[.2cm]
  \bpm h \\ \Delta^0 \epm  = &\
    \bpm \phantom{-}\cos\xi & \sin\xi\\ -\sin\xi & \cos\xi\epm
    \bpm \Re[\varphi^0] \\ \Re[\hat\Delta^0] \epm \ ,
\esp\label{eq:h0}\ee
where the $h-\Delta^0$ mixing angle $\xi$ is defined by
\begin{widetext}
\be
  \tan2\xi = \frac{4 v v_\Delta \Big( (v^2+2v_\Delta^2)(2 \lambda_{h\Delta1}
    + \lambda_{h\Delta2}) - 4 m_{\Delta^\pm}^2 \Big)}
  {(v^2+2v_\Delta^2) \Big(8  m_{\Delta^{\pm\pm}}^2 + (8\lambda_h + \lambda_{h\Delta2})
  v^2 - 8 \lambda_{\Delta1} v_\Delta^2\Big) - 12 m_{\Delta^\pm}^2 v^2} \ .
\ee
\end{widetext}
Moreover, the two
charged states $\varphi^\pm$ and $\hat\Delta^\pm$ mix into a massive physical
charged Higgs boson $\Delta^\pm$ and a Goldstone boson $G^\pm$ that gets
absorbed by the $W$ boson,
\be\label{eq:hp}\renewcommand{\arraystretch}{1.2}
  \bpm G^\pm \\  \Delta^\pm\epm \!=\! \frac{1}{\sqrt{v^2\!+\!2 v_\Delta^2}}
    \bpm \phantom{-}v & \sqrt{2} v_\Delta\\
         -\sqrt{2} v_\Delta & v \epm\!\bpm\varphi^\pm \\ \hat\Delta^\pm \epm \ .
\ee
This notation reflects the hierarchy $v_\Delta \ll v$, originating from
strong constraints on triplet contributions to EW symmetry breaking~\cite{Chen:2005jx,Chen:2008jg,Melfo:2011nx,Kanemura:2012rs,Das:2016bir}.
This implies that the $\Delta$ fields are mostly of a triplet nature, and that the
$h$ field and the Goldstone bosons mostly align with the SM doublet.
As the doubly charged \confirm{gauge} state $\hat\Delta^{\pm\pm}$ is aligned with its \confirm{mass} state $\Delta^{\pm\pm}$,
we henceforth use the latter notion for consistency.

The Higgs sector is thus defined by nine parameters:
seven couplings appearing in the scalar potential of Eq.~\ref{eq:vpot} and the two vacuum
expectation values of the neutral scalar fields. This leads to seven independent
free parameters, after accounting for the minimization of the scalar potential,
which we choose to be
\be
 \bigg\{\lambda_{\Delta1},\ \lambda_{h\Delta1},\ m_h, \
  m_{\Delta^{\pm\pm}},\ m_{\Delta^\pm},\ m_{\Delta^0}, \ v_\Delta \bigg\} \ .
\label{eq:prm1}\ee
This trades four couplings for the masses  of 
the SM Higgs boson $m_h$, 
the neutral $CP$-even triplet $m_{\Delta^0}$,
the singly charged ($m_{\Delta^\pm}$) triplet state,
and the doubly charged ($m_{\Delta^{\pm\pm}}$) state. 
Relations linking the parameters of the theory to these
seven inputs are provided in appendix~\ref{app:scalar}.

The upshot of having most of the tree-level triplet masses (except for $\chi$) as inputs
is to facilitate  parameter scanning with physically meaningful inputs. However, only one
``large'' mass splitting is allowed by the sum rule in Eq.~\ref{eq:sumrule}.
The sum rule being approximate, the model includes departures of the order ${\cal O}(v_\Delta^2)$.
Thus care must be taken when choosing the input masses to avoid nonperturbative
$\lambda$ couplings or an unstable vacuum.
This can be done by defining the model file with exact tree-level mixings and 
keeping $m_{\Delta^0}$ computed internally while controlling $\lambda_{\Delta2}$ as an 
external parameter.\footnote{This implementation is also available at \href{https://feynrules.irmp.ucl.ac.be/wiki/TypeIISeesaw}{feynrules.irmp.ucl.ac.be/ wiki/TypeIISeesaw}.}

After shifting the neutral scalar fields relatively to their vevs,
the new physics contributions to the Yukawa interactions and fermionic mass
terms are given by
\be
  \lag_{{\rm Y}_\Delta} \!=\! \frac{v_\Delta}{2\sqrt{2} }\Big[
   ({\bf Y}_\Delta\! +\! {\bf Y}_\Delta^T)_{ff'} \ \overline {\nu_{Lf}^c}\cdot\nu_{Lf'}
     + {\rm H.c.} \Big] +\ \dots\ ,
\label{eq:lyd}\ee
where the dots stand for scalar-fermion-antifermion interactions and
$f, f'$ are flavor indices. The neutrino mass matrix ${\cal M}_\nu$
originating from the Lagrangian of Eq.~\ref{eq:lyd} is diagonalized by
introducing the unitary Pontecorvo-Maki-Nakagawa-Sakata (PMNS) matrix
$V_{\rm PMNS}$,
\be
  {\cal M}_\nu = V_{\rm PMNS}^*\  m_\nu^{\rm diag}\
  V_{\rm PMNS}^\dagger \ .
\ee
Here $m_\nu^{\rm diag}$ is diagonal and its entries are the three
physical neutrino masses $m_{\nu_1}$, $m_{\nu_2}$ and $m_{\nu_3}$.
The Yukawa matrix hence reads
\be
  {\bf Y}_\Delta = \frac{{\cal M}_\nu}{\sqrt{2} v_\Delta} \ .
\ee

 Adopting a normal hierarchy \confirm{(NH)} for the neutrino mass order, \ie, $m_{\nu_1} < m_{\nu_2} <
m_{\nu_3}$, we express all the free parameters of
the neutrino sector in terms of the neutrino oscillation parameters and the mass
of the lightest neutrino $m_{\nu_1}$:
{\small
\be\label{eq:prm2}
  \bigg\{ m_{\nu_1},\ \Delta m_{21}^2,\ \Delta m_{31}^2,\
    \theta_{12},\ \theta_{13}, \ \theta_{23}, \
    \varphi_{CP},\ \varphi_1, \ \varphi_2 \bigg\} \ .
\ee
}
Here the $\theta_{ij}$ (with $i,j=1,2,3$) stand for the three neutrino mixing
angles, $\varphi_{CP}$ for the Dirac $CP$-violating phase, and $\varphi_1$ and
$\varphi_2$ for the two Majorana phases. 
\confirm{
We take as inputs the neutrino squared mass-differences in the NH,} $\Delta m^2_{31}>0$ and $\Delta m_{21}^2>0$,
defined by
\be
  m_{\nu_2} = \sqrt{m_{\nu_1}^2 + \Delta m^2_{21}} \ , \
  m_{\nu_3} = \sqrt{m_{\nu_1}^2 + \Delta m^2_{31}} \ .
\ee

In the case of inverted hierarchy \confirm{(IH)}, the third neutrino $m_{\nu_3}$ is the lightest, \ie, that
 $m_{\nu_3}< m_{\nu_1} < m_{\nu_2}$,
 and we use it to set the neutrino mass scale.
 The list of input parameters in Eq.~\ref{eq:prm2} is then replaced by
{\small
\be
  \bigg\{ m_{\nu_3},\ \Delta m_{21}^2,\ \Delta m_{32}^2,\
    \theta_{12},\ \theta_{13}, \ \theta_{23}, \
    \varphi_{CP},\ \varphi_1, \ \varphi_2 \bigg\} \ ,
\label{eq:prm2b}\ee
}
so that $\Delta m^2_{21}>0$, $\Delta m^2_{32}<0$, and
\be\bsp
  m_{\nu_1} =&\ \sqrt{m_{\nu_3}^2 - \Delta m^2_{32} - \Delta m^2_{21}} \ , \\
  m_{\nu_2} =&\ \sqrt{m_{\nu_3}^2 - \Delta m^2_{32}} \ .
\esp\ee

The neutrino mass matrix ${\bf Y}_\Delta$ is largely fixed by the global fit of oscillation data~\cite{Esteban:2018azc}, 
apart from the smallest neutrino mass $m_{\nu_{\min}}$, which is set by $m_{\nu_1}(m_{\nu_3})$ in the NH (IH) scenario,
 and the two Majorana phases.
These are not fixed by current (lepton number conserving) oscillation data, but do affect
the leptonic decay modes of $\Delta^{++}$~\cite{Chun:2003ej, Garayoa:2007fw, Kadastik:2007yd}. 
Specifically, the decay rate to a pair of charged leptons is given by
\begin{align} \label{eq:GamDppEllEll}
    \Gamma_{\Delta^{++} \to \ell^+_i \ell^+_j}&= \frac{m_{\Delta^{++}}}{8 \pi \left(1 + \delta_{ij} \right)} 
  \left|\frac{{{\cal M}_\nu}_{ij}}{v_\Delta}\right|^2,
\end{align}
and the resulting branching rates, for $v_\Delta = 10^{-7}\GeV$, are shown in Fig.~\ref{fig:typeIInlo_BrFlavor}. 
We use the central values of the neutrino oscillation data, as reported in Ref.~\cite{Esteban:2018azc}, 
including the hint for the nonzero Dirac phase.
The spread in branching ratios comes exclusively from the unknown Majorana phases. 

While the production cross section does not depend on the flavor structure of ${\bf Y}_\Delta$,
the distribution of events among the six flavor final states in Fig.~\ref{fig:typeIInlo_BrFlavor}
may have a significant impact on the search strategy and on the limit derived from the
experimental searches. In any case, the dependence on neutrino data is implemented in the model 
file and can be controlled externally by the user. 

\subsection{The \libName~UFO Libraries}\label{sec:implementation}
To simulate the hadronic production of Type II scalars up to NLO+PS, we implement the model presented
in section~\ref{sec:TH_lag} into \fr~2.3.35~\cite{Alloul:2013bka},  and generate 
a UFO model~\cite{Degrande:2011ua} suitable for such calculations
within the \mgFull~framework~\cite{Alwall:2014hca}.
In the following, we focus on the case of NH for neutrino mass ordering, 
such that Eq.~\ref{eq:prm2} is used to define the neutrino mass spectrum.
In the appendix~\ref{app:fr_ih} we provide details for the use of the \libName~libraries with the IH neutrino mass ordering.

\begin{table}
\renewcommand{\arraystretch}{1.4}
\setlength\tabcolsep{5pt}
\begin{tabular}{c c c c c c}
  Field & Spin & Rep. & Self-conj. & FR name\\
  \hline\hline
  $L$ & 1/2 & $({\bf 1}, {\bf 2})_{-1/2}$ & no & {\tt LL}\\
  $\varphi$ & 0 & $({\bf 1}, {\bf 2})_{1/2\phantom{+}}$ & no & {\tt Phi}\\
  $\hat\Delta$ & 0 & $({\bf 1}, {\bf 3})_{1\phantom{--}}$ & no & {\tt hatD}\\
\end{tabular}
\caption{
Gauge eigenstates that either supplement the SM or whose definition is
  altered relatively to the SM,
  their spin (second column) and $SU(3)_c\times SU(2)_L \times U(1)_Y$ (third column) representation. We indicate
  whether the fields are self-conjugate (fourth column) and their name in the
\fr\ implementation (last column).
  }
\label{tab:gaugefields}
\begin{tabular}{c c c c c}
  Field & Spin & Self-conj. & FR name & PDG \\
  \hline\hline
  $\nu_i$ ($i=1,2,3$) & 1/2 & yes & {\tt vi}& 12, 14, 16\\
  $\Delta^0$          & 0   & yes & {\tt D0}  & 44\\
  $\Delta^+$          & 0   & no  & {\tt DP}  & 38\\
  $\Delta^{++}$       & 0   & no  & {\tt DPP} & 61\\
  $\chi$              & 0   & yes & {\tt chi} & 62\\
\end{tabular}
\caption{
  Mass eigenstates that either supplement the SM or whose definition is
  altered relatively to the SM, with their spin quantum number
  (second column). We indicate
  whether the fields are self-conjugate (third column),
  the names used in the \fr\ (FR) implementation and the \libName~UFO (fourth column),
   and the associated Particle Data Group (PDG) identification number (last column).
} \label{tab:massfields}
\end{table}

\begin{table}
\renewcommand{\arraystretch}{1.4}
\setlength\tabcolsep{5pt}
\begin{tabular}{c c c c}
  Parameter& FR name & LH block & Counter\\
  \hline\hline
  $\lambda_{h\Delta 1}$ & {\tt lamHD1} & {\tt QUARTIC}& 1\\
  $\lambda_{\Delta 1}$ & {\tt lamD1} & {\tt QUARTIC}& 2\\
  \hline
  $v_\Delta$ & {\tt vevD} & {\tt VEVD}& 1\\
  \hline
  $m_h$                 & {\tt MH}   & {\tt MASS} & 25\\
  $m_{\Delta^\pm}$      & {\tt MDP}  & {\tt MASS} & 38\\
  $m_{\Delta^0}$        & {\tt MD0}  & {\tt MASS} & 44\\
  $m_{\Delta^{\pm\pm}}$ & {\tt MDPP} & {\tt MASS} & 61\\
  \hline
  $m_{\nu_1}$       & {\tt Mv1}    & {\tt MASS} & 12\\
  $\Delta m^2_{21}$ & {\tt dmsq21} & {\tt MNU}  & 2\\
  $\Delta m^2_{31}$ & {\tt dmsq31} & {\tt MNU}  & 3\\
  \hline
  $\theta_{12}$ & {\tt th12} & {\tt PMNS}& 1\\
  $\theta_{23}$ & {\tt th23} & {\tt PMNS}& 2\\
  $\theta_{13}$ & {\tt th13} & {\tt PMNS}& 3\\
  $\varphi_{\rm CP}$ & {\tt delCP} & {\tt PMNS}& 4\\
  $\varphi_1$ & {\tt PhiM1} & {\tt PMNS}& 5\\
  $\varphi_2$ & {\tt PhiM2} & {\tt PMNS}& 6\\
\end{tabular}
\caption{External parameters defining the scalar sector of the model (first two
panels) and the neutrino sector in the context of a normal neutrino mass
hierarchy (last two panels), so that $m_{\nu_1} < m_{\nu_2} < m_{\nu_3}$,
$\Delta m^2_{31}>0$ and $\Delta m_{21}^2>0$. Each parameter is given
together with the symbol employed in the \fr\ (FR) implementation (and therefore
the corresponding UFO libraries) and the Les Houches (LH) block and counter
information allowing to change its numerical value when a Monte Carlo event
generator is used.}
\label{tab:prm}
\end{table}

To build the \libName~UFO, we start with the implementation of the SM shipped with \fr, 
and modify the definition of the physical neutrino states to align them with the mass basis.
The definition of the lepton doublets is also modified such that its upper component in the flavor basis can be decomposed 
into the neutrino mass basis using the PMNS mixing.
We add to the SM scalar sector the definition of the triplet field $\hat\Delta$,
as well as those for the physical states $\chi$, $\Delta^0$,
$\Delta^\pm$, and $\Delta^{\pm\pm}$. We tune the components' definitions
of the gauge eigenstates $\varphi$ and $\hat\Delta$ according to Eqs.~\ref{eq:h0}
and \ref{eq:hp}. The properties of these fields and information on their
\fr\ implementation are collected in Tables~\ref{tab:gaugefields} and
\ref{tab:massfields} for the gauge and mass eigenstates respectively.

The  free parameters associated with our implementation, as given by
Eqs.~\ref{eq:prm1} and \ref{eq:prm2}, are presented in Table~\ref{tab:prm},
together with information on their name and the Les Houches data block~\cite{Skands:2003cj}.
The relations relating those external parameters to all the
other parameters of the model can be found in appendix~\ref{app:scalar}.

We jointly use \fr\ with \mogre~(version 1.1)~\cite{Frixione:2019fxg},
\nloct~(version 1.0.1)~\cite{Degrande:2014vpa} and \fa~(version
3.9)~\cite{Hahn:2000kx} to renormalize the bare Lagrangian
introduced in the previous section with respect to \confirm{$\mathcal{O}(\alpha_s)$} QCD interactions.
This allows one to extract UV counterterms and so-called
$R_2$ Feynman rules that are needed to numerically evaluate one-loop integrals in four dimensions.
Together with a description of tree-level interactions, these counterterms are packaged into a UFO library that can be used by the \mgamc\ event generator for LO and NLO calculations in QCD, 
as well as by {\sc Herwig++}~\cite{Bellm:2015jjp} and {\sc Sherpa}~\cite{Gleisberg:2008ta} at LO.

Our model files are publicly available, both for the NH and IH  neutrino mass hierarchy, from the
\fr\  database \href{http://feynrules.irmp.ucl.ac.be/wiki/TypeIISeesaw}{feynrules.irmp.ucl.ac.be/wiki/TypeIISeesaw}.

\section{Computational Setup}\label{sec:mcSetup}

We now summarize the computational setup used to obtain our numerical results.
After describing our MC tool chain, we list the model inputs considered
for our out-of-the-box computations using \libName.

\subsection{Monte Carlo Tool Chain}\label{sec:mcSetup_MC}

In conjunction with \mgamc~(version 2.6.6)~\cite{Alwall:2014hca}, the \libName~UFO 
\confirm{ allows us to simulate tree-induced processes involving SM and Type II Seesaw particles up to NLO+PS}
and  loop-induced processes involving these states up to LO+PS.
This is possible through the MC@NLO formalism~\cite{Frixione:2002ik} and the packages
\textsc{MadLoop}~\cite{Hirschi:2011pa, Hirschi:2015iia} and \textsc{MadFKS}~\cite{Frixione:1995ms,Frixione:1997np,Frederix:2009yq} 
 as implemented in~\mgamc.

For threshold-resummed and jet veto-resummed computations,
we employ Soft-Collinear Effective Field Theory (SCET)~\cite{Bauer:2000yr, Bauer:2001yt, Beneke:2002ph} in momentum space~\cite{Becher:2006nr}.
N$^3$LL threshold corrections to the GF process,
which capture the leading corrections to the total normalization up to next-to-next-to-leading order (NNLO) in QCD~\cite{Bonvini:2014qga}, 
are obtained following Refs.~\cite{Becher:2006nr,Becher:2006mr,Ahrens:2008nc,Ruiz:2017yyf}.
Cross sections with a static jet veto at NLO in QCD with jet veto resummation matching at NNLL
are computed using the automated \textsc{MadGraph5\_aMC@NLO+SCET} libraries developed in Refs.~\cite{Alwall:2014hca, Becher:2014aya}.

\subsection{Standard Model Inputs}\label{sec:mcSetup_SM}
In the following numerical computations, we assume $n_f=4$ flavors of massless quarks, 
approximate the Cabbibo-Kobayashi-Masakawa (CKM) matrix to be diagonal with unit entries,
and take as the SM inputs
\begin{eqnarray}
 \alpha^{\rm \overline{MS}}(M_{Z})	= 1/127.900,	 				&\quad& M_{Z}=91.1876\GeV, \nonumber\\ 
 						G_F = 1.16637\times10^{-5}{\GeV}^2,	&\quad& m_H = 125\GeV, \nonumber\\
						m_t(m_t)=172\GeV,					&\quad& m_b(m_b) = 4.7\GeV. \qquad
 \label{eq:smInputs}
\end{eqnarray}
For fixed-order computations, we use the NLO MMHT15qed set of parton
densities (\texttt{lhaid=26000})~\cite{Harland-Lang:2019pla}. 
When matching to a resummed result, we use the NNLO MMHT15qed set (\texttt{lhaid=26300}).
Both sets take $\alpha_s(M_Z)=0.118$ and use the LUXqed formalism to match the proton's elastic and inelastic photon PDFs~\cite{Manohar:2016nzj,Manohar:2017eqh}.
Scale evolution for PDFs and $\alpha_s(\mu_r)$ are extracted using the \textsc{LHAPDF} (version 6.2.1) libraries~\cite{Buckley:2014ana}.

For the DY, AF, and VBF processes, we set the reference collinear factorization $(\mu_f)$ and QCD renormalization $(\mu_r)$ scales
dynamically to half the scalar sum of transverse energies $(E_T)$ of all final-state objects 
(\texttt{dynamical\_scale\_choice=3} in \mgamc),
\be\bsp
 \mu_f, \mu_r = \zeta \times \mu_0, \quad\text{where}\quad  \mu_0 \equiv \frac{1}{2}\sum_{k=\Delta,\text{jets},...}E_{T}^k,\\
  \quad E_T^k  = \frac12\sum_{k} \sqrt{m_k^2 + p_{T,k}^2}, \quad\text{and}\quad \zeta = 1.
\esp\label{eq:hardScaleDef}
\ee
For the GF channel at LO, we choose as a central scale the invariant mass $m_{\Delta\Delta}$
of the $\Delta^{++}\Delta^{--}$ system.
These scale choices follow the recommendations of Refs.~\cite{Ruiz:2015zca,Degrande:2016aje,Ruiz:2017yyf},
which investigated scale choices in heavy exotic lepton production via analogous mechanisms.
To summarize: For high-mass, DY and VBF topologies, Eq.~\ref{eq:hardScaleDef} ensures that  
differential $K$-factors at NLO are effectively flat for inclusive observables,
\eg, the transverse momentum ($p_T^\Delta$) or the pseudorapidity
($\eta^\Delta$) of the triplet scalar~\cite{Ruiz:2015zca,Degrande:2016aje}.
For AF, Eq.~\ref{eq:hardScaleDef} allows one to account for the impact on the cross section normalization of hard, initial-state $q\to q\gamma$ splittings 
that are otherwise matched to hard jets when using QED parton showers~\cite{Alva:2014gxa,Degrande:2016aje}.
For GF, the threshold resummation formalism employed~\cite{Becher:2006nr,Becher:2006mr,Ahrens:2008nc} 
is derived with the hard factorization scale set to the hardest scale of the process.
In the present case, we identify this as $m_{\Delta\Delta}$.

As a conventional measure of theoretical uncertainty in cross section normalizations,
we vary the parameter $\zeta$ in Eq.~\ref{eq:hardScaleDef} discretely over the range $\zeta \in [0.5,1.0,2.0]$ to obtain a nine-point uncertainty band.
For the DY jet veto and GF threshold resummation computations, scale uncertainties are determined according to Refs.~\cite{Becher:2014aya,Ruiz:2017yyf}.

\subsection{Type II Seesaw Inputs}\label{sec:mcSetup_Seesaw}
In this study we aim to explore the stability of our production modeling prescriptions across a broad range 
of scalar masses and collider scales.
Hence for conciseness, we consider the mass-degenerate limit, where
\begin{equation}
m_{\dpmpm} = m_{\dpm} = m_{\dxx} \equiv m_{\Delta},
\label{eq:degenerateMasses}
\end{equation}
and scan over a range of $m_{\Delta}$ for $m_{\Delta}>100\GeV$.
While direct searches for doubly charged Higgs rule out $m_{\Delta}\lesssim 200-400\GeV$~\cite{Chatrchyan:2012ya,ATLAS:2012hi,Aaboud:2018qcu},
we consider such masses for completeness and validation against previous results.
In addition, we set the remaining free parameters to
\begin{equation}
  \lambda_{h\Delta1} = \lambda_{\Delta1} = 1, \quad
  v_\Delta = 1\times10^{-8}\GeV,
\end{equation}
and keep the remaining inputs at their default values.
We execute an on-the-fly parameter scan with \mgamc~using the following steering commands
\begin{verbatim}
  set mdpp scan1:range(100,1950,50)
  set mdp  scan1:range(100,1950,50)
  set md0  scan1:range(100,1950,50)
  set lamHD1 1.0
  set lamD1  1.0
  set vevD   1e-8
\end{verbatim}
The first three lines enforce the equality of all the masses in Eq.~\ref{eq:degenerateMasses} when they are varied
in  $\delta m_{\Delta} = 50\GeV$ increments over the range
\begin{equation}
m_{\Delta}\in[100\GeV,~1950\GeV).
\end{equation}

\section{Doubly Charged Higgs Boson Production at Hadron Colliders}\label{sec:results}
\begin{figure*}
 \subfigure[]{\includegraphics[width=.47\textwidth]{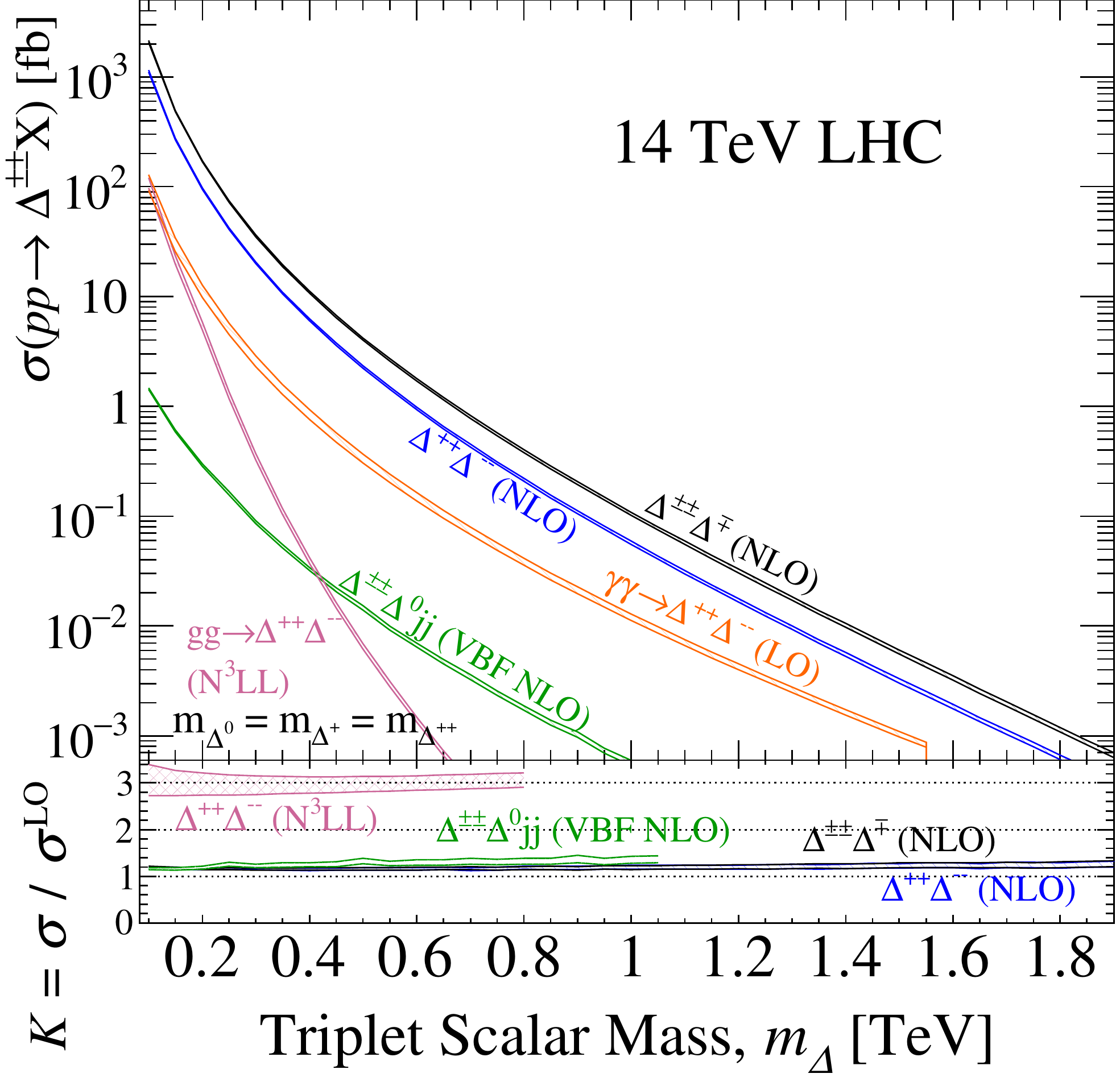}
   \label{fig:typeIInlo_XSec_vs_mD_LHCX14}}
   \hfill
 \subfigure[]{\includegraphics[width=.47\textwidth]{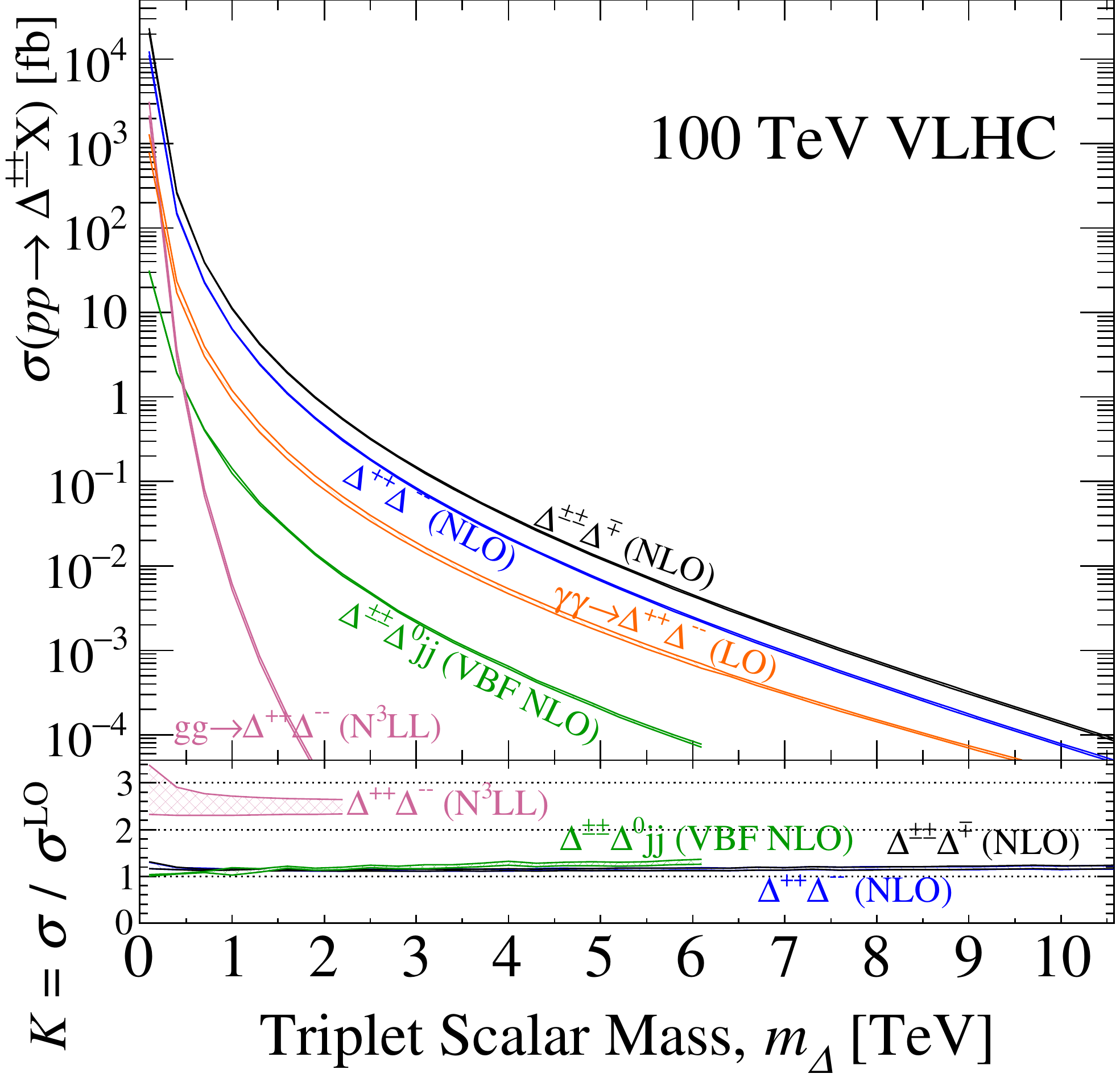}
   \label{fig:typeIInlo_XSec_vs_mD_LHC100}}
\caption{
  Upper panel: Inclusive cross section $\sigma$ (in fb) associated with the
    production of a pair of triplet scalars through the DY currents at NLO in
    QCD, the AF mechanism at LO, and the GF channel at N$^3$LL,
    and the  $\dpmpm\dxx j j$ process with VBF cuts at NLO in QCD. The
    results are given as a function of the degenerate triplet mass $m_\Delta$,
    given in TeV, and at (a) $\sqrt{s}=14\TeV$ and (b) $\sqrt{s}=100\TeV$.
  Lower panel: The associated QCD $K$-factor defined by Eq.~\ref{eq:kfact}.
  For all curves, the band thickness corresponds to the residual scale
  uncertainty.
}
\label{fig:typeIInloXSec}
\end{figure*}

We now present our modeling prescriptions for the production of doubly charged Higgs bosons $(\dpmpm)$ at hadron colliders through various mechanisms,
and compare their relative importance at the $\sqrt{s}=14\TeV$ LHC and a hypothetical $\sqrt{s}=100\TeV$ Very Large Hadron Collider (VLHC).

Explicitly, we consider in Sec.~\ref{sec:results_DYX} the charged current and neutral current DY processes given in Eqs.~\ref{eq:dyAssociated} and \ref{eq:dyPair},
 and shown in Figs.~\ref{fig:diagrams}(a) and (b).
For the DY channels, we give special attention to jet veto cross sections assuming a static jet veto.
In Sec.~\ref{sec:results_AFX}, we turn to the AF channel as shown in Fig.\ref{fig:diagrams}(d) and given in Eq.~\ref{eq:afPair},
and investigate various sources of cross section uncertainties, particularly those stemming from the photon PDF modeling.
In Sec.~\ref{sec:results_GFX}, we consider the GF mechanism shown in Fig.\ref{fig:diagrams}(c) and given in Eq.~\ref{eq:gfPair}.
Finally, we investigate the EW production of $\dpmpm\dxx j j$ in Sec.~\ref{sec:results_VBF}.
For this channel, we introduce a procedure for imposing generator-level VBF cuts at NLO within the \mgamc~formalism,
and use the label ``VBF NLO'' to distinguish it from more inclusive collider signatures.
We also estimate the discovery potential of the
Type II Seesaw at the LHC and a hypothetical VLHC in Sec.~\ref{sec:results_sensitivity}.

The main results of this section are summarized in Fig.~\ref{fig:typeIInloXSec}.
In the upper panel of both subfigures, we present, as a function of the
degenerate triplet mass defined in Eq.~\ref{eq:degenerateMasses} and for a
center-of-mass energy of $\sqrt{s}=14\TeV$ (a) and $100\TeV$ (b), the inclusive
production cross section of the DY currents at NLO, the AF mechanism at
LO, and the GF channel at N$^3$LL, and the $\dpmpm\dxx j j$ process with
VBF cuts at NLO. In the lower panel of the subfigures, we report the
corresponding QCD $K$-factor
\begin{equation}
 K^{\rm NLO}   ~\equiv~ \sigma^{\rm NLO}   / \sigma^{\rm LO}\ , \quad
 K^{\rm N^3LL} ~\equiv~ \sigma^{\rm N^3LL} / \sigma^{\rm LO}\ ,
\label{eq:kfact}\end{equation}
defined with respect to LO cross sections $\sigma^{\rm LO}$ obtained by
convoluting LO matrix elements with NLO PDFs\footnote{This choice can lead to
LO predictions that are systematically lower by $\mathcal{O}(10\%-20\%)$ than
those obtained by considering LO PDFs, due to differences in the PDF
normalizations.}.
For all curves in Fig.~\ref{fig:typeIInloXSec}, the band thickness corresponds to the residual scale uncertainty.
For representative triplet masses, we report the same information in Table~\ref{tb:xSecBenchmark}.

Throughout this section, we detail our usage of the \libName~libraries within \mgamc, including the syntax needed to readily reproduce our results.

\subsection{Triplet Scalars from Drell-Yan Annihilation}\label{sec:results_DYX}

\begin{table*}
  \begin{center}
  \renewcommand{\arraystretch}{1.9}
  \setlength\tabcolsep{10pt}
  \resizebox{\textwidth}{!}{
    \begin{tabular}{ c  c  || c  c  c | c  c  c || c  c  c | c  c  c}
      \multicolumn{2}{c||}{} &
        \multicolumn{6}{c||}{$\sqrt{s}=$14 TeV}&\multicolumn{6}{c}{$\sqrt{s}=$100 TeV}\tabularnewline
      \multicolumn{2}{c||}{}&
        \multicolumn{3}{c|}{$m_\Delta = 550\GeV$}&
        \multicolumn{3}{c||}{$m_\Delta = 1\TeV$}&
        \multicolumn{3}{c|}{$m_\Delta = 550\GeV$}&
        \multicolumn{3}{c}{$m_\Delta = 1\TeV$}\tabularnewline
      \hline
      \multicolumn{2}{c||}{Process}    &
        $\sigma^{\rm LO}$ [fb]  & $\sigma^{\rm NLO}$ [fb]  & $K$ &
        $\sigma^{\rm LO}$ [ab]  & $\sigma^{\rm NLO}$ [ab]  & $K$ &
        $\sigma^{\rm LO}$ [fb]  & $\sigma^{\rm NLO}$ [fb]  & $K$ &
        $\sigma^{\rm LO}$ [fb]  & $\sigma^{\rm NLO}$ [fb]  & $K$  
        \tabularnewline\hline\hline
$\Delta^{\pm\pm}\Delta^\mp$ & DY  			& 2.28 				&	2.66$^{+2\%}_{-2\%}$	&	1.17
									& 89.0				&	106$^{+3\%}_{-4\%}$	&	1.19
									& 79.2				&	90.4$^{+1\%}_{-2\%}$	&	1.14
									& 9.81				&	11.1$^{+1\%}_{-1\%}$	&	1.13
\tabularnewline\hline 
$\Delta^{++}\Delta^{--}$ & DY  	 		& 1.26 				&	1.47$^{+2\%}_{-3\%}$		& 1.16
								& 47.9				&	57.3$^{+3\%}_{-4\%}$		& 1.20
								& 44.9				&	51.8$^{+1\%}_{-2\%}$		& 1.15
								& 5.65				&	6.42$^{+1\%}_{-1\%}$		& 1.14
\tabularnewline\hline 						
$\Delta^{++}\Delta^{--}$ & AF   	& 0.244				& $-$	&	$-$
						& 12.9				& $-$	&	$-$
						& 8.84				& $-$	&	$-$
						& 1.21				& $-$	&	$-$
\tabularnewline\hline 						
$\Delta^{\pm\pm}\Delta^0jj$ & VBF	& 7.54$\times10^{-3}$	&	9.59$\times10^{-3}~^{+5\%}_{-5\%}$		& 1.27
							& 417$\times10^{-3}$	&	557$\times10^{-3}~^{+5\%}_{-6\%}$		& 1.34
							& 752				&	807$^{+1\%}_{-<0.5\%}$				& 1.07
							& 120				&	132$^{+1\%}_{-1\%}$				& 1.10
\tabularnewline\hline 
$\Delta^{++}\Delta^{--}$ & GF 	& 988$\times10^{-6}$	& 3.04$\times10^{-3}~^{+2\%}_{-8\%}$			&	3.07
						& 4.08$\times10^{-3}$	& 13.2$\times10^{-3}~^{+2\%}_{-8\%}$			&	3.24
						& 0.154				& 0.403$^{+8\%}_{-12\%}$					& 	2.61
						& 2.26$\times10^{-3}$	& 5.78$\times10^{-3}~^{+6\%}_{-10\%}$			& 	2.56
\end{tabular}
} 
\caption{For representative triplet masses $m_{\Delta}$ and center-of-mass energy $\sqrt{s}$,
triplet scalar production cross section $\sigma$ [fb] at various accuracies in QCD via the DY, AF, and GF
processes, as well as the EW $\dpmpm\dxx j j$ process with VBF cuts.
For computations beyond LO, the QCD $K$-factors $(K)$ and  the residual scale uncertainties $[\%]$ are also reported.
}
\label{tb:xSecBenchmark}
\end{center}
\end{table*}

As a baseline process for our work and comparison to past work, we start with triplet scalar production via the DY quark-antiquark annihilation mechanism.
In our framework,  $\dpmpm\dmp$ associated production via the CC DY process 
and $\dpp\dmm$ pair production via the NC DY process can be modeled up to NLO in QCD via the respective 
\mgamc~commands\footnote{For a more complete description of the software suite \mgamc, its underlying construction, and its usage, see Ref.~\cite{Alwall:2014hca}.}:
\begin{verbatim}
import model TypeII_NLO_v1_2_UFO
define p = g u c d s u~ c~ d~ s~ 
define dxx = d++ d--
define dx  = d+  d-

generate p p > dxx dx QED=2 QCD=0 [QCD]
output TypeIInlo_DYX_DxxDx_NLO

generate p p > dxx dxx QED=2 QCD=0 [QCD]
output TypeIInlo_DYX_DxxDxx_NLO
\end{verbatim}
The first line in the above syntax imports the \libName~UFO library into \mgamc, and the three subsequent lines defines multiparticle objects to streamline our setup.
Here, we exclude photons \texttt{a} from the multiparticle definition of a proton \texttt{p}.
The two \texttt{generate} commands, together with the associated \texttt{[QCD]}
flag, allow for the generation of the tree-level and loop-level helicity
amplitude routines describing the DY processes at $\mathcal{O}(\alpha^2)$ and
$\mathcal{O}(\alpha^2\alpha_s)$. This moreover includes automatically the
$\mathcal{O}(\alpha_s)$ subtraction terms needed for parton shower-matching
within the MC@NLO formalism.

As a function of triplet masses $(m_{\Delta})$, 
we present in Fig.~\ref{fig:typeIInloXSec} the NLO in QCD production cross section [fb], their residual scale uncertainty (band thickness), and the NLO $K$-factor
 for the CC and NC DY processes at (a) $\sqrt{s}=14\TeV$ and (b) $100\TeV$.
For the triplet mass range  $m_\Delta \approx 100-2000\GeV~(0.1-10\TeV)$, the NLO production rates span at 14 (100) TeV:
\begin{eqnarray}
 \DYCC	&:& {2.1\pb-0.67\ab	~\quad(22\pb-140\zb)},\\
 \DYNC	&:& {1.1\pb-0.38\ab	~\quad(12\pb-76\zb)},
\end{eqnarray}
with corresponding scale uncertainties of about
\begin{eqnarray}
\text{CC,~NC DY} : {\pm 1\%-\pm5\%~\quad(\pm1\%-\pm7\%)},
\label{eq:dyScaleUnc}
\end{eqnarray}
and likewise nearly identical $K$-factors
\begin{eqnarray}
\DYCC 	&:& {1.15-1.27~(1.13-1.26)},\\
\DYNC  	&:& {1.15-1.26~(1.12-1.25)}.
\end{eqnarray}
At LO, we successfully reproduce the Tevatron and LHC cross section predictions for the DY channels as reported in Refs.~\cite{Muhlleitner:2003me,Han:2007bk}.
At NLO, we report agreement with the well-known calculation of Ref.~\cite{Muhlleitner:2003me}.
As a further check, we have also computed the NC DY process at NLO using the phase space slicing method~\cite{Harris:2001sx}, as implemented in Ref.~\cite{Ruiz:2015zca}.
We find good agreement across all three NLO computations.
We briefly remark that the $\mathcal{O}(100)$ zb cross sections at $\sqrt{s}=100\TeV$ are incredibly tiny.
However, the proposed integrated luminosity goals for such potential machines
are of $\mathcal{L}=30-50\invab$~\cite{Arkani-Hamed:2015vfh,CEPC-SPPCStudyGroup:2015csa,Golling:2016gvc,Benedikt:2018csr,Abada:2019ono}, 
indicating that $\mathcal{O}(3-5)$ triplet pairs with masses of
$\mathcal{O}(10)\TeV$ could be produced in such a collider.

\subsubsection{Flavor Scheme Dependence}\label{sec:results_DYFS}

Throughout this study, we work in the $n_f=4$ active quark flavor scheme (4FS) with variable flavor scheme PDFs.
We assume that all third generation fermions are massive (see Eq.~\ref{eq:smInputs}),
and do so to consistently describe decays of $\tau$ leptons and $b$-flavored hadrons when using the \libName~libraries.
However, for the $m_\Delta$ under consideration, $(m_b/m_\Delta)^2\ll1$.
Hence, modeling the proton with $n_f=5$ massless quark flavors (5FS), \ie, with a $b$ quark PDF, 
is arguably more appropriate.

We investigate the impact of this assumption by simulating the CC and NC DY processes in the 5FS.
For representative $m_\Delta$, we report in Table~\ref{tb:xSecBenchmarkFS} the same information 
as reported in Table~\ref{tb:xSecBenchmark} for the CC DY and NC DY channels in the 4FS and 5FS.
At $\sqrt{s}=14\TeV$ and for $m_\Delta = 550\GeV$ and $1\TeV$, we find subpercent differences that 
are consistent with MC uncertainties.
At $\sqrt{s}=100\TeV$, we observe the same quantitative behavior for the CC DY mode;
for the NC DY process, we find a larger cross section in the 5FS by $\delta\sigma/\sigma =1-2\%$.

The origin of this difference is physical and can be attributed to the relevant diagrams contributing to 
the NC DY process: In the 4FS, the heavy quark fusion subprocess 
$b\overline{b}\to \gamma^*/Z^* \to \dpp\dmm$ does not occur at LO and arises only at 
$\mathcal{O}(\alpha_s^2)$ as a component of the $gg \to \dpp\dmm b\overline{b}$ production.
Such contributions from initial-state $g\to b\overline{b}$ splittings, however, 
are not power suppressed and are precisely factorized into $b$ quark PDFs in the 
5FS~\cite{Aivazis:1993pi,Collins:1998rz,Collins:2011zzd}.
Hence, $b\overline{b}\to \gamma^*/Z^* \to \dpp\dmm$ is a LO channel in the 5FS.
The apparently negligible impact of $b$ quark PDFs at $\sqrt{s}=14\TeV$ follows from the need for 
large-$x$ partons to make TeV-scale $\dpp\dmm$ pairs.
High-mass DY processes are driven by valence quark-sea antiquark annihilation
and since $b$ quark PDFs are generated perturbatively, \ie, are not intrinsic,
they mirror the gluon PDF, which is more concentrated at Bjorken $x\ll1$. 
In short, $b$ PDFs are negligible at $x\lesssim1$.
For fixed $\dpmpm$ masses, the requisite Bjorken $x$ are smaller at $\sqrt{s}=100\TeV$, 
leading to $b$ PDFs playing a more relevant role.
The $\delta\sigma/\sigma =1-2\%$ larger rates we find in the 5FS sit within the scale uncertainties of the 4FS results at NLO,
consistent with $\mathcal{O}(\alpha_s^2)$ corrections.
At a differential level, we anticipate larger changes  than those found for inclusive cross sections.
However, such checks are beyond our present scope.

For the CC DY process, we do not observe differences in total cross sections between the 4FS and 5FS  
because we assume the CKM matrix to be diagonal.
This means that $b$ quark PDFs are only relevant at tree level when they can be paired up with their isospin 
doublet partner, \ie, when $t$ quark PDFs are relevant. 
For $m_\Delta$ under consideration, the $t\overline{b}$ and $\overline{t}b$ parton luminosities are either not applicable or tiny~\cite{Dawson:2014pea,Han:2014nja}.
For the AF and VBF production mechanisms, we anticipate similar differences between working in the 4FS or 5FS, and therefore only report results in the 4FS.
The impact on the GF mode is briefly assessed in Sec.~\ref{sec:results_GFFS}. 
To aid future investigations, a 5FS version of the \libName~libraries is also publicly available.

\begin{table*}
  \begin{center}
  \renewcommand{\arraystretch}{1.9}
  \setlength\tabcolsep{10pt}
  \resizebox{\textwidth}{!}{
    \begin{tabular}{ c  c || c  c  c | c  c  c || c  c  c | c  c  c}
      \multicolumn{2}{c||}{} &
        \multicolumn{6}{c||}{$\sqrt{s}=$14 TeV}&\multicolumn{6}{c}{$\sqrt{s}=$100 TeV}\tabularnewline
      \multicolumn{2}{c||}{}&
        \multicolumn{3}{c|}{$m_\Delta = 550\GeV$}&
        \multicolumn{3}{c||}{$m_\Delta = 1\TeV$}&
        \multicolumn{3}{c|}{$m_\Delta = 550\GeV$}&
        \multicolumn{3}{c}{$m_\Delta = 1\TeV$}\tabularnewline
      \hline
      	\multicolumn{2}{c||}{Process}    &
        $\sigma^{\rm LO}$ [fb]  & $\sigma^{\rm NLO}$ [fb] 	 $\quad\delta\sigma_{\rm PDF}$ & $K$ &
        $\sigma^{\rm LO}$ [ab]  & $\sigma^{\rm NLO}$ [ab] 	 $\quad\delta\sigma_{\rm PDF}$ & $K$ &
        $\sigma^{\rm LO}$ [fb]  & $\sigma^{\rm NLO}$ [fb] 	 $\quad\delta\sigma_{\rm PDF}$ & $K$ &
        $\sigma^{\rm LO}$ [fb]  & $\sigma^{\rm NLO}$ [fb] 	 $\quad\delta\sigma_{\rm PDF}$ & $K$  
        \tabularnewline\hline\hline
$\Delta^{\pm\pm}\Delta^\mp$ & DY-4FS 		& 2.28	&	2.65$^{+2\%}_{-2\%}$  $\quad^{+4\%}_{-5\%}$	&	1.16
									& 89.0	&	106$^{+3\%}_{-4\%}$  $\quad^{+5\%}_{-5\%}$	&	1.19
									& 79.2	&	90.4$^{+1\%}_{-2\%}$  $\quad^{+2\%}_{-2\%}$	&	1.14
									& 9.81	&	11.1$^{+1\%}_{-1\%}$  $\quad^{+2\%}_{-2\%}$	&	1.13
\tabularnewline\hline 
$\Delta^{++}\Delta^{--}$ & DY-4FS 	& 1.26 				&	1.47$^{+2\%}_{-3\%}$  $\quad^{+4\%}_{-4\%}$		& 1.16
							& 47.9		&	57.3$^{+3\%}_{-4\%}$  $\quad^{+6\%}_{-5\%}$		& 1.20
							& 44.9		&	51.8$^{+1\%}_{-2\%}$  $\quad^{+2\%}_{-2\%}$		& 1.15
							& 5.65		&	6.42$^{+1\%}_{-1\%}$  $\quad^{+2\%}_{-2\%}$		& 1.14
\tabularnewline\hline 						
$\Delta^{\pm\pm}\Delta^\mp$ & DY-5FS 		& 2.28	&	2.65$^{+2\%}_{-2\%}$  $\quad^{+4\%}_{-4\%}$	&	1.16
									& 89.0	&	106$^{+3\%}_{-4\%}$  $\quad^{+5\%}_{-5\%}$	&	1.19
									& 79.0	&	90.7$^{+1\%}_{-2\%}$  $\quad^{+2\%}_{-2\%}$	&	1.15
									& 9.81	&	11.1$^{+1\%}_{-1\%}$  $\quad^{+2\%}_{-2\%}$	&	1.13
\tabularnewline\hline 						
$\Delta^{++}\Delta^{--}$ & DY-5FS 		& 1.27	&	1.47$^{+2\%}_{-2\%}$  $\quad^{+4\%}_{-4\%}$		& 1.16
								& 48.0	&	57.4$^{+3\%}_{-3\%}$  $\quad^{+6\%}_{-5\%}$		& 1.20
								& 45.8	&	52.8$^{+1\%}_{-2\%}$  $\quad^{+2\%}_{-2\%}$		& 1.15
								& 5.72	&	6.49$^{+1\%}_{-1\%}$  $\quad^{+2\%}_{-2\%}$		& 1.13
\tabularnewline\hline 
$\Delta^{++}\Delta^{--}$ & GF-4FS		& 988$\times10^{-6}$	& 3.04$\times10^{-3}~^{+2\%}_{-8\%}$ 	$^{+6\%}_{-6\%}$ 			&	3.07
								& 4.08$\times10^{-3}$	& 13.2$\times10^{-3}~^{+2\%}_{-8\%}$ 	$^{+13\%}_{-11\%}$			&	3.24
								& 0.154				& 0.403$^{+8\%}_{-12\%}$		    	$^{+1\%}_{-1\%}$			& 	2.61
								& 2.26$\times10^{-3}$	& 5.78$\times10^{-3}~^{+6\%}_{-10\%}$ 	$^{+2\%}_{-2\%}$ 			& 	2.56
\tabularnewline\hline 
$\Delta^{++}\Delta^{--}$ & GF-5FS		& 983$\times10^{-6}$	& 3.02$\times10^{-3}~^{+2\%}_{-8\%}$ 	$^{+6\%}_{-6\%}$ 			&	3.07
								& 4.05$\times10^{-3}$	& 13.1$\times10^{-3}~^{+2\%}_{-8\%}$ 	$^{+13\%}_{-11\%}$			&	3.24								
								& 0.154				& 0.401$^{+8\%}_{-12\%}$		    	$^{+1\%}_{-1\%}$			& 	2.61
								& 2.25$\times10^{-3}$	& 5.75$\times10^{-3}~^{+6\%}_{-10\%}$ 	$^{+2\%}_{-2\%}$ 			& 	2.56
\end{tabular}
} 
\caption{
Same as Table~\ref{tb:xSecBenchmark} but for the CC DY, NC DY, and GF processes, in the  4FS and 5FS and with PDF uncertainties.
}
\label{tb:xSecBenchmarkFS}
\end{center}
\end{table*}

\subsubsection{PDF Dependence}\label{sec:results_DYPDF}

Due to the smallness of the residual scale uncertainties found for the CC and NC DY cross sections (see Eq.~\ref{eq:dyScaleUnc}), 
it is helpful to also investigate the relative magnitude of uncertainties associated with our PDF choices.
To quantify this uncertainty, we follow the prescription~\cite{Harland-Lang:2019pla} for the MMHT15qed set,
which is based on the Hessian method~\cite{Pumplin:2001ct} as implemented~\cite{Martin:2009iq} in \textsc{LHAPDF}~\cite{Buckley:2014ana}.
For the same triplet masses $(m_\Delta)$ and collider energies $(\sqrt{s})$ in Table~\ref{tb:xSecBenchmark},
we report in Table~\ref{tb:xSecBenchmarkFS} the PDF uncertainties [\%] for the two channels (in both the 4FS and 5FS).

Overall, we find that PDF uncertainties roughly span
\begin{equation}
  \text{CC, NC DY} \quad:\quad \pm2\% - \pm6\%,
\end{equation}
are symmetric and slightly larger than the reported scale uncertainties.
We find the PDF uncertainties for the CC and NC modes to be essentially the same.
Likewise, the uncertainties for the 4FS and 5FS are comparable.

We extrapolate to other triplet masses and collider energies by noting 
that, due to scale invariance, PDFs are functions of (dimensionless) momentum fractions $(\xi_1,\xi_2)$.
Due to momentum conservation, the $\xi_i$ must satisfy,
\begin{equation}
\min(\xi_1 \xi_2) = \tau_{\min}\equiv (2m_\Delta)^2/s.
\end{equation}
Here, $\tau_{\min}$ is the (dimensionless) threshold below which $\dpp\dmm$ pair production is kinematically forbidden.
For the $m_\Delta$ in Table~\ref{tb:xSecBenchmark}, this corresponds to 
\begin{equation}
  \tau_{\min} \approx 1\cdot10^{-4}, ~4\cdot10^{-4}, ~6\cdot10^{-3}, ~0.2.
\end{equation}
Now, for such thresholds, the associated (geometric) averages of the momentum fractions $(\xi_\star)$ are
\begin{equation}
\xi_\star = \sqrt{\tau_{\min}} \approx 0.01, ~0.02, ~0.08, ~0.1,
\label{eq:results_PDFXi}
\end{equation}
indicating that for the range of $\xi_\star$ that we consider, and hence the range of $m_\Delta$,
PDF uncertainties span:
\begin{eqnarray}
\text{DY-Low mass} 				\quad&:&\quad \pm2\%,  \\
\text{DY-Intermediate mass} 		\quad&:&\quad \pm4\%-\pm5\%,  \\
\text{DY-High mass} 				\quad&:&\quad \pm5\%-\pm6\%.
\end{eqnarray}
For much lower and much higher masses than those assumed, 
we anticipate  larger uncertainties due to poorer constraining power of PDF fits.

Since the VBF channel features parton luminosities at LO and NLO that are comparable to the DY luminosity, 
we expect similar PDF uncertainties and do not  investigate the matter further.
For the AF and GF mechanisms, we report PDF uncertainties in Secs.~\ref{sec:results_photonPDF} and \ref{sec:results_GFFS}.

\subsubsection{Triplet Scalar Production with Jet Vetoes at NLO+NNLL}\label{sec:results_DYXveto}

\begin{figure*}
  \subfigure[]{	\includegraphics[width=0.47\textwidth]{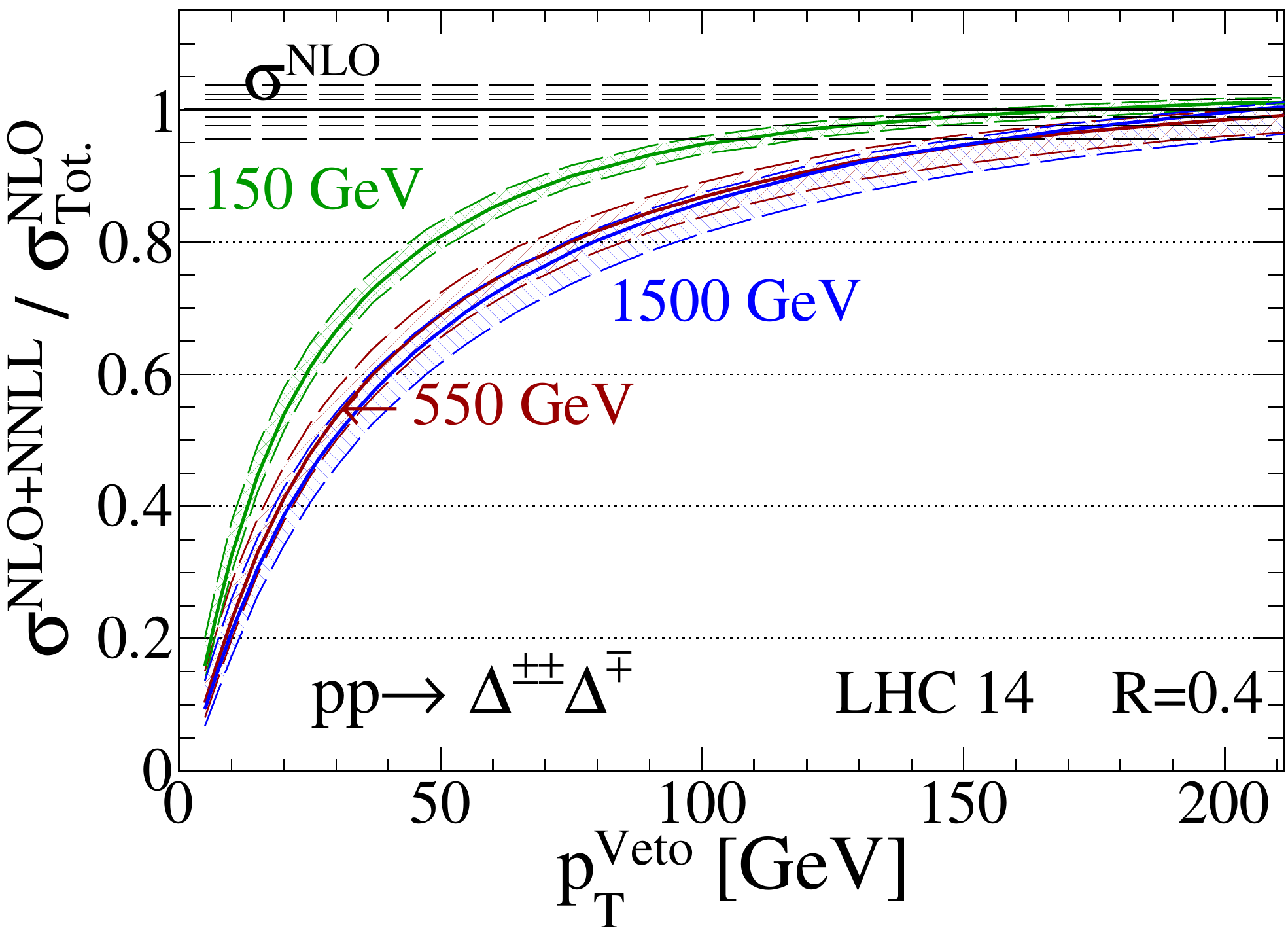}		\label{fig:vetoXSec_DxxDx}	}
  \hfill
  \subfigure[]{	\includegraphics[width=0.47\textwidth]{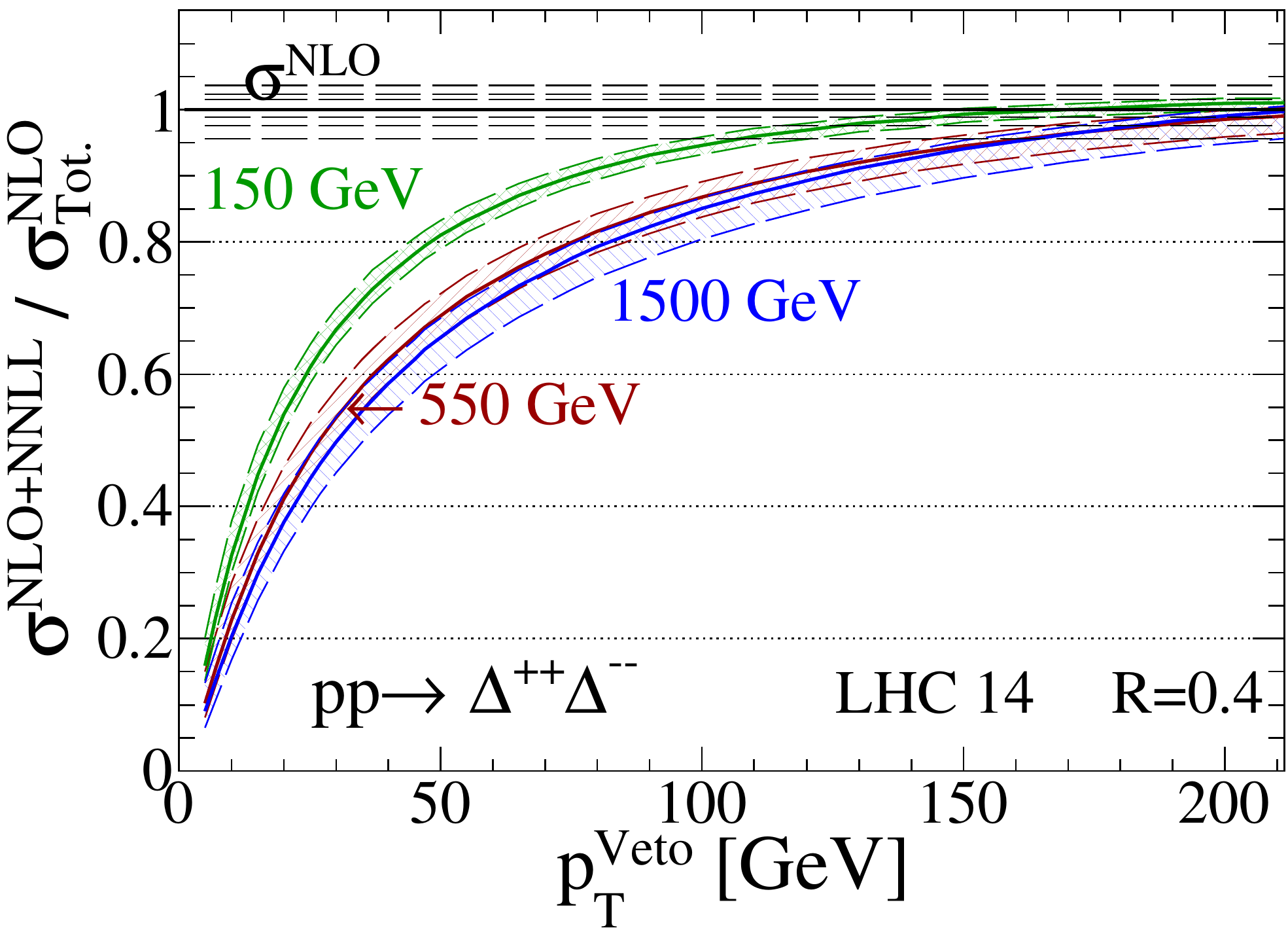}		\label{fig:vetoXSec_DxxDxx}	}
\caption{
  Jet veto efficiency at NLO+NNLL, as defined in Eq.~\ref{eq:vetoEffDef} for
  the (a) charged current process $pp\to \dpmpm\dmp$ and (b) neutral current
  process $pp\to\dpp\dmm$, as a function of jet veto threshold $p_T^{\rm Veto}$,
  for several representative triplet masses and at $\sqrt{s}=14$ TeV. The
  results are normalized to the NLO inclusive cross section and the error bands
  refer to scale uncertainties.
}
\label{fig:vetoXSec}
\end{figure*}

Following significant improvements in modeling soft jet activity in both color-singlet and QCD processes,
the use of jet vetoes in direct searches for new phenomena is becoming increasingly standard at the LHC, 
as shown, for example, in Refs.~\cite{Tackmann:2016jyb,Fuks:2017vtl,Pascoli:2018rsg,Pascoli:2018heg,Fuks:2019iaj,Arpino:2019fmo}.
With the \libName~libraries it is possible to predict jet veto cross sections for triplet scalars up to NLO+NNLL(veto)
using the \mgFull\textsc{-SCET} formalism of Refs.~\cite{Becher:2012qa,Becher:2013xia,Becher:2014aya}.

While a systematic study of soft QCD activity in triplet scalar production at the LHC is beyond the scope of this work,
we nevertheless briefly explore the impact of (static) jet vetoes in the CC and NC DY processes.
For the charged current process, we do so explicitly by steering \mgamc~using the following commands,
\begin{verbatim}
  launch TypeIInlo_DYX_DxxDx_NLO
  order=NLO
  fixed_order=ON
  set req_acc_fo 0.001
  set LHC 14
  set pdlabel lhapdf
  set lhaid 26300
  set fixed_ren_scale False
  set fixed_fac_scale False
  set dynamical_scale_choice -1
  set no_parton_cut
  set jetalgo    -1
  set jetradius 0.4
  set ptj X
  set ickkw -1
\end{verbatim}
This procedure differs from other cross section evaluations in this study in a few aspects.
First, in order to avoid any potential $\mathcal{O}(\alpha_s^2)$ double counting, we match the NLO and NLO+NNLL(veto) computations to the MMHT2015 QED NNLO PDF set (\texttt{lhaid=26300}).
Second, we define jets using the anti-$k_T$ algorithm~\cite{Cacciari:2008gp}, a radius parameter $R=0.4$, and a variable transverse momentum ($p_T$) threshold spanning $p_T^{\rm Veto}=5-250\GeV$.
This serves as the threshold above which jets are vetoed (see Ref.~\cite{Becher:2014aya} for details).
No rapidity cut on jets is imposed here as the NNLL(veto) computation is only singly differential in $p_T$ and hence is inclusive over the observable~\cite{Becher:2014aya}.
An independent study by Ref.~\cite{Michel:2018hui} shows that this is a reasonable approximation for jet veto windows that extend out to $\eta^{\max}\sim4.5$.
For jet vetoes limited to $\eta^{\max}\lesssim2.5$, the computation grows more sensitive to higher order QCD radiation~\cite{Michel:2018hui}.

In Fig.~\ref{fig:vetoXSec}, we present, as a function of the jet veto $p_T$
threshold $p_T^{\rm Veto}$, the jet veto efficiency defined
by
\begin{equation}
\varepsilon^{\rm NLO+NNLL}(p_T^{\rm Veto}) = \cfrac{\sigma^{\rm NLO+NNLL}(p_T^j < p_T^{\rm Veto})}{\sigma^{\rm NLO}_{\rm Total}},
\label{eq:vetoEffDef}
\end{equation}
at $\sqrt{s}=14$ TeV, 
for (a) $\dpmpm\dmp$~associated production and (b) $\dpp\dmm$~pair production
via the DY process at representative triplet scalar masses of 150, 550 and
1500~GeV. Our predictions include theoretical uncertainties originating from
scale variations.
In Eq.~\ref{eq:vetoEffDef}, $\sigma^{\rm NLO+NNLL}(p_T^j < p_T^{\rm Veto})$ is the jet veto cross section for triplet scalar production 
with precisely zero jets above $p_T^{\rm Veto}$ but is inclusive with respect to hadronic clusters possessing a transverse momentum below $p_T^{\rm Veto}$.
Similarly, $\sigma^{\rm NLO}_{\rm Total}$ is the total inclusive rate.
Also shown as overlapping lines at unity are the normalized $\sigma^{\rm NLO}_{\rm Total}$ curves and their respective scale uncertainties.

For both DY processes, we observe similar qualitative dependence on $p_T^{\rm Veto}$ and triplet masses.
For \confirm{$m_{\Delta}=150~(550)\GeV$,} we see that a static jet veto of $p_T^{\rm Veto}=20-40\GeV$ results in a signal efficiency of about 
\begin{equation}
\confirm{\varepsilon^{\rm NLO+NNLL} \approx 55\%-75\%~(40\%-60\%),}
\end{equation}
with uncertainties roughly reaching \confirm{just over 10\% for $p_T^{\rm Veto}=20\GeV$ to just under 10\% for $p_T^{\rm Veto}=40\GeV$.}
For significantly larger $m_{\Delta}$, we observe only small decreases in signal efficiencies and comparable uncertainties relative to $m_{\Delta}=550\GeV$.
For $m_{\Delta}=150~(550)\GeV$, we find that $\varepsilon^{\rm NLO+NNLL}$ surpasses the \confirm{80\% threshold only for $p_T^{\rm Veto}\gtrsim 50~(75)\GeV$}.
Taken together, we see that typical~\cite{Aaboud:2018xdt} experimental jet veto thresholds of $p_T^{\rm Veto}=20-40\GeV$, 
reduce triplet production cross sections severely and potentially discourage their use.
However, we note that the criteria needed to employ a so-called dynamic (or event-based) jet veto~\cite{Pascoli:2018rsg,Pascoli:2018heg,Fuks:2019iaj}, 
which can significantly improve jet veto efficiencies and background rejection rates, appear to be satisfied.
Investigations into dynamic jet vetoes in triplet scalar searches is left to future work.
Lastly, while we use $R=0.4$, previous studies report that larger (smaller) choices of $R$ can decrease 
(increase) perturbative QCD uncertainties, but at the potential cost of increasing the impact of 
nonperturbative uncertainties~\cite{Dasgupta:2007wa,Dasgupta:2014yra,Banfi:2015pju,Dasgupta:2016bnd,Fuks:2017vtl,Pascoli:2018rsg,Pascoli:2018heg}.

\subsection{Triplet Scalar Pairs from Photon Fusion}\label{sec:results_AFX}

Due to its large electromagnetic charge, the production of $\dpp\dmm$ pairs in $\gamma
\gamma$ scattering, as shown in Fig.~\ref{fig:diagrams}(d) and given in
Eq.~\ref{eq:afPair}, has long been discussed in the literature as a means to test the Type II Seesaw mechanism. 
Disagreements about the relative importance of the photon fusion (AF) channel, 
however, have also appeared regularly due to the evolution of modeling prescriptions of initial-state photons, and hence photon PDFs.
While the dominance of the DY channels over the AF process was thought to have been settled by Ref.~\cite{Han:2007bk},
 recent claims to the contrary~\cite{Babu:2016rcr,Ghosh:2017jbw,Ghosh:2018drw}  have reignited the issue.
In recent years, though, schemes to systematically combine the elastic and inelastic components of the photon PDF in the proton~\cite{Martin:2014nqa,Alva:2014gxa,Manohar:2016nzj,Manohar:2017eqh}
and their subsequence implementation into state-of-the-art PDF sets~\cite{Schmidt:2015zda,Manohar:2017eqh,Bertone:2017bme,Harland-Lang:2019pla},
have drastically reduced the uncertainty in photon PDF modeling.
As a result, we are in a position to definitively address the relative importance of the AF fusion channel at hadron colliders.

To model AF at LO, we use the \mgamc~commands
\begin{verbatim}
  generate a a > dxx dxx QCD=0 QED=2
  output TypeIInlo_aaF_DxxDxx_XLO
  launch
  shower=OFF
  madspin=OFF
  analysis=OFF
  set run_card nevents 100k
  set LHC 14
  set dynamical_scale_choice 3
  set use_syst True
  set no_parton_cut
  set pdlabel lhapdf
  set lhaid 26000
\end{verbatim}
As described in Sec.~\ref{sec:mcSetup_SM}, we use the MMHT+LUXqed NLO PDF set.
We explore the AF channel by first presenting in Fig.~\ref{fig:typeIInloXSec} the LO production cross section [fb] as a function of triplet masses $(m_{\Delta})$
with its scale uncertainty at (a) $\sqrt{s}=14\TeV$ and (b) $100\TeV$.	
For $m_\Delta \approx 100-1500\GeV~(0.1-6\TeV)$, the rates
and $\mu_f$ scale uncertainties span at $\sqrt{s}=14$ (100) TeV about
\confirm{
\begin{eqnarray}
\sigma_\AF		&:	110\fb-1.0\ab 		&~\quad(1.0\pb-65\ab), \\
\delta\sigma/\sigma 	&: \pm 15\%-\pm5\% 	&~\quad(\pm25\%-\pm6\%).
\end{eqnarray}}
We report that the AF cross section is much smaller than the NC DY channel.
Over the  $m_\Delta$ mass range we consider,
the rate differs at $\sqrt{s}=14$ (100) TeV by 
\confirm{
\begin{eqnarray}
\sigma_{\rm NC~DY}^{\rm NLO}/\sigma_{\rm AF}^{\rm LO} ~&\approx&~ 10\!-\!3~(12\!-\!2).
\end{eqnarray}}
As the DY NLO $K$-factors reach only values of $K^{\rm NLO}\approx 1.2-1.3$,
for the considered $m_\Delta$ values (see Sec.~\ref{sec:results_DYX}),
the absence of AF-dominance cannot be attributed to our inclusion NLO corrections.
To further investigate the claims of Ref.~\cite{Babu:2016rcr,Ghosh:2017jbw,Ghosh:2018drw},
we consider the uncertainties associated with our  photon PDF choice.

\begin{figure*}
  \subfigure[]{	\includegraphics[width=0.47\textwidth]{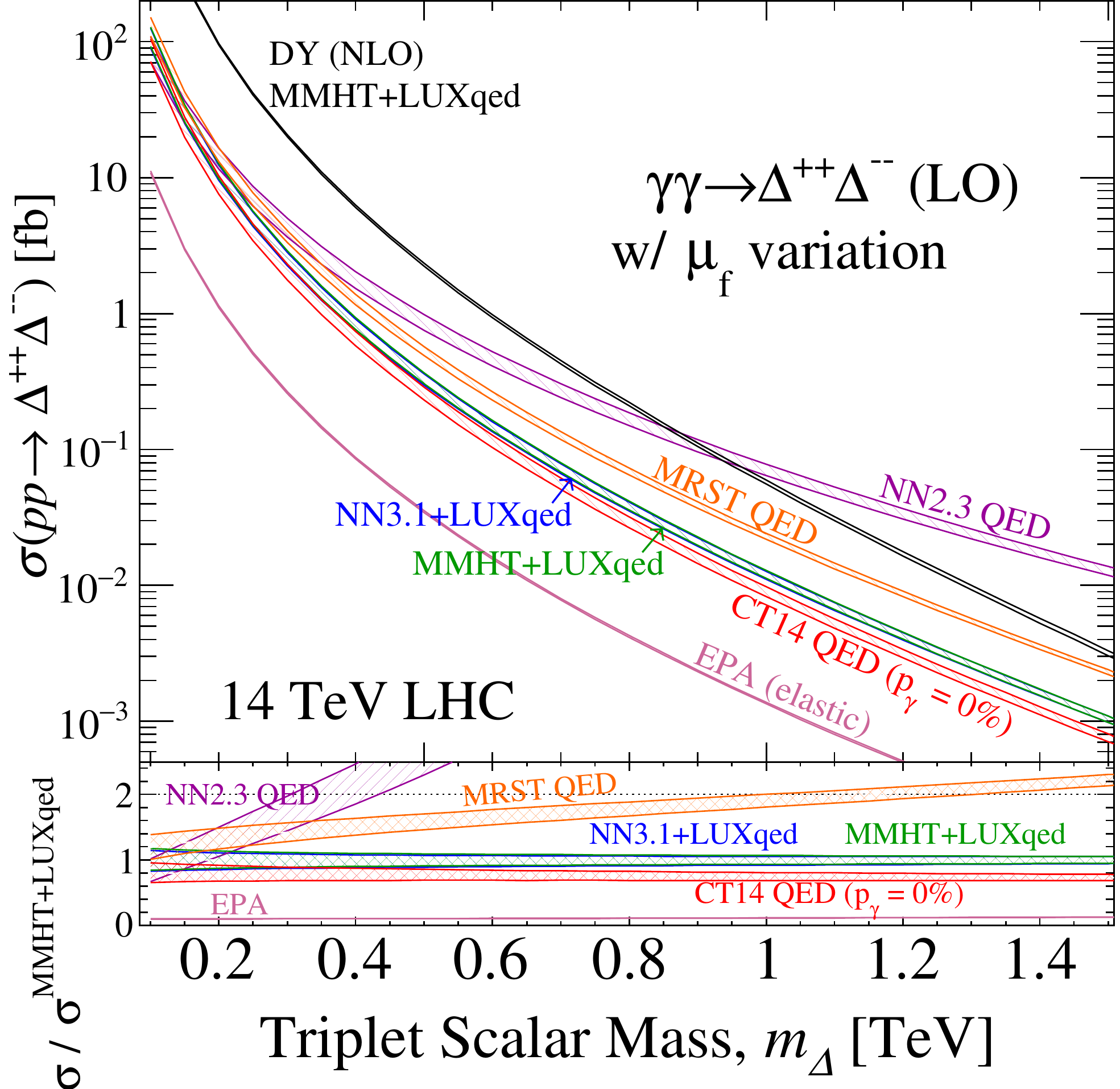}		\label{fig:typeIInlo_AFX_XSec_ScaleUnc_vs_mD_LHCX14}	}
  \hfill
  \subfigure[]{	\includegraphics[width=0.47\textwidth]{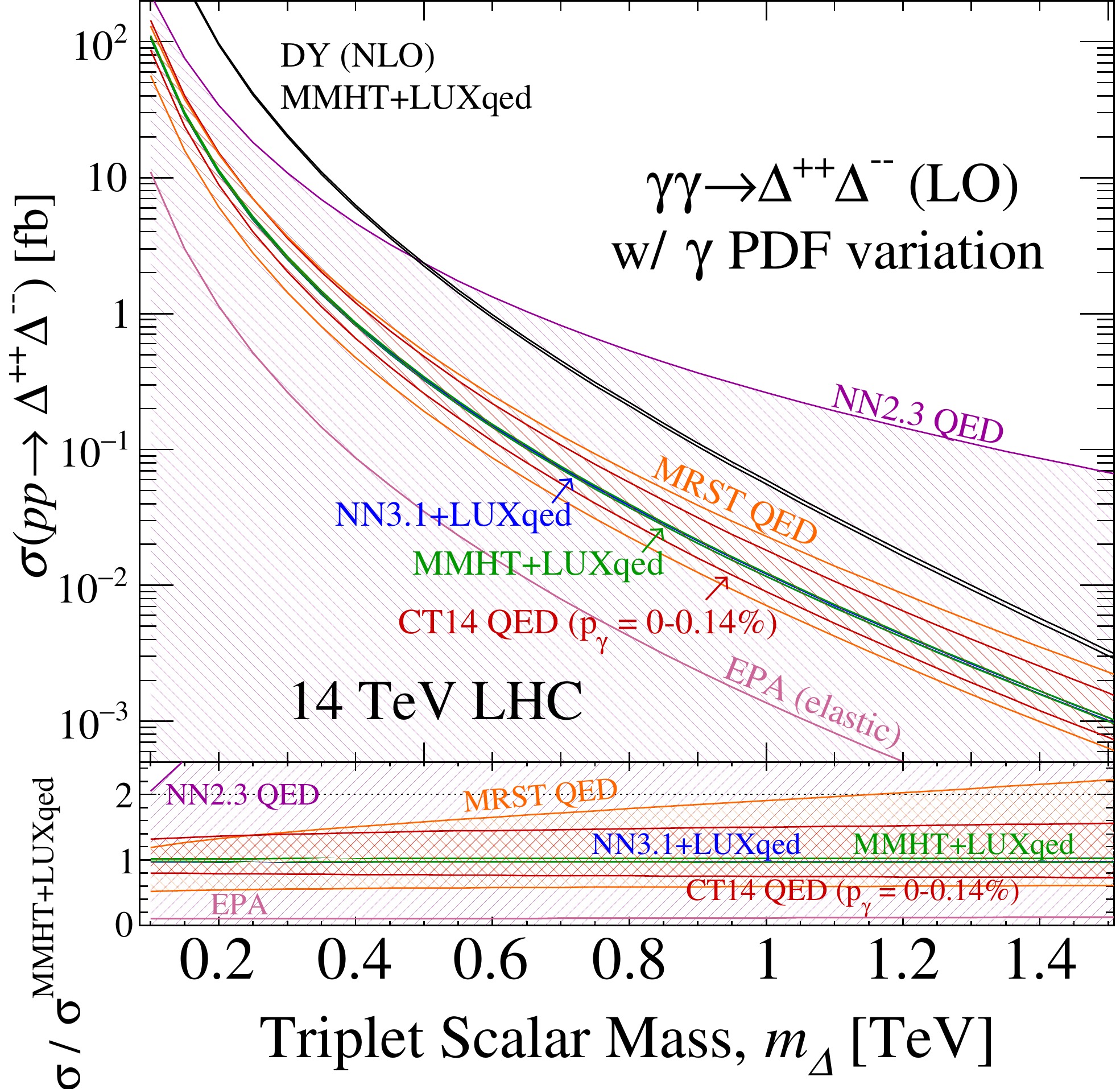}		\label{fig:typeIInlo_AFX_XSec_PDFUnc_vs_mD_LHCX14}	}
\caption{
  Upper panel: Total $\dpp\dmm$ production cross section [fb] from AF at
  $\sqrt{s}=14\TeV$ as a function of the triplet mass $m_{\Delta}$ and
  assuming various options for the modeling of the photon density. The
  uncertainty bands are originating from (a) variations of the factorization
  scale $\mu_f$ and (b) variations of the PDF fit. The results are compared with
  the NC DY production rate at NLO (as computed with the MMHT+LUXqed PDF set).
  Lower panel: Ratio of the AF cross section relatively to the reference results
  obtained with the MMHT+LUXqed PDF set.}
\label{fig:typeIInloXSec_photon}
\end{figure*}

\subsubsection{Photon PDF Uncertainties}\label{sec:results_photonPDF}
We explore differences in photon PDF modeling by 
considering those available from NN3.1+LUXqed NLO (\texttt{lhaid=324900})~\cite{Bertone:2017bme} and CT14 QED NLO (\texttt{lhaid=13400})~\cite{Schmidt:2015zda}.
Whereas the MMHT and NN3.1 sets use the LUXqed formalism~\cite{Manohar:2016nzj,
Manohar:2017eqh} to describe the elastic component of the $\gamma$ density in
the proton, CT14 uses the well-established Equivalent Photon Approximation
(EPA)~\cite{Budnev:1974de}. Notably, the CT14 global fit does not identify a
nonzero, inelastic photon momentum fraction ($p_\gamma$ in their model) at
\confirm{$\mu_f = 1.295\GeV$}, \ie, the evolution scale at which the inelastic  photon PDF starts.
Instead, only an upper limit of \confirm{$p_\gamma<0.14\%$} at 90\% CL is set~\cite{Schmidt:2015zda}.
Subsequently, as our reference CT14 PDF, we take the default member PDF, which assumes $p_\gamma=0\%$.

Beyond the above state-of-the-art photon PDFs,
we also consider the NN2.3 NLO QED set (\texttt{lhaid=244600})~\cite{Ball:2013hta} and MRST 2004 NLO QED (\texttt{lhaid=20463})~\cite{Martin:2004dh}.
While both sets omit an explicit description of the elastic component of the
photon density in the proton, they are still widely used\footnote{In principle, the NNPDF inclusive methodology can account
for the existence of the elastic component of the photon
density~\cite{Ball:2013hta}.}.
For all of the above photon PDFs, the inelastic component is renormalization group-evolved using mixed QCD-QED DGLAP evolution, and assumes current masses for QED radiation off quarks.
For completeness, we also consider the EPA elastic photon PDF itself.
This can be called within \mgamc~at runtime using the commands,
\begin{verbatim}
  set lpp1 2
  set lpp2 2
\end{verbatim}

To determine the scale dependence on the AF production cross sections, 
we compute the three-point $\mu_f$ variation as described in Sec.~\ref{sec:mcSetup_SM}  for all photon PDFs.
The exception is the EPA elastic PDF, where we conservatively assign a flat \confirm{$^{+1\%}_{-2\%}$} uncertainty.
This corresponds to the full uncertainty envelope extracted in Ref.~\cite{Alva:2014gxa}.

To determine the PDF modeling uncertainty, we use the standard replica method for the MMHT+LUXqed, NN3.1+LUXqed, and NN2.3 PDF sets.
For CT14, we follow Ref.~\cite{Schmidt:2015zda} and consider the envelope spanned by 
assuming inelastic photon momentum fractions \confirm{$p_\gamma=0\%$ and $p_\gamma=0.14\%$}.
Similarly, for MRST, we follow Ref.~\cite{Martin:2004dh} and use the envelope spanned 
by assuming current quark masses (default member PDF \texttt{MRST2004qed\_proton\_0000.dat})
and constituent quark masses (\texttt{MRST2004qed\_proton\_0001.dat})
We do not assign a modeling uncertainty for the EPA as it takes  well-measured input parameters such as the proton mass, and therefore exhibits only a small input variation~\cite{Budnev:1974de}.

In the upper panels of Fig.~\ref{fig:typeIInloXSec_photon}, we depict, as a
function of the triplet mass $m_{\Delta}$, the total $\dpp\dmm$ production cross
section [fb] from AF at $\sqrt{s}=14\TeV$,
assuming various photon PDFs and estimating the uncertainties from (a)
$\mu_f$-scale dependence and (b) photon PDF modeling.
We also show $\dpp\dmm$ production rate via the NC DY process at NLO with $\mu_f$ variation (no PDF variation).
In the lower panels, we present the ratio of the AF cross section relative to the reference rate as obtained when using the  central
MMHT+LUXqed photon density.

Starting with the two LUXqed-based computations, we find remarkable agreement of the total normalization and small uncertainties across the entire mass range.
We find at most a \confirm{1\%-2\%}  difference in total normalization at low-to-intermediate masses, with NN3.1+LUXqed being smaller.
This closeness is attributed to the smallness of the LUXqed uncertainty itself~\cite{Manohar:2016nzj,Manohar:2017eqh}.
For CT14, we find that taking $p_\gamma=0\%$ leads to cross sections that are consistently  \confirm{20\%-30\%} smaller across the entire mass range, 
with larger differences appearing at higher masses.
At lower masses, the CT14, MMHT+LUXqed, and NN3.1+LUXqed scale bands overlap.
The worsening disagreement at higher masses is consistent with the $p_\gamma=0\%$ hypothesis.
The corresponding initial-state photons most likely originate from QED radiation
off valence quarks, so that setting $p_\gamma=0\%$ suppresses such contributions
to the scattering rates and therefore yields smaller predictions.
The MMHT+LUXqed uncertainty band sits centrally within the CT14 $p_\gamma=0\%--\!0.14\%$ band.
Explicit checks find that taking  \confirm{$p_\gamma=0.05\%--0.06\%$} reproduces the rates obtained with MMHT+LUXqed.
Similar agreement between CT14 and LUXqed-based photon PDFs has been reported elsewhere~\cite{Dittmaier:2017bnh}.

Turning to the MRST photon PDF, we find that the central member PDF set predicts $\dpp\dmm$ production rates that are about \confirm{a factor of $1.3-2.2$ times larger} than our reference set.  
At the same time though, the MRST curve exhibits a comparable scale uncertainty.
Despite its larger normalization, the MRST result remains below the DY rate for the triplet masses that we consider.
We find that the MRST photon PDF uncertainty captures completely the MMHT+LUXqed, NN3.1+LUXqed, and CT14 predictions as well as their respective PDF uncertainties.
The current (constituent) mass hypothesis for logarithmic evolution as used in the default (second) member of the MRST photon PDF 
overestimates (underestimates) the photon density in the proton.
An exact agreement though is not to be expected because both neglect the elastic component of the proton's photon PDF.
It follows then that the prescription of averaging the two results, as suggested\footnote{See the \texttt{MRST2004qed\_proton.info} PDF file for details.} by Ref.~\cite{Martin:2004dh}, 
provides a reliable estimate of three aforementioned photon PDFs.

We now consider the $\dpp\dmm$ rate obtained using the NN2.3 photon PDF. First,
we find that the NN2.3 rate surpasses the NC DY rate at $m_\Delta\sim900\GeV$ and sits above the default MRST rate, and thereby reproduce the findings of
Refs.~\cite{Babu:2016rcr,Ghosh:2017jbw,Ghosh:2018drw}.
While the scale uncertainty is also comparable to the other photon PDFs with inelastic components,
we find that for \confirm{$m_\Delta > 250\GeV$, the NN2.3 prediction exceeds the reference rate by a factor of $1.5\!-\!13$.}
For \confirm{$m_\Delta = 100- 250\GeV$, the two $\mu_f$ bands overlap.}
The discrepancy is clarified in light of the PDF modeling uncertainty, which shows that the NN2.3 uncertainty is unbounded from below.
As reported by the NNPDF collaboration itself~\cite{Ball:2013hta}, the NN2.3 photon PDF uncertainty \confirm{exceeds 100\%} due to the lack of available data feeding into the PDF fit.
This is larger than the \confirm{few-to-20\%} uncertainty quoted in Ref.~\cite{Babu:2016rcr}, which applies to the quark and antiquark PDFs,
and ultimately leads to an unreliable estimation of the $\gamma\gamma$ luminosity.

The AF rate obtained from the EPA itself, and hence only the elastic component of the proton's photon density, 
sits at about \confirm{10\%-13\%} of our reference result.
This indicates that the inelastic component is the driving factor behind the $\dpp\dmm$ production rate, which is consistent with similar investigations~\cite{Alva:2014gxa}.
Interestingly, we find that taking the sum of MRST constituent-mass rate and the EPA rate can largely reproduce the rate obtained with the CT14 $p_\gamma=0\%$
set.
Omitting the semielastic EPA-MRST contribution, we find that the EPA+MRST(constituent) rate sits about \confirm{22\%-0\%} below the $p_\gamma=0\%$ rate over the entire $m_\Delta$ range.

\subsubsection{Summary and Recommendations}

In summary, we find that, in principle, all photon PDFs including an inelastic component studied here give consistent predictions for the AF process.
However, this is undercut by the fact that the default and/or recommended central member PDFs are not always representative of central values of now-accepted photon PDFs.
For future computations, we recommend using MMHT+LUXqed, NN3.1+LUXqed, or CT14 QED with $p_\gamma=0.05\%-0.06\%$.
Summing the EPA and MRST constituent-mass rates is a possible option, but perhaps unnecessary due to more modern options. 

We moreover note that when the above procedure is matched to parton showers that support QED and photon PDF evolution, such as {\sc Pythia~8}\footnote{
Details are provided on the URL \href{http://home.thep.lu.se/~torbjorn/pythia81html/SpacelikeShowers.html}{\texttt{home.thep.lu.se/$\sim$torbjorn/ pythia81html/SpacelikeShowers.html}}
}, then initial-state photons can be matched to initial-state $q\to q\gamma*$ splittings.
This permits one to study forward jet-tagging and central jet vetoes in the context of photon fusion.

\subsection{Triplet Scalar Pairs from $gg$ Fusion}\label{sec:results_GFX}
With the \libName~libraries, it is possible to simulate the production of triplet scalars from loop-induced processes up to $\mathcal{O}(\alpha_s)$.
For the specific case of $\dpp\dmm$ pair production from GF, as shown in Fig.~\ref{fig:diagrams}(c) and given in Eq.~\ref{eq:afPair},
we use the following \mgamc~syntax,
\begin{verbatim}
  generate g g > dxx dxx [QCD]
  output TypeIInlo_ggF_DxxDxx_XLO
  launch
  shower=OFF
  madspin=OFF
  analysis=OFF
  set run_card nevents 100k
  set LHC 14
  set dynamical_scale_choice 4
  set use_syst True
  set no_parton_cut
  set pdlabel lhapdf
  set lhaid 26300
\end{verbatim}
For the GF channel, we use of the MMHT+LUXqed NNLO PDF set (\texttt{lhaid=26300}),
but defer momentarily the reason for the choice.
For the mass range $m_\Delta = 100-600\GeV~(100-1300\GeV)$, we find that the LO GF rates at $\sqrt{s}=14~(100)\TeV$ span approximately
\confirm{
\begin{eqnarray}
\GF	&:& 36\fb-0.47\ab ~\quad (930\pb-0.32\ab).
\end{eqnarray}}
We find that our computations match the results reported in Ref.~\cite{Hessler:2014ssa}.
In particular, we can identify \confirm{the doublet-triplet Higgs couplings
$f_3$ and $f_4$ in the scalar potential of Ref.~\cite{Hessler:2014ssa} with our
$\lambda_{h\Delta1}$ and $\lambda_{h\Delta2}$ parameters respectively.}
Given their benchmark values, Ref.~\cite{Hessler:2014ssa} obtains an effective $h\dpp\dmm$ coupling of $\Lambda_{+2}=-2f_3 + f_4T^3=-3.9$, with weak isospin $T^3=1$,
for their analytic expression of the Born-level  $gg\to \dpp\dmm$ cross section.
For input choices listed in Sec.~\ref{sec:mcSetup_Seesaw}, we obtain
\begin{eqnarray}
\Lambda_{+2}&=&-2f_3 + f_4T^3= -2\lambda_{h\Delta1} + \lambda_{h\Delta2} T^3 \\
&\approx&  -2\lambda_{h\Delta1} = -2.
\end{eqnarray}
This implies that the results reported in Ref.~\cite{Hessler:2014ssa} can be obtained from our results by using the scale factor
\begin{equation}
 \sigma_{\GF}^{\text{Ref.~\cite{Hessler:2014ssa}}}  \approx \left(\frac{-3.9}{-2}\right)^2 ~\times~ \sigma_{\GF}^{\rm \libName~Default}.
 \label{eq:GFconversion}
\end{equation}

We do not report scale uncertainties at LO because processes initiated by gluon fusion are poorly approximated at LO, as detailed for
example in Refs.~\cite{Dawson:1990zj,Anastasiou:2015ema,Dawson:1998py,deFlorian:2013jea}.
For the production of heavy, color-singlet states, however, soft gluon threshold resummation can capture the leading contributions to the total cross section normalization,
as demonstrated for instance in Refs.~\cite{Bonvini:2014qga,
Bonvini:2014joa}\footnote{Matching threshold resummation to fixed-order
computations, however, is necessary to fully describe particle kinematics.}.
Hence, in addition to the LO predictions obtained with \mgamc, we also compute
directly the soft-gluon threshold corrections for $gg\to\dpp\dmm$ up to N$^3$LL.

To do so, we follow very closely the calculation of Ref.~\cite{Ruiz:2017yyf}, which reports the same resummation for the heavy neutrino production process $gg\to N\nu$.
The resummation is based on the SCET~\cite{Bauer:2000yr,Bauer:2001yt,Beneke:2002ph} threshold formalism developed in Refs.~\cite{Becher:2007ty,Ahrens:2008nc}, 
works directly in momentum space~\cite{Becher:2006nr,Becher:2006mr}, and uses the Born-level, partonic  $gg\to \dpp\dmm$ cross section given in Ref.~\cite{Hessler:2014ssa}.
Using the scale prescriptions of Ref.~\cite{Becher:2006nr,Becher:2006mr,Ruiz:2017yyf},
we are able to obtain a total normalization that captures the principle terms at NNLO in QCD. 
We thus match our PDF accordingly and adopt the NNLO set.

For each choice of $m_\Delta$, we compute the N$^3$LL $K$-factor and associated uncertainties. We then rescale the LO result of \mgamc.
Subsequently, we present in Fig.~\ref{fig:typeIInloXSec}, the $\dpp\dmm$ production rate via GF at N$^3$LL.
For the mass range $m_\Delta = 100-600\GeV~(100-1300\GeV)$, we find that the GF rates, residual scale uncertainties, and $K$-factors at $\sqrt{s}=14~(100)\TeV$ are approximately 
\confirm{
\begin{eqnarray}
\GF					:& \quad110\fb-1.4\ab\quad 	& (2.7\pb-0.81\ab), \quad
\\
\delta\sigma/\sigma		:& \pm10\%-\pm5\% 			& (\pm18\%-\pm7\%),
\\
K^{\rm N^3LL} 			:& 3.04-3.15 				& (2.54-2.90).
\end{eqnarray}}%
As described in Ref.~\cite{Bonvini:2014qga,Ruiz:2017yyf}, the sizable uncertainties associated with a 
calculation at this order of perturbation theory are an artifact of the scale choice in the resummation evolution factors.
The precise scale choice aims to minimize further shifts in the total normalization 
by compensating for missing contributions from hard, initial state radiation with additional renormalization-group running.
This leads to a large $\mu_f$ dependence that can be reduced by matching to fixed-order computations.

In comparison to other $\dpmpm$ production modes, we find that the GF channel contribution is quite small.
For our choice of doublet-triplet Higgs couplings, the GF rate is comparable to the AF rate for $m_\Delta\lesssim200~(500)\GeV$ at $\sqrt{s}=14~(100)\TeV$,
but quickly falls below for higher masses.
This rapid falloff can be attributed to the suppression of the $gg$ luminosity at large invariant masses.
We caution that our results are sensitive to our choice of doublet-triplet Higgs couplings.
As shown in Eq.~\ref{eq:GFconversion}, using the inputs of
Ref.~\cite{Hessler:2014ssa} leads to an enhancement of the total rate by a factor of \confirm{4}.

\subsubsection{Flavor Scheme and PDF Dependence}\label{sec:results_GFFS}

As a final comment for the GF channel, we investigate the role of the active quark flavor scheme and PDF dependence
on total cross sections across triplet mass scales and collider energies.
As for the DY cases, we summarize our results in Table~\ref{tb:xSecBenchmarkFS} for representative $m_\Delta$ and $\sqrt{s}$.

To start, we first draw attention to the fact that the GF mechanism proceeds at lowest order through heavy quark loops.
While we work consistently in the 4FS, we noted in Sec.~\ref{sec:results_DYFS}
that the assumed triplet scalar masses suggest that we should instead work in the 5FS.
In the 5FS, however, a massless $b$ quark does not directly contribute to the $gg\to\dpp\dmm$ cross section, 
and therefore leads to a modeling ambiguity.

To quantify this flavor scheme uncertainty, we evaluate the GF rates at LO and with threshold resummation at N$^3$LL, assuming $m_b=0\GeV$, \ie, in the 5FS.
We observe that at $\sqrt{s}=14\TeV$, the 5FS cross section are about $\delta\sigma/\sigma = 0.5\%-0.8\%$ smaller than in the 4FS.
At $\sqrt{s}=100\TeV$, we see that the drop in rate is milder and more uniform, with 5FS rates being about $\delta\sigma/\sigma = 0.5\%$ smaller than their 4FS counterparts.
For both schemes, the scale uncertainties are comparable but are nevertheless much larger than the differences in central rates.
In summary, in light of the size of scale uncertainties, we find unimportant differences between the 4FS and 5FS.

To determine PDF uncertainties, we repeat the procedure reported in Sec.~\ref{sec:results_DYPDF}.
Unlike the DY cases, we find large differences in PDF uncertainties as a function of triplet mass and collider energy.
At $\sqrt{s}=14\TeV$ and $\sqrt{s}=100\TeV$, we find that uncertainties roughly span:
\begin{eqnarray}
\text{GF-14\TeV} 	\quad &:& \quad 	\pm6\% - \pm13\%,\\
\text{GF-100\TeV} 	\quad &:& 	\quad	\pm1\% - \pm2\%.
\end{eqnarray}
Relative to the scale uncertainties, PDF uncertainties are comparable in magnitude at $\sqrt{s}=14\TeV$
but are a factor of several-to-an order of magnitude smaller at $\sqrt{s}=100\TeV$.
We observe no significant difference in the PDF uncertainty between the 4FS and 5FS.

To extrapolate to other masses and collider energies, we again 
consider the geometric average of the momentum fractions $(\xi_\star)$, as done in Eq.~\ref{eq:results_PDFXi}
for the triplet masses and collider energies in Table~\ref{tb:xSecBenchmarkFS}.
For the considered range of $m_\Delta$, we find that PDF uncertainties reach:
\begin{eqnarray}
  \text{GF-Low mass} 		\quad&:&\quad \pm1\%,  		\\
  \text{GF-Intermediate mass}	\quad&:&\quad \pm2\%-\pm6\%,  	\\
  \text{GF-High mass} 		\quad&:&\quad \pm6\%-\pm13\%.
\end{eqnarray}
As for the DY cases, for much lower and much higher masses, the
uncertainties generally grow due to poorer constraining power of fixed-order PDF fits.

\subsection{Triplet Scalars from Weak Boson Fusion}\label{sec:results_VBF}
We now discuss the ability to model VBF production of Type II scalars at NLO using the \libName~libraries with \mgamc.
We focus on the  channel
\begin{eqnarray}
pp\to\dpm \dxx jj,
\label{eq:vbfProc}
\end{eqnarray}
which is shown in Fig.~\ref{fig:diagrams}(e) and is mediated by same-sign $W^\pm W^\pm$ scattering. 
We do not consider the well-studied, $W^\pm W^\pm\to \dpmpm$ single production process
as the  $\dpp W^\mp W^\mp$ coupling is proportional to the vev ratio $v_\Delta/v\ll1$, and is therefore suppressed.
The treatment at NLO of  $\dpp\dmm j j$, as shown in Fig.~\ref{fig:diagrams}(f),
requires care due to the presence of $t$-channel photon diagrams,
and will be visited in a {dedicated publication}~\cite{photonNLO}.

Unlike more inclusive processes considered elsewhere in this work, our modeling of $\dpm \dxx jj$ production involves three subtleties.
First, in this study, we consider all interfering diagrams at $\mathcal{O}(\alpha^4)$.
We do not make the commonly used \textit{Vector Boson Fusion Approximation,}
which takes into account only VBF diagrams, \ie, those characterized by the scattering of two $t$-channel EW bosons.
Aside from gauge invariance concerns, such approximations are now known to poorly describe the kinematics of 
leading, subleading, and trailing jets~\cite{Campanario:2013fsa,Campanario:2018ppz,Ballestrero:2018anz}, which are crucial to identifying VBF signatures.
Second, we apply generator-level phase space cuts in order to
identify the VBF topology over the one of the interfering subprocesses.
These include nonresonant diboson, nonresonant DY, and resonant $\dpmpm\dxx W^\mp$ configurations.
We emphasize that loose, generator-level cuts that mirror tight, analysis-level cuts are, as a rule, important for efficient MC event generation.
Third, we work at NLO in QCD in order to reliably describe ``third jet'' kinematics after parton shower-matching.
Doing so enables one to reliably impose a central jet veto, which in turn is crucial to reducing multiboson and top quark backgrounds.

The nuance of the above is our desire to also model a central jet veto. This
requires one to \textit{start} modeling predictions at NLO in QCD. However,
within \mgamc, phase space cuts for NLO(+PS) computations cannot be applied at
the parton level. Cuts on partons are infrared (IR) unsafe and hence ill-defined
in perturbation theory. Generator-level cuts are, on the other hand, necessary
 to suppress interfering diagrams at $\mathcal{O}(\alpha^4)$ that must be
accounted to reliably describe the event's jet activity.

Subsequently, we report developing a prescription for introducing generator-level, VBF cuts within \mgamc.
The method exploits the fact that jets in \mgamc~are defined using \textit{IR-safe} hadron clusters.
As long as care is given not to disrupt the cancellation of IR poles at $\mathcal{O}(\alpha_s)$, 
then one can construct observables from IR-safe quantities and impose appropriate phase space cuts.

\subsubsection{VBF Cuts at NLO in~\mgamc}\label{sec:results_VBFcuts}
We now describe our procedure for imposing generator-level VBF cuts at NLO in
\mgamc~for the process in Eq.~\ref{eq:vbfProc}.
We start by using the usual commands,
\begin{verbatim}
  generate p p > dxx d0 j j QCD=0 QED=4 [QCD]
  output TypeIInlo_VBF_DxxD0_NLO
\end{verbatim}
to build the event-generation working directory.
In the \mgamc\ framework, generator-level cuts on kinematic observables are
imposed by the routines available from the file \texttt{SubProcesses/cuts.f},
right after a phase space point is populated by MC sampling.
In this file, all final-state QCD partons are eventually passed through 
{\sc FastJet}~\cite{Cacciari:2005hq, Cacciari:2011ma} 
according to some IR-safe sequential clustering algorithm. 
Clusters are promoted to jets if they further satisfy basic transverse momentum
($p_T^j > p_T^{\min}$) and pseudorapidity ($\vert\eta^j\vert < \eta^{\max}$)
requirements.
In our case, jets are defined as anti-$k_T$  clusters with
\begin{equation}
R=1, \quad  p_T^{\min}=25\GeV, \quad\text{and}\quad  \eta^{\max}=4.5.
\label{eq:genJetCuts}
\end{equation}
At  runtime, this can be set with the syntax
\begin{verbatim}
  set jetalgo    -1
  set jetradius 1.0
  set ptj   25
  set etaj 4.5
\end{verbatim}
After building jets, \mgamc~evaluates the jet multiplicity of the event
(\texttt{njet}).
If an event contains fewer than the number of jets specified at the Born level
or contains at least two more jets than at the Born level, then the phase space point is rejected.
For Eq.~\ref{eq:vbfProc}, the Born jet multiplicity is two and events with fewer than two jets or more than three jets are rejected.
We note that pathological phase space regions, such as resonant  $q\overline{q'}\to\dpmpm\dxx W^\mp$ production with $W\to q\overline{q'}$ 
forming a collimated cluster, are removed at LO by the \texttt{njet} cut.

At this point in \texttt{SubProcesses/cuts.f},
for phase space points satisfying  \texttt{njet>1}
we impose on the two \textit{jets} with highest $p_T$ (labeled $j_1, j_2$)
the following VBF cuts:
\begin{eqnarray}
M(j_1,j_2) &>& M_{jj}^{\min} = 200\GeV \quad\text{and}\quad 
\label{eq:genVBFCutMjj}
\\
\vert \Delta\eta(j_1,j_2)\vert &>& \Delta\eta^{\min}=2.0.
\label{eq:genVBFCutEta}
\end{eqnarray}
Our precise additions to  \texttt{SubProcesses/cuts.f} and affiliated files are provided in appendix~\ref{app:vbfCuts}.

\begin{figure}
  \subfigure[]{	\includegraphics[width=0.3\textwidth]{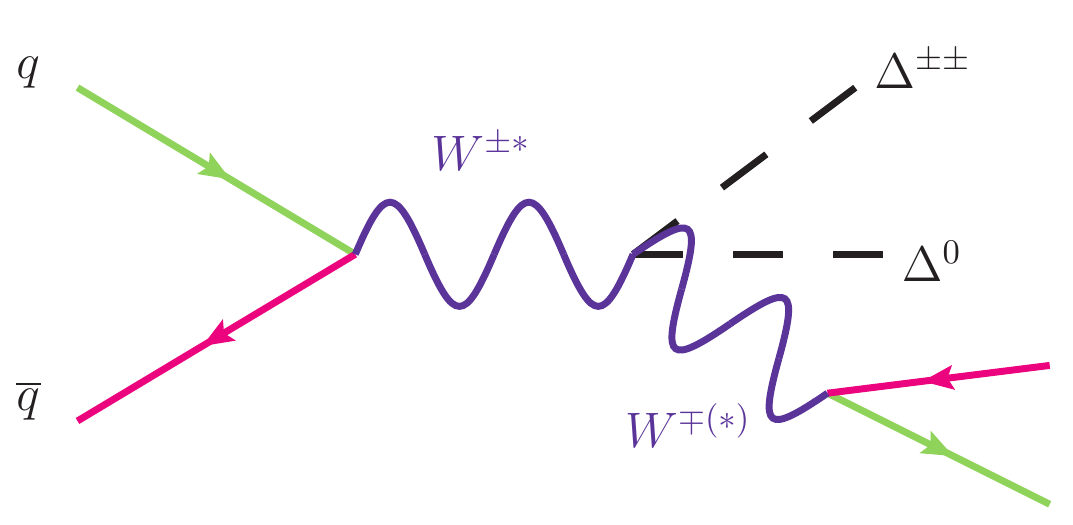}		\label{fig:diagrams_TypeIInlo_Associated}	}
  \subfigure[]{	\includegraphics[width=0.3\textwidth]{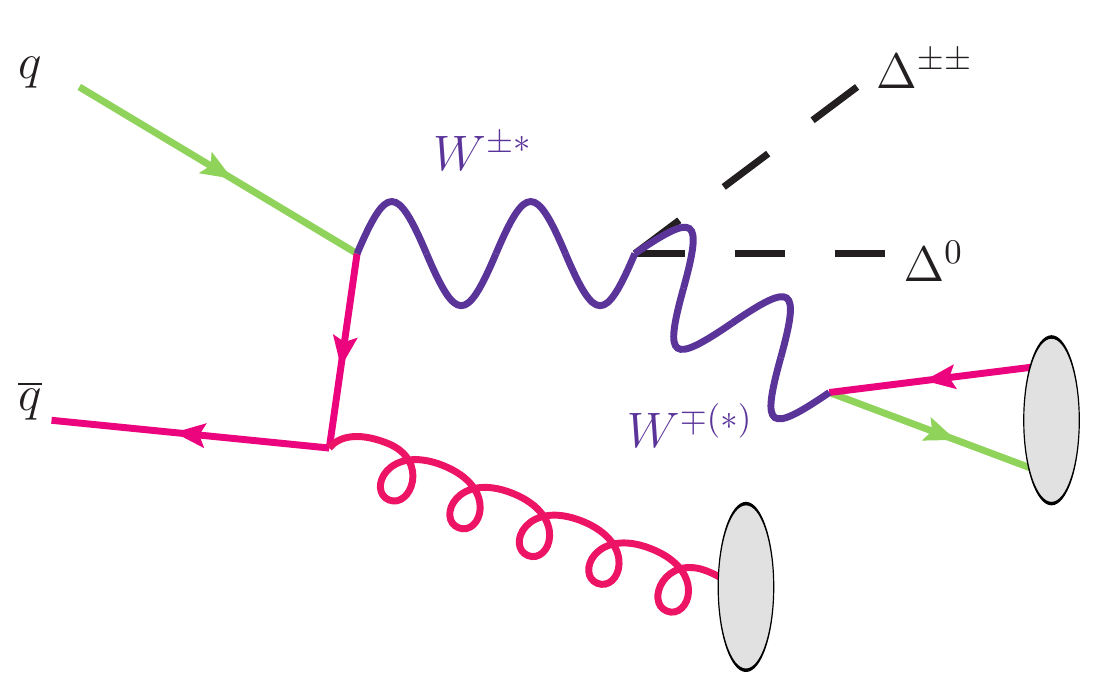}		\label{fig:diagrams_TypeIInlo_Associated1j}	}
\caption{
Representative diagrams contributing to associated $pp\to\dpmpm\dxx W^\mp$ with a $W\to jj$ splitting at (a) LO and (b) NLO in QCD.
}
\label{fig:diagramsAssociated}
\end{figure}

For Born and virtual diagrams, Eq.~\ref{eq:genVBFCutMjj} has the important effect  of removing resonant $W\to q\overline{q'}$ splittings,
which appear in diagrams such as Fig.~\ref{fig:diagrams_TypeIInlo_Associated}.
These contributions dominate the EW process $pp\to\dpmpm\dxx j j$ at
$\mathcal{O}(\alpha^4)$ because of the $W$ mass pole\footnote{Under the narrow
width approximation, $pp\to\dpmpm\dxx W^\mp$ is an $\mathcal{O}(\alpha^3)$
process.}.
At NLO, however, the rejection of  $W\to q\overline{q'}$ splittings is not guaranteed due to pathological phase space configurations.
For example, at $\mathcal{O}(\alpha_s)$, a high-$p_T$ initial-state radiation can push the parton-level, 
$(q\overline{q'})$ system into a more collimated configuration, as shown in Fig.~\ref{fig:diagrams_TypeIInlo_Associated1j}.
When QCD partons are sequentially clustered, this pair would be clustered while the gluon is clustered with only itself, resulting in \texttt{njet=2}.
In turn, the invariant mass cut of Eq.~\ref{eq:genVBFCutMjj} is applied to the three-parton invariant mass $M(q,\overline{q'},g)$,
allowing the  $(q\overline{q'})$ system to go onto the $W$ boson mass shell.

\begin{table*}
  \begin{center}
  \renewcommand{\arraystretch}{1.6}
  \setlength\tabcolsep{10pt}
\begin{tabular}{ c  c || c c | c  c | c  c  | c  c}
  \multicolumn{2}{c||}{} & \multicolumn{4}{c|}{$\sqrt{s} = 14\TeV$} &
    \multicolumn{4}{c}{$\sqrt{s} = 100\TeV$} \tabularnewline
  \multicolumn{2}{c||}{} &
    \multicolumn{2}{c|}{$m_\Delta = 550\GeV$}&
    \multicolumn{2}{c|}{$m_\Delta =1\TeV$} &
    \multicolumn{2}{c|}{$m_\Delta = 550\GeV$}&
    \multicolumn{2}{c}{$m_\Delta =1\TeV$}\tabularnewline
  $M_{jj}^{\min} $ & $\Delta\eta^{\min}$		& ~$\sigma^{\rm NLO}$ [ab]~ & ~$K$~  & ~$\sigma^{\rm NLO}$ [ab]~ & ~$K$~ 
									& ~$\sigma^{\rm NLO}$ [ab]~ & ~$K$~   & ~$\sigma^{\rm NLO}$ [ab]~ & ~$K$	
	\tabularnewline\hline\hline
$0\GeV$		& $0$	&	19.0$^{+5\%}_{-5\%}$			&	1.33
					&	0.983$^{+6\%}_{-6\%}$		&	1.38
					&	1300$^{+2\%}_{-1\%}$		&	1.22
					&	210$^{+2\%}_{-2\%}$			&	1.29
	\tabularnewline					
$100\GeV$	& $2$	&	9.92$^{+5\%}_{-5\%}$		&	1.30
					&	0.574$^{+6\%}_{-6\%}$		&	1.37
					&	826$^{+1\%}_{-1\%}$			&	1.08
					&	132$^{+1\%}_{-1\%}$			&	1.11
	\tabularnewline					
$200\GeV$	& $2$	&	9.59$^{+5\%}_{-5\%}$		&	1.27
					&	0.557$^{+5\%}_{-6\%}$		&	1.34
					&	807$^{+1\%}_{<-0.5\%}$			&	1.07
					&	132$^{+1\%}_{-1\%}$			&	1.10
	\tabularnewline					
$300\GeV$	& $2$	&	9.08$^{+4\%}_{-4\%}$		&	1.23
					&	0.533$^{+6\%}_{-6\%}$		&	1.30
					&	786$^{+1\%}_{<-0.5\%}$		&	1.04
					&	132$^{+1\%}_{-1\%}$		&	1.10
	\tabularnewline					
$400\GeV$	& $2$	&	8.57$^{+3\%}_{-4\%}$		&	1.22
					&	0.515$^{+5\%}_{-5\%}$		&	1.31
					&	760$^{+1\%}_{<-0.5\%}$		&	1.02
					&	126$^{+1\%}_{-1\%}$		&	1.06
	\tabularnewline					
$500\GeV$	& $2$	&	7.98$^{+3\%}_{-4\%}$		&	1.20
					&	0.489$^{+5\%}_{-5\%}$		&	1.32
					&	732$^{+1\%}_{<-0.5\%}$		&	0.99
					&	124$^{+1\%}_{-1\%}$		&	1.03
	\tabularnewline					
$750\GeV$	& $2$	&	6.68$^{+3\%}_{-4\%}$		&	1.21
					&	0.411$^{+5\%}_{-6\%}$		&	1.32
					&	688$^{+1\%}_{-1\%}$		&	0.98
					&	116$^{+1\%}_{<-0.5\%}$		&	1.02
\end{tabular}
\caption{
  NLO QCD cross section $(\sigma^{\rm NLO})$ [ab] for the EW
  $pp\to\dpmpm\dxx jj$ process, with $\mu_f,\mu_r$ scale variation [\%] and the
  associated NLO $K$-factor, and for representative triplet mass $m_\Delta$ at
  $\sqrt{s}=14$ and $100\TeV$.
  Cross sections assume jets defined according to Eq.~\ref{eq:genJetCuts}, the
  cluster veto of Sec.~\ref{sec:results_VBFcuts}, as well as the VBF cuts
  $M_{jj}^{\min}$ and $\Delta\eta^{\min}$ on the two leading jets.
}
\label{tb:xSec_VBFcut}
\end{center}
\end{table*}

As resonant $W$ bosons are not present at LO, the emergence of an on-shell $W$ boson at $\mathcal{O}(\alpha_s)$ is a new kinematic configuration
and can lead to numerically giant $K$-factors~\cite{Rubin:2010xp} for production cross sections.
Aside from numerical instabilities, this obfuscates perturbative convergence as well as the importance of QCD corrections to VBF rates, 
which are expected to be minor~\cite{Han:1991ia,Bolzoni:2010xr,Dreyer:2016oyx}.

To remove resonant $W$ bosons at $\mathcal{O}(\alpha_s)$, we introduce a phase space cut that vetoes the clustering of weak isospin partners.
To do this, we rely on clusters having at most two QCD partons in \texttt{njet=2} events.
For such clusters, we check if the PDG identifiers of the two constituents
(\texttt{PID}) correspond to weak isospin partners that would come from
$W$ boson decays, \eg, $(\overline{c},s)$ pairs.
Events possessing a cluster with isospin partners are rejected.
This is equivalent to double flavor-tagging jets seeded by light constituents. 
Physically, the veto corresponds to imposing an upper bound on the $p_T$ of resonant $W$ bosons.
For a generic $i \to j + k$ splitting, the opening angle between $j$ and $k$ scales inversely with the Lorentz boost $(\gamma)$ of the $(jk)$ system,
\begin{equation}
\theta^{jk} \approx \frac{2}{\gamma^{jk}} = \frac{2 M(j,k)}{E^{jk}} < \frac{2 M(j,k)}{p_T^{jk}}.
\end{equation}
As the cluster veto restricts isospin partners to be angularly separated by the
jet radius parameter $R$, the veto effectively requires, for resonant $W$
bosons,
\begin{equation}
p_T^W < 2 M_W /  R \approx 160\GeV.
\end{equation}
For a nonresonant $(jk)$ systems, the veto acts to also impose a minimum mass cut on clustered isospin pairs of
\begin{equation}
M_{\rm cluster} = M(j,k) > p_T^j R / 2 = 12.5\GeV.
\end{equation}
This would interestingly regulate  $s$-channel poles if applied to collimated $\gamma^*\to q\overline{q}$ splittings.

This prescription does not impact IR pole cancellations at $\mathcal{O}(\alpha_s)$ because for our process such contributions are, by definition, 
\texttt{njet=2} configurations with each jet containing one quark or one antiquark.
Events with \texttt{njet=2} with one jet containing both a quark and antiquark are hard, wide-angle emission  configurations.

\subsubsection{Triplet Scalars from VBF at NLO}\label{sec:results_VBFxsec}
At both LO and NLO in QCD, we define the VBF contribution of the EW process $pp\to \dpmpm\dxx j j$ 
as the full $\mathcal{O}(\alpha^4)$ process with jets defined according to Eq.~\ref{eq:genJetCuts},
VBF cuts applied according to Eqs.~\ref{eq:genVBFCutMjj}-\ref{eq:genVBFCutEta}, 
and the cluster veto of the previous section.
For $m_\Delta\approx100-1000\GeV~(0.1-6\TeV)$ at $\sqrt{s}=14~(100)\TeV$, 
we present in Fig.~\ref{fig:typeIInloXSec} the VBF cross section at NLO, residual scale uncertainties, and QCD $K$-factor.
For the mass range under consideration, these quantities span roughly
\begin{eqnarray}
\sigma_\VBF^{\rm NLO}	:& 1.5\fb-0.57\ab ~&~ (31\fb-74\zb),
\\
K^{\rm NLO} 			:& 1.17-1.35 ~&~ (1.02-1.32),
\\
\delta\sigma/\sigma		:& \pm2\%-\pm6\% ~&~ (\pm1\%-\pm8\%).
\end{eqnarray}
In comparison to the CC DY channel, the VBF process rate is
quite small for our benchmark choices of triplet self-couplings and doublet-triplet couplings.
In particular, the ratio of the two cross sections at the two collider energies respectively span
\begin{equation}
\sigma_{\DYCC}^{\rm NLO} /\sigma_{\VBF}^{\rm NLO} \sim 1500\!-\!200~(700\!-\!60).
\end{equation}

Most notably we find rather mild QCD $K$-factors that remain about $K^{\rm NLO}\sim 1-1.2$ for triplet scalar masses at or around the EW scale,
and that grow mildly to $K^{\rm NLO}\sim 1.3-1.4$ for the largest masses considered at $\sqrt{s}=14\TeV$.
The size of these corrections are consistent with having an admixture of finite, $\mathcal{O}(\alpha_s)$ virtual corrections 
from DY-like (see above) and VBF-like topologies (see Refs.~\cite{Han:1991ia,Bolzoni:2010xr,Dreyer:2016oyx}),
and suggests a somewhat stable perturbative convergence with modest residual scale dependence.
At $\sqrt{s}=100\TeV$, we observe that the $K$-factor drops to $K^{\rm NLO}\sim 0.9-1.1$ for much of the mass range considered,
but again grows to $K^{\rm NLO}\sim 1.3-1.4$ for the largest masses considered.
At both colliders, however, the largest $K$-factors are found when $\tau_0 = (2m_\Delta)^2/s\sim0.1-0.2$, 
indicating that these numbers may also be tied to sizable PDF uncertainties at momentum fractions near unity.
The range of these $K$-factors are comparable to those found
in the production scalars in Georgi-Machacek models~\cite{Degrande:2015xnm} and heavy neutrinos~\cite{Degrande:2016aje} from VBF at NLO.

To explore the stability of our computation and the robustness of imposing VBF cuts at NLO, 
we report in table~\ref{tb:xSec_VBFcut} for the EW process $pp\to \dpmpm\dxx jj$, the NLO in QCD cross section $(\sigma^{\rm NLO})$ [ab] with $\mu_f,\mu_r$ scale variation [\%] 
and the NLO $K$-factor, for representative triplet scalar masses at $\sqrt{s}=14$ and $100\TeV$.
All cross sections here assume  jets defined according to Eq.~\ref{eq:genJetCuts}, the cluster veto of Sec.~\ref{sec:results_VBFcuts},
as well as several different choices of $M_{jj}^{\min}$ and $\Delta\eta^{\min}$ cuts on the two leading jets.
For increasing $M_{jj}^{\min}$ we observe that $K$-factors are reduced from $K^{\rm NLO}\sim1.3~(1.1)$ for $M_{jj}^{\min}=100\GeV$ at $\sqrt{s}=14~(100)\TeV$ 
to $K^{\rm NLO}\sim1.2~(1.0)$ for $M_{jj}^{\min}=500\GeV$, and in line with a more pure VBF sample.
At even larger $M_{jj}^{\min}$ we find that $K^{\rm NLO}$ mildly grow (reduce) at  $\sqrt{s}=14~(100)\TeV$.
This further supports possible PDF instability at $\sqrt{s}=14\TeV$ associated with very large momentum fractions,
which for $M_{jj}^{\min}=750\GeV$ implies a kinematic threshold of $Q\sim 2m_\Delta + M_{jj}^{\min}\sim 1900-2800\GeV.$

\subsection{LHC and VLHC Discovery Potential}\label{sec:results_sensitivity}

As an outlook for the sensitivity at the $\sqrt{s}=14\TeV$ HL-LHC and a hypothetical $\sqrt{s}=100\TeV$ VLHC,
we estimate the combined discovery potential  of searches via the CC DY, NC DY, and AF channels.
To do so, we first assume a standard benchmark in which the $\dpmpm$ and $\dpm$ scalars decay purely to leptonic final states:
\begin{equation}
{\rm BR}(\dpmpm\to \ell^\pm_i \ell^\pm_j), \quad {\rm BR}(\dpm\to \ell^\pm_i \nu_j) =1. 
\end{equation}
For the $pp\to 4\ell^\pm$ and $pp\to 3\ell^\pm$ plus missing transverse energy channels, we assume that the acceptance rate for charged lepton particle identification (PID) 
is uniform $\mathcal{A}_\ell = 70\%$ per charged lepton~\cite{Pascoli:2018heg}.
For $\dpmpm$ decays, we assume a dilepton invariant mass cut efficiency of $\varepsilon_{\rm Inv}=95\%$ per dilepton pair. 
For $\dpm$ decays, we assume a transverse mass cut efficiency of $\varepsilon_{M_T}=70\%$~\cite{Perez:2008ha}.
(While $\dpmpm$ and $\dpm$ widths are naturally narrow, they nevertheless grow with increasing $m_\Delta$, and hence mass cuts require tuning to maintain these acceptances.)
Residual backgrounds after such cuts include diboson and triboson processes as well as SM DY production in association with ``fake'' charged leptons~\cite{Perez:2008ha}.
Such processes are especially sensitive to dynamic jet vetoes (DJV)  whereas the signal processes considered are robust~\cite{Pascoli:2018rsg}.
Therefore, to suppress these residual backgrounds, we suppose also a DJV with a 
$\varepsilon_{\rm DJV}=95\%$ efficiency~\cite{Pascoli:2018heg, Fuks:2019iaj}.
Subsequently, for the two final states, the net analysis-level efficiencies are:
\begin{eqnarray}
\mathcal{A}_{4\ell} 	&=& \mathcal{A}_\ell^4 \times \varepsilon_{\rm Inv}^2 \times \varepsilon_{\rm DJV} 	~\approx~ 20.6\%, 
\\
\mathcal{A}_{3\ell} 	&=& \mathcal{A}_\ell^3 \times \varepsilon_{\rm Inv}\times \varepsilon_{M_T} \times \varepsilon_{\rm DJV} ~\approx~ 21.2\%. \quad
\end{eqnarray}

\begin{figure}
  \includegraphics[width=\columnwidth]{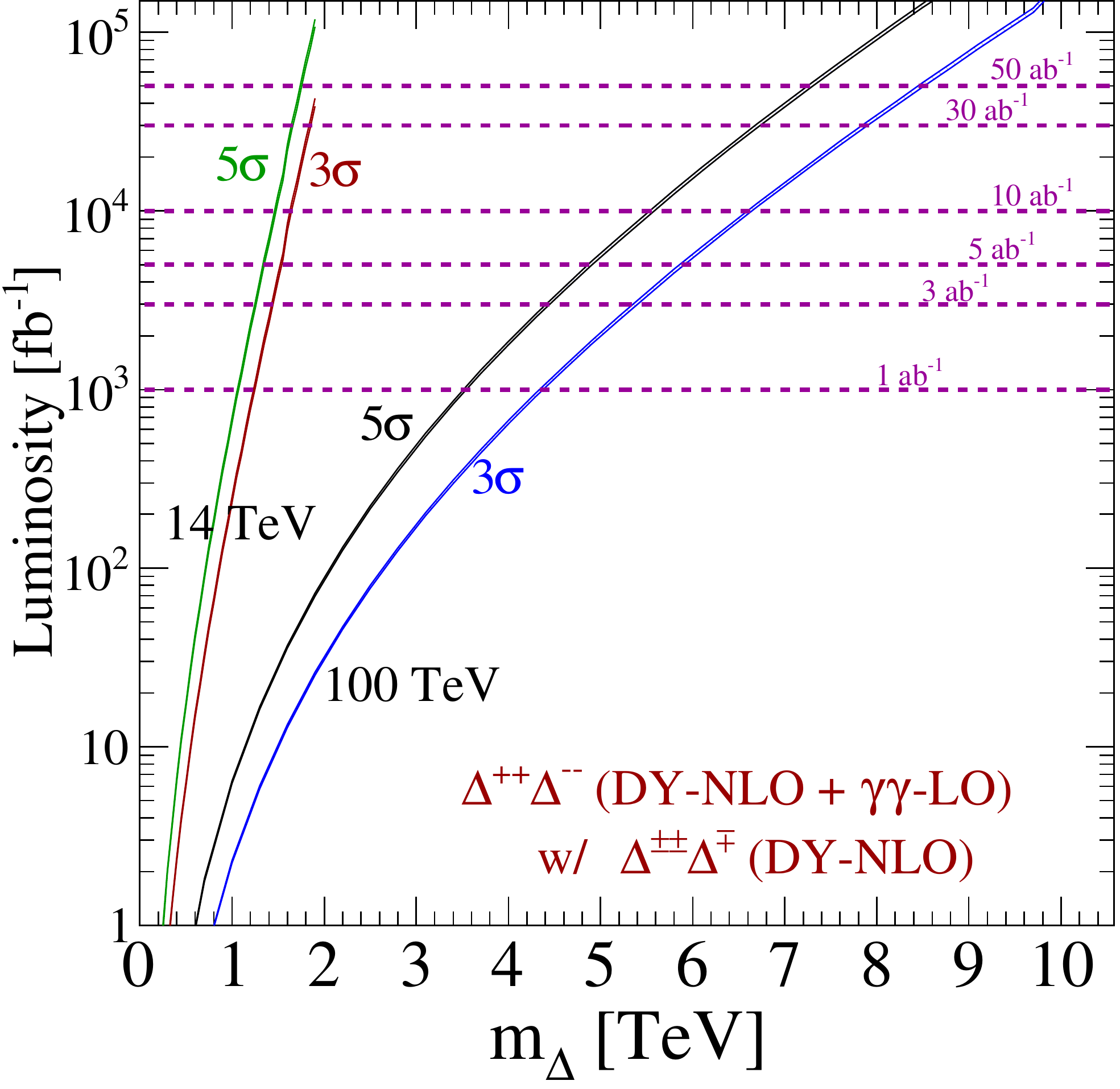}		
\caption{
As a function of $m_\Delta=m_\dpmpm=m_\dpm$, the estimated integrated luminosity [fb$^{-1}$]
required for $3\sigma$ sensitivity and $5\sigma$ discovery at the $\sqrt{s}=14\TeV$ HL-LHC and $100\TeV$ VLHC,
after combining the CC DY, NC DY, and AF channels.}
\label{fig:typeIInlo_Lumi_vs_mD_LHCMulti}
\end{figure}

For an integrated luminosity $\mathcal{L}$, the estimated number of observed events for each production mechanism is then
\begin{eqnarray}
N_{\rm CC~DY} 	&=& \sigma_{\rm CC~DY}^{\rm NLO} \times {\rm BR}_{\dpmpm}\times{\rm BR}_{\dpm}\times \mathcal{A}_{3\ell} \times \mathcal{L}, \nonumber
\\
N_{\rm NC~DY} 	&=& \sigma_{\rm NC~DY}^{\rm NLO} \times {\rm BR}_{\dpmpm}^2\times \mathcal{A}_{4\ell} \times \mathcal{L}, \nonumber
\\
N_{\rm AF} 		&=& \sigma_{\rm AF}^{\rm LO} \times {\rm BR}_{\dpmpm}^2\times \mathcal{A}_{4\ell} \times \mathcal{L}, \nonumber
\end{eqnarray}
with the total number of signal events given by the sum, $N^{\rm Tot}_S = N_{\rm CC~DY} + N_{\rm NC~DY}  + N_{\rm AF}$.
With the many assumptions on strong cuts, we make a zero-background hypothesis, $N_B^{\rm Tot}\approx0$, 
and conservatively quantify the signal significance $\mathcal{S}$ over the null hypothesis by
\begin{equation}
\mathcal{S} = \frac{N^{\rm Tot}_S }{\sqrt{N^{\rm Tot}_S +N^{\rm Tot}_B}} \approx \sqrt{N^{\rm Tot}_S }.
\label{eq:significance}
\end{equation}

We define the 95\% CL sensitivity ($\mathcal{S}^{95}$) and discovery  ($\mathcal{S}^D$) thresholds to be when $\mathcal{S}^{95~(D)}=3~(5)$.
As such, we can invert Eq.~\ref{eq:significance} and solve, as a function of $m_\Delta$ (assuming degenerate triplet masses), 
for the $\mathcal{L}$ that corresponds to $\mathcal{S}^{95~(D)}=3~(5)$.
Subsequently, we show in Fig.~\ref{fig:typeIInlo_Lumi_vs_mD_LHCMulti} 
the integrated luminosity required for $\mathcal{S}^{95~(D)}=3~(5)$, labeled by $3\sigma~(5\sigma)$,
as a function of triplet scalar mass.

At the LHC and with $\mathcal{L}=3-5\invab$, we report $3\sigma~(5\sigma)$ sensitivity to $m_\Delta \approx 1.2-1.5~(1.1-1.4) \TeV$.
At the VLHC and with $\mathcal{L}=30-50\invab$, we report $3\sigma~(5\sigma)$ sensitivity to $m_\Delta \approx 7.8-8.5~(6.7-7.3) \TeV$.
A dedicated study, now more readily with the availability of the \libName~libraries, can likely improve on this discovery potential estimation.

\section{Summary and Conclusion}\label{sec:Conclusions}
The canonical Type II Seesaw mechanism remains one of the best explanations for the origin of the tiny neutrino masses,
and arguably is the leading candidate that does not require hypothesizing the existence of new fermionic states.
In this regard, searches for the scalars predicted by the Type II Seesaw are well-motivated and 
remain an active component of the present and future collider search programs.

In this study, we investigate the production of doubly charged Higgs
bosons $(\dpmpm)$ from a variety of leading production mechanisms at the
$\sqrt{s}=14\TeV$ LHC and a hypothetical $\sqrt{s}=100\TeV$ VLHC. We
systematically take into account QCD corrections using state-of-the-art
formalisms.
For most channels, this is the first presentation of cross sections beyond LO,
and required the development of a new, publicly available, \textsc{FeynRules}
model implementation of the Type II Seesaw, dubbed the \libName~UFO libraries.
The UFO encoding the new interactions
is available from the URL \href{https://feynrules.irmp.ucl.ac.be/wiki/TypeIISeesaw}{feynrules.irmp.ucl.ac.be/wiki/TypeIISeesaw}.
In conjunction with \mgamc, the new \libName~UFO enables tree-induced processes to be simulated up to NLO+PS and
$\mathcal{O}(\alpha_s)$ loop-induced processes up to LO+PS.
The construction of these libraries is described in Sec.~\ref{sec:theory}.

We report that for $m_\Delta = 100-2000\GeV$, the CC DY (associated production), NC DY (pair production), and AF (photon fusion pair production)
processes are the leading three production channels considered at $\sqrt{s}=14\TeV$.
The same is found for $m_\Delta = 0.1-10\TeV$ at $\sqrt{s}=100\TeV$.
At both colliders, the AF mode remains suppressed by about an order of magnitude with respect to DY production.
However, in conjunction with a static jet veto of $p_T^{\rm Veto}=20-50\GeV$, the DY rates at NLO+NNLL(veto) and the AF rates are comparable, as detailed in
Sec.~\ref{sec:results_DYX}.
We find that state-of-the-art photon PDFs give a consistent and reliable description.
However, as described in Sec.~\ref{sec:results_AFX},
some alternative options possess prohibitively large uncertainties 
with recommended central member PDFs are not always representative of central values of now-accepted photon PDFs.

Finally, the GF and VBF channels are subdominant compared to the AF mode.
When accounting for QCD corrections up to N$^3$LL, we find that the GF rate is comparable to the AF rate for lower triplet Higgs masses and our benchmark doublet-triplet scalar couplings,
but that the rate quickly reduces at higher masses.
For the GF channel, we report large QCD $K$-factors of $K^{\rm N^3LL}\sim2.5-3.1$, which is consistent with $K$-factors at this accuracy in other gluon-initiated processes. Details are given in Sec.~\ref{sec:results_VBF}.
Further below the GF rate is the $\dpmpm\dxx j j$ VBF channel,
which necessitates the implementation of important generator-level VBF cuts for
a satisfactory modeling, as shown in Sec.~\ref{sec:results_VBF}.
We find remarkable stability in the $\dpmpm\dxx j j$ VBF channel, with NLO $K$-factors maintaining a stable $K^{\rm NLO}\sim1-1.3$.
We attribute the modest normalization shifts to interfering DY-type diagrams.
Further investigation of generator-level VBF cuts at NLO within \mgamc~is ongoing.

We conclude in Sec.~\ref{sec:results_sensitivity} with a brief comment on the discovery potential of the $\sqrt{s}=14\TeV$ HL-LHC and $100\TeV$ VLHC,
and report our findings in Fig.~\ref{fig:typeIInlo_Lumi_vs_mD_LHCMulti}.

\noindent
\acknowledgements{
\small
\textit{
The authors are grateful to Olivier Skywalker Mattelaer for enlightsabering  discussions on VBF MC generation. 
RR thanks Elisabetta Barberio and Federico Scutti for discussions that motivated this work;
Joshua Isaacson and Carl Schmidt for helpful discussions regarding CT PDF uncertainties;
Co Co~Tamarit for useful suggestions; and the Jo{\v z}ef Stefan Institute for its generous hospitality.
The authors also thank the organizers of the Chennai MadGraph School where some of this work was completed.
RR is supported under the UCLouvain fund ``MOVE-IN Louvain,'' and the F.R.S.-FNRS ``Excellence of Science'' EOS be.h Project No. 30820817.
The authors acknowledge the contribution of the VBSCan COST Action CA16108.
MN was supported by the Slovenian Research Agency under the
research core funding No. P1-0035 and in part by the research grant J1-8137.
MN acknowledges the support of the COST actions 
CA15108 - ``Connecting insights in fundamental physics'', 
CA16201 - ``Unraveling new physics at the LHC through the precision frontier''.
MN is grateful to the Mainz Institute for Theoretical Physics (MITP) of the DFG Cluster of
Excellence PRISMA+ (Project ID 33083149), for its hospitality and its partial support during which some of this work was done.
Computational resources have been provided by the Consortium des \'Equipements de Calcul Intensif (C\'ECI), 
funded by the Fonds de la Recherche Scientifique de Belgique (F.R.S.-FNRS) under Grant No. 2.5020.11 and by the Walloon Region. 
This work received funding from the EU's Horizon 2020 research and innovation programme as part of the Marie Sklodowska-Curie Innovative Training Network MCnetITN3 (grant No. 722104). 
}}

\medskip

\appendix
\section{Diagonalization of the scalar sector}\label{app:scalar}
In this appendix, we provide technical details to derivations in Sec.~\ref{sec:theory}.
We first impose the minimization condition of the full scalar potential, which can reduce the number of degrees of freedom of the model by two:
\be\bsp
 \mu_h^2 = &\ \lambda_h v^2
  - \frac{v_\Delta^2}{2 v^2}
      \Big[\lambda_{h\Delta} v^2 + 4 m_{\hat{\Delta}}^2\Big]
  - 2 \frac{v_\Delta^4}{v^4} \lambda_{\Delta}v^2 \ ,\\
  \mu_{h\Delta} = &\ \sqrt{2} v_\Delta \Big[
     \frac{m_{\hat{\Delta}}^2}{v^2} + \frac12 \lambda_{h\Delta}
      + \lambda_{\Delta} \frac{v_\Delta^2}{v^2}\Big]\ ,
\esp\label{eq:minV}\ee
where we have introduced the reduced parameters
\be
  \lambda_{h\Delta} = \lambda_{h\Delta 1} + \lambda_{h\Delta 2} \ ,\qquad
  \lambda_\Delta = \lambda_{\Delta 1} + \lambda_{\Delta 2} \ .
\ee
As evident from the above relations, the triplet vev $v_\Delta$ is controlled by the $\mu_{h \Delta}$ parameter. 
It constrained to be below a few GeV~\cite{Chen:2005jx,Chen:2008jg,Kanemura:2012rs,Das:2016bir}
and must satisfy
\be
  v^2 + v_\Delta^2 =  (\sqrt{2}G_F)^{-1}   \approx (246\GeV)^2\ .
\ee

After electroweak symmetry breaking, the squared mass of the doubly charged
scalar $\hat\Delta^{\pm\pm} \equiv \Delta^{\pm\pm}$ reads
\be
  m_{\Delta^{\pm\pm}}^2 = m_{\hat{\Delta}}^2 + \frac12 \lambda_{h\Delta1} v^2 +
    \lambda_{\Delta1} v_\Delta^2 \ .
\label{eq:dppm}\ee
This allows us to trade $m_{\hat{\Delta}}$ for $m_{\Delta^{\pm\pm}}$ as a  parameter.

In the $(\varphi^+, \hat\Delta^+)$ basis,
the squared mass matrix associated with the charged degrees of freedom is,
{\small
\be
 ({\cal M}^\pm)^2 \!=\! \frac{\lambda_{h\Delta2}v^2 \!+\!
   4 (m_{\Delta^{\pm\pm}}^2 \!+\! \lambda_{\Delta2} v_\Delta^2)}{2 v^2}
 \bpm
    v_\Delta^2 & -\frac{ v v_\Delta}{\sqrt{2}}\\[.2cm]
    -\frac{ v v_\Delta}{\sqrt{2}}& \frac12 v^2
  \epm \ .
\ee}
Such a matrix is diagonalized by the rotation of Eq.~\ref{eq:hp}, so that the
mass of the physical triplet-like $\Delta^\pm$ field reads
\be
  m_{\Delta^\pm}^2 = \frac{v^2+2 v_\Delta^2}{v^2} \Big[ m_{\Delta^{\pm\pm}}^2
    + \frac14 \lambda_{h\Delta2} v^2 + \lambda_{\Delta2} v_\Delta^2\Big] \ .
\ee
Using this last equation, we promote
the mass of the singly charged Higgs boson to an external parameter, making
$\lambda_{\Delta2}$ internal instead.

\begin{table}
  \begin{center}
  \renewcommand{\arraystretch}{1.5}
  \setlength\tabcolsep{3pt}
  \begin{tabular}{cc}
    Name & Strength \\ \hline\hline
      $c_{hhh}$ & $6 \lambda_h $\\
      $c_{h \Delta^0 \Delta^0}$ &  $\lambda_{h \Delta}$ \\ 
    \hline
      $c_{h\Delta^+\Delta^-}$ &
        $\lambda_{h \Delta 1} + \frac{\lambda_{h \Delta 2}}{2} $ \\
  $c_{\Delta^0 \Delta^+ \Delta^-}$ & 
  	$( 2 \lambda_{\Delta} - \lambda_{h \Delta2}) v_\Delta/v -
	\theta_{h\Delta} \left(2 \lambda_{h \Delta1} + \frac{\lambda_{h \Delta2}}{2} \right)$\\
    \hline
      $c_{h\Delta^{++}\Delta^{--}}$ & $\lambda_{h \Delta 1}$\\
  $c_{\Delta^0 \Delta^{++} \Delta^{--}}$ & 
  	$2 \lambda_{\Delta 1} v_\Delta/v - \theta_{h \Delta} \lambda_{h \Delta 1}$
\end{tabular}
\caption{
  Triple Higgs vertices at tree level up to $\mathcal O(v_\Delta/v)^2$,
  normalized to $v$, such that the corresponding Feynman rules read $c_{xyz}v$.
  The three panels respectively focus on neutral, singly charged and
  doubly charged interactions.}
\label{tb:3hv}
\end{center}
\end{table}

In the
$(\Im[\varphi^0], \Im[\hat\Delta^0])$ and $(\Re[\varphi^0], \Re[\hat\Delta^0])$
bases,
the squared mass matrices of the neutral scalar and pseudoscalar states $({\cal M}^\Re)^2$ and $({\cal M}^\Im)^2$ are respectively 
\be\bsp
 & ({\cal M}^\Re)^2 = \\
  & \bpm
    2 \lambda_h v^2 &
    v\big(\lambda_{h\Delta} v_\Delta \!-\! \sqrt{2} \mu_{h\Delta}\big) \\[.3cm]
    v\big(\lambda_{h\Delta} v_\Delta \!-\! \sqrt{2} \mu_{h\Delta}\big) &
    m_\Delta^2 \!+\! \frac12 v^2 \lambda_{h\Delta}
      \!+\! 3 v_\Delta^2 \lambda_{\Delta} \epm\ , \\[.5cm]
 & ({\cal M}^\Im)^2 = \\
 & \bpm
    \lambda_{h\Delta2} v_\Delta^2 \!+\! \frac{4 v_\Delta^2}{v^2+2 v_\Delta^2}
      m_{\Delta^\pm}^2 &
    -\sqrt{2} \mu_{h\Delta} v\\[.3cm]
    -\sqrt{2} \mu_{h\Delta} v&
    \frac14\lambda_{h\Delta2} v^2 \!+\! \frac{v^2}{v^2+2 v_\Delta^2}
      m_{\Delta^\pm}^2 \epm\ .
\esp\ee
These two matrices can be diagonalized via the rotations of Eq.~\ref{eq:h0},
so that the squared masses of the SM Higgs boson $h$ and of the triplet states
$\Delta^0$ and $\chi$ are given by
\be\bsp
  m_h^2 \!= & \frac12 m^2_{\Delta^{\pm\pm}} \!+\! \big(\lambda_h
    \!-\! \frac14 \lambda_{h\Delta2}\big) v^2 \!+\! \big(\lambda_{\Delta1}
    \!+\! \frac32 \lambda_{\Delta2}\big) v_\Delta^2 \!-\! \tilde{x}\ , \\
  m_{\Delta^0}^2 \!= & \frac12 m^2_{\Delta^{\pm\pm}} \!+\! \big(\lambda_h
    \!-\! \frac14 \lambda_{h\Delta2}\big) v^2 \!+\! \big(\lambda_{\Delta1}
    \!+\! \frac32 \lambda_{\Delta2}\big) v_\Delta^2 \!+\! \tilde{x}\ , \\
  m_\chi^2 \!= & \frac{v^2+4v_\Delta^2}{v^2} \Big[ m_{\Delta^{\pm\pm}}^2 +
    \frac12 \lambda_{h\Delta2} v^2 + \lambda_{\Delta2} v_\Delta^2\Big] \ ,
\esp\ee
in which we have introduced the abbreviation
\be\bsp
  & \tilde x^2 = 4 \frac{v_\Delta^6}{v^2} \lambda_{\Delta2}^2
  + \frac14 \frac{v_\Delta^4}{v^2}\Big[ v^2 \big(4 \lambda_{\Delta1}^2
    + 9\lambda_{\Delta2}^2\big)\\ &\
    + 4 \lambda_{\Delta2} \big(8 m^2_{\Delta^{\pm\pm}} +
    (3\lambda_{\Delta1} - 4 \lambda_{h\Delta1}) v^2\big) \Big]\\ &\
    + \frac14 \frac{v_\Delta^2}{v^2} \Big[16 m^4_{\Delta^{\pm\pm}} +
      2 m^2_{\Delta^{\pm\pm}} v^2
    \big(2\lambda_{\Delta1} + 3 \lambda_{\Delta2} \\&\
     - 8 \lambda_{h\Delta1}\big)
      + v^4 \big(4
      \lambda_{h\Delta1}^2 + 2 \lambda_{\Delta1}\lambda_{h\Delta2} \!+\!
      3 \lambda_{\Delta2} \lambda_{h\Delta2}\big)\\ &\
    -4 (2 \lambda_{\Delta1} + 3 \lambda_{\Delta2}) \lambda_h \Big]
    + \frac{1}{16} \Big[2 m^2_{\Delta^{\pm\pm}} + (\lambda_{h\Delta2}\\&\
    - 4 \lambda_h) v^2\Big]^2 \ .
\esp\ee
Those expressions allow us to render the $m_h$ and $m_{\Delta^0}$ masses
external, instead of the $\lambda_h$ and $\lambda_{h\Delta2}$ parameters. It can
be seen that for $v_\Delta \ll v$, one recovers the well-known sum rule
\be \label{eq:sumrule}
  m_{\Delta^0}^2 - m_{\Delta^+}^2 = m_{\Delta^+}^2 - m_{\Delta^{++}}^2 =
    \frac{\lambda_{h \Delta 2}}{4} v^2,
\ee
which implies that the $\Delta$ mass eigenstates possess mass splitting ranging
up to the electroweak scale. The scalar sector is finally defined by the
parameters of Eq.~\ref{eq:prm1}. In table~\ref{tb:3hv}, we present the
strength of the trilinear scalar interactions, after taking the limit of $v_\Delta/v\to0$, 
that are especially relevant for the VBF processes studied here.

\begin{table}[t]
\renewcommand{\arraystretch}{1.1}
\setlength\tabcolsep{5pt}
\begin{tabular}{c c c c}
  Parameter& FR name & LH block & Counter\\
  \hline\hline
  $\lambda_{h\Delta 1}$ & {\tt lamHD1} & {\tt QUARTIC}& 1\\
  $\lambda_{\Delta 1}$ & {\tt lamD1} & {\tt QUARTIC}& 2\\
  \hline
  $v_\Delta$ & {\tt vevD} & {\tt VEVD}& 1\\
  \hline
  $m_h$                 & {\tt MH}   & {\tt MASS} & 25\\
  $m_{\Delta^\pm}$      & {\tt MDP}  & {\tt MASS} & 38\\
  $m_{\Delta^0}$        & {\tt MD0}  & {\tt MASS} & 44\\
  $m_{\Delta^{\pm\pm}}$ & {\tt MDPP} & {\tt MASS} & 61\\
  \hline
  $m_{\nu_3}$       & {\tt Mv3}    & {\tt MASS} & 16\\
  $\Delta m^2_{21}$ & {\tt dmsq21} & {\tt MNU}  & 2\\
  $\Delta m^2_{32}$ & {\tt dmsq32} & {\tt MNU}  & 3\\
  \hline
  $\theta_{12}$ & {\tt th12} & {\tt PMNS}& 1\\
  $\theta_{23}$ & {\tt th23} & {\tt PMNS}& 2\\
  $\theta_{13}$ & {\tt th13} & {\tt PMNS}& 3\\
  $\varphi_{\rm CP}$ & {\tt delCP} & {\tt PMNS}& 4\\
  $\varphi_1$ & {\tt PhiM1} & {\tt PMNS}& 5\\
  $\varphi_2$ & {\tt PhiM2} & {\tt PMNS}& 6\\
\end{tabular}
\caption{Same as Table~\ref{tab:prm} but for the IH neutrino mass spectrum.
}
\label{tab:prmb}
\end{table}

\section{\libName-Model Libraries in the Case of an Inverted Neutrino Mass
Hierarchy} \label{app:fr_ih}
The implementation of the Type II Seesaw model of section~\ref{sec:theory} in
\fr\ when an inverted neutrino mass hierarchy is at stake proceeds similarly as
in section~\ref{sec:implementation}. The only difference is related to the
replacement of Eq.~\ref{eq:prm2} by Eq.~\ref{eq:prm2b} for the free
parameters defining the neutrino sector. Correspondingly, the external
parameters of the model implementation are given by Table~\ref{tab:prmb}
instead of Table~\ref{tab:prm}.

\newpage
\onecolumngrid
\section{VBF Cuts at NLO in QCD with \mgFull}\label{app:vbfCuts}
Implementing the generator-level VBF cuts on jets in \mgamc, as described in Sec.~\ref{sec:results_VBFcuts}, 
requires nontrivial modifications to several files within an event-generation directory of \mgamc.
In this appendix we describe the additions and modifications used to produce the results  reported in Sec.~\ref{sec:results_VBFxsec}.
The structure of our changes has four components:
\begin{enumerate}
  \item We implement a ``tag and probe'' protocol to veto any clustered, 
  weak isospin  $q\overline{q'}$ pairs, \eg $(\overline{u},d)$.
  \item We implement phase space restrictions (cuts) on the two highest-$p_T$ jets in an event.
  \item We turn off hard-coded, change-of-variables routines that are automatically called to optimize phase space integration when resonant particles appear in $s$-channel propagators.
\item We use the minimum dilepton invariant mass variable \texttt{mll} and minimum dilepton separation variable \texttt{drll} to pass our $M_{jj}^{\min}$ and $\Delta\eta^{\min}$ cuts at runtime. 
This last modification is used to further improve the efficiency of our computations and is not necessarily needed to implement the VBF cuts. 
\end{enumerate}
We start by assuming that an event-generation directory has been generated with the usual commands,
\begin{verbatim}
  generate p p > dxx d0 j j QCD=0 QED=4 [QCD]
  output TypeIInlo_VBF_DxxD0_NLO
\end{verbatim}
and that one is working directly in the \texttt{TypeIInlo\_VBF\_DxxD0\_NLO} directory.
Near line \confirm{20} of the file \texttt{Source/cuts.inc} file, we introduce the following global declarations:
\begin{verbatim}
      logical doVBFcuts
      integer vbfMinNrJets
      double precision vbfTiny,vbfMinMjj,vbfMinDEtajj
      parameter (vbfTiny=1.0d-05)
      parameter (vbfMinNrJets=2)
      parameter (doVBFcuts = .true.)
\end{verbatim}
The boolean \texttt{doVBFcuts} is a global switch set by hand that activates/deactivates the following VBF cuts.
The parameter \texttt{vbfTiny} is a tiny number used to protect against miscancellations between floating-point numbers.
The quantities \texttt{vbfMinNrJets}, \texttt{vbfMinMjj}, and \texttt{vbfMinDEtajj} are, respectively,
the minimum number of jets a phase space point must have,  $M_{jj}^{\min}$, and $\Delta\eta^{\min}$.
As two jets are always needed, we declare the parameter \texttt{vbfMinNrJets=2}.

The file \texttt{SubProcesses/cuts.f} is where phase space cuts are applied and where a bulk of our additions and modifications are made.
We start at around line \confirm{70} where we add the following local declarations:
\begin{verbatim}
      logical is_a_qqxPair, tmp_is_a_qqxPair
      integer jj,nrPairs,tmpPID
      double precision tmpMjj2,tmpPZjj
      double precision ppPairs(0:3,nexternal),m2Pairs(nexternal)
\end{verbatim}
The boolean \texttt{is\_a\_qqxPair} is a flag to check if an event possibly possesses a weak isospin  $q\overline{q'}$ pair,
and helps ensure that the cluster veto is not applied to events without weak isospin pairs.
The number of pairs is denoted by \texttt{nrPairs}.
The four-momentum and invariant mass (squared) of each weak isospin pair are stored by the arrays \texttt{ppPairs} and \texttt{m2Pairs},
Around \confirm{line 171}, just after the properties of the external particles with \texttt{istatus=1} are investigated to check whether these particles include a QCD parton,
we collect the four-momentum and invariant mass of each weak isospin pair in an event:
\begin{verbatim}
      if(doVBFcuts) then
         nrPairs=0
         is_a_qqxPair = .false.
         do i=nincoming+1,nexternal
            do j=i+1,nexternal
               tmp_is_a_qqxPair = .false.
               if(is_a_j(i).and.is_a_j(j)) then		      ! are i and j QCD partons?                                                                                          
                  if(abs(ipdg(i)+ipdg(j)).eq.1) then	 ! is Delta Isospin = +/-1?                                                                                          
                     tmpPID = max(abs(ipdg(i)),abs(ipdg(j)))
                     if( mod(tmpPID,2) .eq. 0 ) tmp_is_a_qqxPair = .true. ! is u,c, or t?                                                                                             
                  endif
               endif
               if(tmp_is_a_qqxPair) then
                  tmpMjj2 = invm2_04(p(0,i),p(0,j),1d0) ! (i+j)^2                                                                                                                             
                  if(tmpMjj2.gt.0d0) then       ! if m(ij)^2 > 0                                                                                                                              
                     is_a_qqxPair = .true.      !   then at least one qqBar pair exists                                                                                                              
                     nrPairs=nrPairs+1          !   update counter                                                                                                                        
                     m2Pairs(nrPairs) = tmpMjj2 !   collect mass                                                                                                                          
                     do jj=0,3                  !   collect momentum                                                                                                                   
                        ppPairs(jj,nrPairs) = p(jj,i) + p(jj,j)
                     enddo
                  endif
               endif
            enddo
         enddo
      endif ! if(doVBFcuts)  
\end{verbatim}

In the same file, all QCD partons are eventually passed through \textsc{FastJet} for clustering at around \confirm{line 290}.
Clusters are built according to the algorithm \texttt{jetalgo} with radius parameter \texttt{jetradius}, which are set in the file \texttt{Cards/run\_card.dat}.
Likewise, jets are defined as clusters with $p_T$ greater than \texttt{ptj} and $\vert\eta\vert$ smaller than \texttt{etaj}, which are also set in \texttt{Cards/run\_card.dat},
and the momentum of each jet is contained in the array \texttt{pjet}. 
We note that jet momenta contained by \texttt{pjet} are ordered in $p_T$, such that for jets $k$ and $k+1$, one has $p_T^{k}>p_T^{k+1}$.
Immediately after this, \mgamc~checks if the number of jets \texttt{njet} is equal to the number of jets specified in the Born-level process (\texttt{nQCD-1}) 
or one more than this number (\texttt{nQCD}). In the present case, if \texttt{njet} is smaller than 2 or larger than 3, the phase space point is rejected.
At this point, we impose generator-level,  $M_{jj}^{\min}$, and $\Delta\eta^{\min}$ cuts
on the two highest $p_T$:
\begin{verbatim}
         if(doVBFcuts) then
            vbfMinMjj    = mll
            vbfMinDEtajj = drll
            if((invm2_04(pjet(0,1),pjet(0,2),1d0).lt.vbfMinMjj**2)
     &         .or.(njet.lt.vbfMinNrJets)
     &         .or.(abs(eta_04(pjet(0,1))-eta_04(pjet(0,2))).lt.vbfMinDEtajj)                                                                                                                 
     &         ) then
               passcuts_user=.false.
               return
            endif
         endif
\end{verbatim}
As described at the beginning of this appendix, we pass the minimum dilepton invariant mass variable \texttt{mll} and minimum dilepton separation variable \texttt{drll},
which can both be set in \texttt{Cards/run\_card.dat}, to \texttt{vbfMinMjj} and \texttt{vbfMinDEtajj} in order to make our computations more efficient and more easily tunable.
More specifically, setting \texttt{mll} nonzero has the consequence in the file \texttt{SubProcesses/setcuts.f} of modifying the lower limit of phase space integration over PDFs.
Instead of integrating from a minimum kinematic threshold of $Q_0 = \min(p_{\dpp}+p_{\dmm}+p_{j_1}+p_{j_2})=2m_{\Delta} + 2p_T^{j~\min}$,
the lower threshold is set appropriately  to $Q_0=2m_{\Delta} + 2p_T^{j~\min}+M_{jj}^{\min}$. 
As the $M_{jj}^{\min}$ cuts we consider are quite stringent, this adjustment ultimately improves MC integration efficiency.

After imposing cuts on the two leading jets, we introduce the following at around \confirm{line 310}:
\begin{verbatim}
         if(doVBFcuts.and.njet.eq.vbfMinNrJets.and.is_a_qqxPair) then
            do i=1,njet
               tmpMjj2 = invm2_04(pjet(0,i),pjet(0,i),0d0)
               if(tmpMjj2.gt.0d0) then
                  do j=1,nrPairs
                     tmpMjj2 = invm2_04(pjet(0,i),pjet(0,i),0d0)
                     tmpMjj2 = abs(tmpMjj2   - m2Pairs(j))
                     tmpPZjj = abs(pjet(3,i) - ppPairs(3,j))
                     if(tmpMjj2.lt.vbfTiny .and. tmpPZjj.lt.vbfTiny) then                                                                                                                     
                        passcuts_user=.false.
                        return
                     endif
                  enddo
               endif
            enddo
         endif
\end{verbatim}
The above imposes the veto on weak isospin partners that have been clustered together.
We first check if the event possesses a pair of isospin partners (\texttt{is\_a\_qqxPair.eq.true}) and whether the event has an \texttt{njet=2} or \texttt{njet=3} topology; 
the clustering pathology outlined in Sec.~\ref{sec:results_VBFcuts} does not occur in \texttt{njet=3} events.
To identify if an isospin pair has been clustered, we compare the invariant mass and $p_z$ momentum of the isospin system to those of each jet.
If the invariant mass and $p_z$ match, we tag the jet as a clustered weak isospin pair and reject the phase space point.

Our last set of changes are to the file \texttt{SubProcesses/genps\_fks.f}. This controls the population of phase space points in \mgamc.
In particular, the subroutine \texttt{trans\_x} near \confirm{line 2206} effectively performs a change of variable on Bret-Wigner 
propagators to significantly improve MC integration efficiency for processes with $s$-channel resonances, such as the $pp\to \dpmpm\dxx W^{\mp} \to \dpmpm\dxx j j $ subprocess.
When activated, the MC integration over $t$-channel momentum transfers becomes much less efficiency.
As we explicitly remove $s$-channel resonances with phase space cuts, we turn off these changes of variables.
To do this, we introduce the following local declarations near \confirm{line 2226}:
\begin{verbatim}
      integer tmpitype
      logical doVBFcuts
      parameter (doVBFcuts = .true.)
\end{verbatim}
which are set independent of \texttt{Source/cuts.inc}.
We then catch the change-of-variable flag \texttt{itype}, and bypass the change itself for the relevant cases with the following commands (and approximate line numbers \texttt{LXXXX}):
\begin{verbatim}
  L2231      tmpitype=itype
  L2232      if(doVBFcuts.and.(itype.gt.2.or.itype.gt.7)) tmpitype=2
  L2233      if (tmpitype.eq.1) then
  L2239      elseif (tmpitype.eq.2) then
  L2267      elseif(tmpitype.eq.3) then
  L2275      elseif(tmpitype.eq.4) then
  L2307      elseif(tmpitype.eq.5) then
  L2365      elseif(tmpitype.eq.6) then
  L2415      elseif (tmpitype.eq.7) then
\end{verbatim}
Event generation can then be steered with a script like the following:
\begin{verbatim}
  launch TypeIInlo_VBF_DxxD0_NLO
  order=NLO
  fixed_order=ON
  set mdpp scan1:[550,1000]
  set mdp  scan1:[550,1000]
  set md0  scan1:[550,1000]
  set lamHD1 1.0
  set lamD1  1.0
  set vevD   1e-8
  set run_card reweight_scale true
  set run_card reweight_PDF false
  set LHC 14
  set pdlabel lhapdf
  set lhaid 26000
  set fixed_ren_scale False
  set fixed_fac_scale False
  set dynamical_scale_choice -1
  set no_parton_cut
  set jetalgo    -1
  set jetradius 1.0
  set ptj   25
  set etaj 4.5
  set mll 200
  set drll 2.0
\end{verbatim}

\twocolumngrid


\begin{thebibliography}{99}

\bibitem{Magg:1980ut} 
  M.~Magg and C.~Wetterich,
  Phys.\ Lett.\  {\bf 94B}, 61 (1980).
  doi:10.1016/0370-2693(80)90825-4



\bibitem{Schechter:1980gr} 
  J.~Schechter and J.~W.~F.~Valle,
  Phys.\ Rev.\ D {\bf 22}, 2227 (1980).
  doi:10.1103/PhysRevD.22.2227



\bibitem{Cheng:1980qt} 
  T.~P.~Cheng and L.~F.~Li,
  Phys.\ Rev.\ D {\bf 22}, 2860 (1980).
  doi:10.1103/PhysRevD.22.2860



\bibitem{Mohapatra:1980yp} 
  R.~N.~Mohapatra and G.~Senjanovic,
  Phys.\ Rev.\ D {\bf 23}, 165 (1981).
  doi:10.1103/PhysRevD.23.165



\bibitem{Lazarides:1980nt} 
  G.~Lazarides, Q.~Shafi and C.~Wetterich,
  Nucl.\ Phys.\ B {\bf 181}, 287 (1981).
  doi:10.1016/0550-3213(81)90354-0



\bibitem{Minkowski:1977sc} 
  P.~Minkowski,
  Phys.\ Lett.\  {\bf 67B}, 421 (1977).
  doi:10.1016/0370-2693(77)90435-X



\bibitem{Yanagida:1979as} 
  T.~Yanagida,
  Conf.\ Proc.\ C {\bf 7902131}, 95 (1979).



\bibitem{GellMann:1980vs} 
  M.~Gell-Mann, P.~Ramond and R.~Slansky,
  Conf.\ Proc.\ C {\bf 790927}, 315 (1979)
  [arXiv:1306.4669 [hep-th]].



\bibitem{Glashow:1979nm} 
  S.~L.~Glashow,
  NATO Sci.\ Ser.\ B {\bf 61}, 687 (1980).
  doi:10.1007/978-1-4684-7197-7\_15



\bibitem{Mohapatra:1979ia} 
  R.~N.~Mohapatra and G.~Senjanovic,
  Phys.\ Rev.\ Lett.\  {\bf 44}, 912 (1980).
  doi:10.1103/PhysRevLett.44.912



\bibitem{Shrock:1980ct} 
  R.~E.~Shrock,
  Phys.\ Rev.\ D {\bf 24}, 1232 (1981).
  doi:10.1103/PhysRevD.24.1232



\bibitem{Foot:1988aq} 
  R.~Foot, H.~Lew, X.~G.~He and G.~C.~Joshi,
  Z.\ Phys.\ C {\bf 44}, 441 (1989).
  doi:10.1007/BF01415558



\bibitem{Cai:2017jrq} 
  Y.~Cai, J.~Herrero-García, M.~A.~Schmidt, A.~Vicente and R.~R.~Volkas,
  Front.\ in Phys.\  {\bf 5}, 63 (2017)
  doi:10.3389/fphy.2017.00063
  [arXiv:1706.08524 [hep-ph]].



\bibitem{Cai:2017mow} 
  Y.~Cai, T.~Han, T.~Li and R.~Ruiz,
  Front.\ in Phys.\  {\bf 6}, 40 (2018)
  doi:10.3389/fphy.2018.00040
  [arXiv:1711.02180 [hep-ph]].



\bibitem{Chun:2003ej} 
  E.~J.~Chun, K.~Y.~Lee and S.~C.~Park,
  Phys.\ Lett.\ B {\bf 566}, 142 (2003)
  doi:10.1016/S0370-2693(03)00770-6
  [hep-ph/0304069].



\bibitem{Garayoa:2007fw} 
  J.~Garayoa and T.~Schwetz,
  JHEP {\bf 0803}, 009 (2008)
  doi:10.1088/1126-6708/2008/03/009
  [arXiv:0712.1453 [hep-ph]].



\bibitem{Kadastik:2007yd} 
  M.~Kadastik, M.~Raidal and L.~Rebane,
  Phys.\ Rev.\ D {\bf 77}, 115023 (2008)
  doi:10.1103/PhysRevD.77.115023
  [arXiv:0712.3912 [hep-ph]].



\bibitem{Perez:2008zc} 
  P.~Fileviez Perez, T.~Han, G.~Y.~Huang, T.~Li and K.~Wang,
  Phys.\ Rev.\ D {\bf 78}, 071301 (2008)
  doi:10.1103/PhysRevD.78.071301
  [arXiv:0803.3450 [hep-ph]].



\bibitem{Perez:2008ha} 
  P.~Fileviez Perez, T.~Han, G.~y.~Huang, T.~Li and K.~Wang,
  Phys.\ Rev.\ D {\bf 78}, 015018 (2008)
  doi:10.1103/PhysRevD.78.015018
  [arXiv:0805.3536 [hep-ph]].



\bibitem{Rizzo:1981xx} 
  T.~G.~Rizzo,
  Phys.\ Rev.\ D {\bf 25}, 1355 (1982)
  Addendum: [Phys.\ Rev.\ D {\bf 27}, 657 (1983)].
  doi:10.1103/PhysRevD.27.657, 10.1103/PhysRevD.25.1355



\bibitem{Gunion:1996pq} 
  J.~F.~Gunion, C.~Loomis and K.~T.~Pitts,
  eConf C {\bf 960625}, LTH096 (1996)
  [hep-ph/9610237].



\bibitem{Huitu:1996su} 
  K.~Huitu, J.~Maalampi, A.~Pietila and M.~Raidal,
  Nucl.\ Phys.\ B {\bf 487}, 27 (1997)
  doi:10.1016/S0550-3213(97)87466-4
  [hep-ph/9606311].



\bibitem{Muhlleitner:2003me} 
  M.~Muhlleitner and M.~Spira,
  Phys.\ Rev.\ D {\bf 68}, 117701 (2003)
  doi:10.1103/PhysRevD.68.117701
  [hep-ph/0305288].



\bibitem{Akeroyd:2005gt} 
  A.~G.~Akeroyd and M.~Aoki,
  Phys.\ Rev.\ D {\bf 72}, 035011 (2005)
  doi:10.1103/PhysRevD.72.035011
  [hep-ph/0506176].



\bibitem{Han:2007bk} 
  T.~Han, B.~Mukhopadhyaya, Z.~Si and K.~Wang,
  Phys.\ Rev.\ D {\bf 76}, 075013 (2007)
  doi:10.1103/PhysRevD.76.075013
  [arXiv:0706.0441 [hep-ph]].



\bibitem{Akeroyd:2007zv} 
  A.~G.~Akeroyd, M.~Aoki and H.~Sugiyama,
  Phys.\ Rev.\ D {\bf 77}, 075010 (2008)
  doi:10.1103/PhysRevD.77.075010
  [arXiv:0712.4019 [hep-ph]].



\bibitem{Akeroyd:2011zza} 
  A.~G.~Akeroyd and H.~Sugiyama,
  Phys.\ Rev.\ D {\bf 84}, 035010 (2011)
  doi:10.1103/PhysRevD.84.035010
  [arXiv:1105.2209 [hep-ph]].



\bibitem{Melfo:2011nx} 
  A.~Melfo, M.~Nemevsek, F.~Nesti, G.~Senjanovic and Y.~Zhang,
  Phys.\ Rev.\ D {\bf 85}, 055018 (2012)
  doi:10.1103/PhysRevD.85.055018
  [arXiv:1108.4416 [hep-ph]].



\bibitem{Alloul:2013raa} 
  A.~Alloul, M.~Frank, B.~Fuks and M.~Rausch de Traubenberg,
  Phys.\ Rev.\ D {\bf 88}, 075004 (2013)
  doi:10.1103/PhysRevD.88.075004
  [arXiv:1307.1711 [hep-ph]].



\bibitem{Li:2018jns} 
  T.~Li,
  JHEP {\bf 1809}, 079 (2018)
  doi:10.1007/JHEP09(2018)079
  [arXiv:1802.00945 [hep-ph]].



\bibitem{Dev:2018sel} 
  P.~S.~B.~Dev, M.~J.~Ramsey-Musolf and Y.~Zhang,
  Phys.\ Rev.\ D {\bf 98}, no. 5, 055013 (2018)
  doi:10.1103/PhysRevD.98.055013
  [arXiv:1806.08499 [hep-ph]].



\bibitem{Crivellin:2018ahj} 
  A.~Crivellin, M.~Ghezzi, L.~Panizzi, G.~M.~Pruna and A.~Signer,
  Phys.\ Rev.\ D {\bf 99}, no. 3, 035004 (2019)
  doi:10.1103/PhysRevD.99.035004
  [arXiv:1807.10224 [hep-ph]].



\bibitem{Du:2018eaw} 
  Y.~Du, A.~Dunbrack, M.~J.~Ramsey-Musolf and J.~H.~Yu,
  JHEP {\bf 1901}, 101 (2019)
  doi:10.1007/JHEP01(2019)101
  [arXiv:1810.09450 [hep-ph]].



\bibitem{Antusch:2018svb} 
  S.~Antusch, O.~Fischer, A.~Hammad and C.~Scherb,
  JHEP {\bf 1902}, 157 (2019)
  doi:10.1007/JHEP02(2019)157
  [arXiv:1811.03476 [hep-ph]].



\bibitem{Primulando:2019evb} 
  R.~Primulando, J.~Julio and P.~Uttayarat,
  JHEP {\bf 1908}, 024 (2019)
  doi:10.1007/JHEP08(2019)024
  [arXiv:1903.02493 [hep-ph]].

\bibitem{deMelo:2019asm} 
  T.~B.~de Melo, F.~S.~Queiroz and Y.~Villamizar,
  Int.\ J.\ Mod.\ Phys.\ A {\bf 34}, no. 27, 1950157 (2019)
  doi:10.1142/S0217751X19501574
  [arXiv:1909.07429 [hep-ph]].
  
  \bibitem{Padhan:2019jlc} 
  R.~Padhan, D.~Das, M.~Mitra and A.~Kumar Nayak,
  arXiv:1909.10495 [hep-ph].

\bibitem{Deppisch:2015qwa} 
  F.~F.~Deppisch, P.~S.~Bhupal Dev and A.~Pilaftsis,
  New J.\ Phys.\  {\bf 17}, no. 7, 075019 (2015)
  doi:10.1088/1367-2630/17/7/075019
  [arXiv:1502.06541 [hep-ph]].



\bibitem{Chatrchyan:2012ya} 
  S.~Chatrchyan {\it et al.} [CMS Collaboration],
  Eur.\ Phys.\ J.\ C {\bf 72}, 2189 (2012)
  doi:10.1140/epjc/s10052-012-2189-5
  [arXiv:1207.2666 [hep-ex]].



\bibitem{ATLAS:2012hi} 
  G.~Aad {\it et al.} [ATLAS Collaboration],
  Eur.\ Phys.\ J.\ C {\bf 72}, 2244 (2012)
  doi:10.1140/epjc/s10052-012-2244-2
  [arXiv:1210.5070 [hep-ex]].



\bibitem{Aaboud:2018qcu} 
  M.~Aaboud {\it et al.} [ATLAS Collaboration],
  Eur.\ Phys.\ J.\ C {\bf 79}, no. 1, 58 (2019)
  doi:10.1140/epjc/s10052-018-6500-y
  [arXiv:1808.01899 [hep-ex]].



\bibitem{Binosi:2008ig} 
  D.~Binosi, J.~Collins, C.~Kaufhold and L.~Theussl,
  Comput.\ Phys.\ Commun.\  {\bf 180}, 1709 (2009)
  doi:10.1016/j.cpc.2009.02.020
  [arXiv:0811.4113 [hep-ph]].



\bibitem{delAguila:2013mia} 
  F.~del Águila and M.~Chala,
  JHEP {\bf 1403}, 027 (2014)
  doi:10.1007/JHEP03(2014)027
  [arXiv:1311.1510 [hep-ph]].



\bibitem{Sjostrand:2006za} 
  T.~Sjostrand, S.~Mrenna and P.~Z.~Skands,
  JHEP {\bf 0605}, 026 (2006)
  doi:10.1088/1126-6708/2006/05/026
  [hep-ph/0603175].



\bibitem{Sjostrand:2014zea} 
  T.~Sjöstrand {\it et al.},
  Comput.\ Phys.\ Commun.\  {\bf 191}, 159 (2015)
  doi:10.1016/j.cpc.2015.01.024
  [arXiv:1410.3012 [hep-ph]].



\bibitem{Pukhov:2004ca} 
  A.~Pukhov,
  hep-ph/0412191.



\bibitem{Belyaev:2012qa} 
  A.~Belyaev, N.~D.~Christensen and A.~Pukhov,
  Comput.\ Phys.\ Commun.\  {\bf 184}, 1729 (2013)
  doi:10.1016/j.cpc.2013.01.014
  [arXiv:1207.6082 [hep-ph]].



\bibitem{Dasgupta:2018nvj} 
  M.~Dasgupta, F.~A.~Dreyer, K.~Hamilton, P.~F.~Monni and G.~P.~Salam,
  JHEP {\bf 1809}, 033 (2018)
  doi:10.1007/JHEP09(2018)033
  [arXiv:1805.09327 [hep-ph]].



\bibitem{Pascoli:2018rsg} 
  S.~Pascoli, R.~Ruiz and C.~Weiland,
  Phys.\ Lett.\ B {\bf 786}, 106 (2018)
  doi:10.1016/j.physletb.2018.08.060
  [arXiv:1805.09335 [hep-ph]].



\bibitem{Pascoli:2018heg} 
  S.~Pascoli, R.~Ruiz and C.~Weiland,
  JHEP {\bf 1906}, 049 (2019)
  doi:10.1007/JHEP06(2019)049
  [arXiv:1812.08750 [hep-ph]].



\bibitem{Fuks:2019iaj} 
  B.~Fuks, K.~Nordström, R.~Ruiz and S.~L.~Williamson,
  Phys.\ Rev.\ D {\bf 100}, no. 7, 074010 (2019)
  doi:10.1103/PhysRevD.100.074010
  [arXiv:1901.09937 [hep-ph]].



\bibitem{Hessler:2014ssa} 
  A.~G.~Hessler, A.~Ibarra, E.~Molinaro and S.~Vogl,
  Phys.\ Rev.\ D {\bf 91}, no. 11, 115004 (2015)
  doi:10.1103/PhysRevD.91.115004
  [arXiv:1408.0983 [hep-ph]].



\bibitem{Arkani-Hamed:2015vfh} 
  N.~Arkani-Hamed, T.~Han, M.~Mangano and L.~T.~Wang,
  Phys.\ Rept.\  {\bf 652}, 1 (2016)
  doi:10.1016/j.physrep.2016.07.004
  [arXiv:1511.06495 [hep-ph]].



\bibitem{CEPC-SPPCStudyGroup:2015csa} 
  M.~Ahmad {\it et al.},
  IHEP-CEPC-DR-2015-01, IHEP-TH-2015-01, IHEP-EP-2015-01.



\bibitem{Golling:2016gvc} 
  T.~Golling {\it et al.},
  CERN Yellow Rep.\ , no. 3, 441 (2017)
  doi:10.23731/CYRM-2017-003.441
  [arXiv:1606.00947 [hep-ph]].



\bibitem{Benedikt:2018csr} 
  A.~Abada {\it et al.} [FCC Collaboration],
  Eur.\ Phys.\ J.\ ST {\bf 228}, no. 4, 755 (2019).
  doi:10.1140/epjst/e2019-900087-0



\bibitem{Abada:2019ono} 
  A.~Abada {\it et al.} [FCC Collaboration],
  Eur.\ Phys.\ J.\ ST {\bf 228}, no. 5, 1109 (2019).
  doi:10.1140/epjst/e2019-900088-6



\bibitem{Frixione:2002ik} 
  S.~Frixione and B.~R.~Webber,
  JHEP {\bf 0206}, 029 (2002)
  doi:10.1088/1126-6708/2002/06/029
  [hep-ph/0204244].



\bibitem{Babu:2016rcr} 
  K.~S.~Babu and S.~Jana,
  Phys.\ Rev.\ D {\bf 95}, no. 5, 055020 (2017)
  doi:10.1103/PhysRevD.95.055020
  [arXiv:1612.09224 [hep-ph]].



\bibitem{Ghosh:2017jbw} 
  K.~Ghosh, S.~Jana and S.~Nandi,
  JHEP {\bf 1803}, 180 (2018)
  doi:10.1007/JHEP03(2018)180
  [arXiv:1705.01121 [hep-ph]].



\bibitem{Ghosh:2018drw} 
  T.~Ghosh, S.~Jana and S.~Nandi,
  Phys.\ Rev.\ D {\bf 97}, no. 11, 115037 (2018)
  doi:10.1103/PhysRevD.97.115037
  [arXiv:1802.09251 [hep-ph]].



\bibitem{Degrande:2011ua} 
  C.~Degrande, C.~Duhr, B.~Fuks, D.~Grellscheid, O.~Mattelaer and T.~Reiter,
  Comput.\ Phys.\ Commun.\  {\bf 183}, 1201 (2012)
  doi:10.1016/j.cpc.2012.01.022
  [arXiv:1108.2040 [hep-ph]].



\bibitem{Alwall:2014hca} 
  J.~Alwall {\it et al.},
  JHEP {\bf 1407}, 079 (2014)
  doi:10.1007/JHEP07(2014)079
  [arXiv:1405.0301 [hep-ph]].



\bibitem{Esteban:2018azc} 
  I.~Esteban, M.~C.~Gonzalez-Garcia, A.~Hernandez-Cabezudo, M.~Maltoni and T.~Schwetz,
  JHEP {\bf 1901}, 106 (2019)
  doi:10.1007/JHEP01(2019)106
  [arXiv:1811.05487 [hep-ph]].



\bibitem{Chen:2005jx} 
  M.~C.~Chen, S.~Dawson and T.~Krupovnickas,
  Int.\ J.\ Mod.\ Phys.\ A {\bf 21}, 4045 (2006)
  doi:10.1142/S0217751X0603388X
  [hep-ph/0504286].



\bibitem{Chen:2008jg} 
  M.~C.~Chen, S.~Dawson and C.~B.~Jackson,
  Phys.\ Rev.\ D {\bf 78}, 093001 (2008)
  doi:10.1103/PhysRevD.78.093001
  [arXiv:0809.4185 [hep-ph]].



\bibitem{Kanemura:2012rs} 
  S.~Kanemura and K.~Yagyu,
  Phys.\ Rev.\ D {\bf 85}, 115009 (2012)
  doi:10.1103/PhysRevD.85.115009
  [arXiv:1201.6287 [hep-ph]].



\bibitem{Das:2016bir} 
  D.~Das and A.~Santamaria,
  Phys.\ Rev.\ D {\bf 94}, no. 1, 015015 (2016)
  doi:10.1103/PhysRevD.94.015015
  [arXiv:1604.08099 [hep-ph]].



\bibitem{Alloul:2013bka} 
  A.~Alloul, N.~D.~Christensen, C.~Degrande, C.~Duhr and B.~Fuks,
  Comput.\ Phys.\ Commun.\  {\bf 185}, 2250 (2014)
  doi:10.1016/j.cpc.2014.04.012
  [arXiv:1310.1921 [hep-ph]].



\bibitem{Skands:2003cj} 
  P.~Z.~Skands {\it et al.},
  JHEP {\bf 0407}, 036 (2004)
  doi:10.1088/1126-6708/2004/07/036
  [hep-ph/0311123].



\bibitem{Frixione:2019fxg} 
  S.~Frixione, B.~Fuks, V.~Hirschi, K.~Mawatari, H.~S.~Shao, P.~A.~Sunder and M.~Zaro,
  JHEP {\bf 1912}, 008 (2019)
  doi:10.1007/JHEP12(2019)008
  [arXiv:1907.04898 [hep-ph]].



\bibitem{Degrande:2014vpa} 
  C.~Degrande,
  Comput.\ Phys.\ Commun.\  {\bf 197}, 239 (2015)
  doi:10.1016/j.cpc.2015.08.015
  [arXiv:1406.3030 [hep-ph]].



\bibitem{Hahn:2000kx} 
  T.~Hahn,
  Comput.\ Phys.\ Commun.\  {\bf 140}, 418 (2001)
  doi:10.1016/S0010-4655(01)00290-9
  [hep-ph/0012260].



\bibitem{Bellm:2015jjp} 
  J.~Bellm {\it et al.},
  Eur.\ Phys.\ J.\ C {\bf 76}, no. 4, 196 (2016)
  doi:10.1140/epjc/s10052-016-4018-8
  [arXiv:1512.01178 [hep-ph]].



\bibitem{Gleisberg:2008ta} 
  T.~Gleisberg, S.~Hoeche, F.~Krauss, M.~Schonherr, S.~Schumann, F.~Siegert and J.~Winter,
  JHEP {\bf 0902}, 007 (2009)
  doi:10.1088/1126-6708/2009/02/007
  [arXiv:0811.4622 [hep-ph]].



\bibitem{Hirschi:2011pa} 
  V.~Hirschi, R.~Frederix, S.~Frixione, M.~V.~Garzelli, F.~Maltoni and R.~Pittau,
  JHEP {\bf 1105}, 044 (2011)
  doi:10.1007/JHEP05(2011)044
  [arXiv:1103.0621 [hep-ph]].



\bibitem{Hirschi:2015iia} 
  V.~Hirschi and O.~Mattelaer,
  JHEP {\bf 1510}, 146 (2015)
  doi:10.1007/JHEP10(2015)146
  [arXiv:1507.00020 [hep-ph]].



\bibitem{Frixione:1995ms} 
  S.~Frixione, Z.~Kunszt and A.~Signer,
  Nucl.\ Phys.\ B {\bf 467}, 399 (1996)
  doi:10.1016/0550-3213(96)00110-1
  [hep-ph/9512328].



\bibitem{Frixione:1997np} 
  S.~Frixione,
  Nucl.\ Phys.\ B {\bf 507}, 295 (1997)
  doi:10.1016/S0550-3213(97)00574-9
  [hep-ph/9706545].



\bibitem{Frederix:2009yq} 
  R.~Frederix, S.~Frixione, F.~Maltoni and T.~Stelzer,
  JHEP {\bf 0910}, 003 (2009)
  doi:10.1088/1126-6708/2009/10/003
  [arXiv:0908.4272 [hep-ph]].



\bibitem{Bauer:2000yr} 
  C.~W.~Bauer, S.~Fleming, D.~Pirjol and I.~W.~Stewart,
  Phys.\ Rev.\ D {\bf 63}, 114020 (2001)
  doi:10.1103/PhysRevD.63.114020
  [hep-ph/0011336].



\bibitem{Bauer:2001yt} 
  C.~W.~Bauer, D.~Pirjol and I.~W.~Stewart,
  Phys.\ Rev.\ D {\bf 65}, 054022 (2002)
  doi:10.1103/PhysRevD.65.054022
  [hep-ph/0109045].



\bibitem{Beneke:2002ph} 
  M.~Beneke, A.~P.~Chapovsky, M.~Diehl and T.~Feldmann,
  Nucl.\ Phys.\ B {\bf 643}, 431 (2002)
  doi:10.1016/S0550-3213(02)00687-9
  [hep-ph/0206152].



\bibitem{Becher:2006nr} 
  T.~Becher and M.~Neubert,
  Phys.\ Rev.\ Lett.\  {\bf 97}, 082001 (2006)
  doi:10.1103/PhysRevLett.97.082001
  [hep-ph/0605050].



\bibitem{Bonvini:2014qga} 
  M.~Bonvini, S.~Forte, G.~Ridolfi and L.~Rottoli,
  JHEP {\bf 1501}, 046 (2015)
  doi:10.1007/JHEP01(2015)046
  [arXiv:1409.0864 [hep-ph]].



\bibitem{Becher:2006mr} 
  T.~Becher, M.~Neubert and B.~D.~Pecjak,
  JHEP {\bf 0701}, 076 (2007)
  doi:10.1088/1126-6708/2007/01/076
  [hep-ph/0607228].



\bibitem{Ahrens:2008nc} 
  V.~Ahrens, T.~Becher, M.~Neubert and L.~L.~Yang,
  Eur.\ Phys.\ J.\ C {\bf 62}, 333 (2009)
  doi:10.1140/epjc/s10052-009-1030-2
  [arXiv:0809.4283 [hep-ph]].



\bibitem{Ruiz:2017yyf} 
  R.~Ruiz, M.~Spannowsky and P.~Waite,
  Phys.\ Rev.\ D {\bf 96}, no. 5, 055042 (2017)
  doi:10.1103/PhysRevD.96.055042
  [arXiv:1706.02298 [hep-ph]].



\bibitem{Becher:2014aya} 
  T.~Becher, R.~Frederix, M.~Neubert and L.~Rothen,
  Eur.\ Phys.\ J.\ C {\bf 75}, no. 4, 154 (2015)
  doi:10.1140/epjc/s10052-015-3368-y
  [arXiv:1412.8408 [hep-ph]].



\bibitem{Harland-Lang:2019pla} 
  L.~A.~Harland-Lang, A.~D.~Martin, R.~Nathvani and R.~S.~Thorne,
  Eur.\ Phys.\ J.\ C {\bf 79}, no. 10, 811 (2019)
  doi:10.1140/epjc/s10052-019-7296-0
  [arXiv:1907.02750 [hep-ph]].



\bibitem{Manohar:2016nzj} 
  A.~Manohar, P.~Nason, G.~P.~Salam and G.~Zanderighi,
  Phys.\ Rev.\ Lett.\  {\bf 117}, no. 24, 242002 (2016)
  doi:10.1103/PhysRevLett.117.242002
  [arXiv:1607.04266 [hep-ph]].



\bibitem{Manohar:2017eqh} 
  A.~V.~Manohar, P.~Nason, G.~P.~Salam and G.~Zanderighi,
  JHEP {\bf 1712}, 046 (2017)
  doi:10.1007/JHEP12(2017)046
  [arXiv:1708.01256 [hep-ph]].



\bibitem{Buckley:2014ana} 
  A.~Buckley, J.~Ferrando, S.~Lloyd, K.~Nordström, B.~Page, M.~Rüfenacht, M.~Schönherr and G.~Watt,
  Eur.\ Phys.\ J.\ C {\bf 75}, 132 (2015)
  doi:10.1140/epjc/s10052-015-3318-8
  [arXiv:1412.7420 [hep-ph]].



\bibitem{Ruiz:2015zca} 
  R.~Ruiz,
  JHEP {\bf 1512}, 165 (2015)
  doi:10.1007/JHEP12(2015)165
  [arXiv:1509.05416 [hep-ph]].



\bibitem{Degrande:2016aje} 
  C.~Degrande, O.~Mattelaer, R.~Ruiz and J.~Turner,
  Phys.\ Rev.\ D {\bf 94}, no. 5, 053002 (2016)
  doi:10.1103/PhysRevD.94.053002
  [arXiv:1602.06957 [hep-ph]].



\bibitem{Harris:2001sx} 
  B.~W.~Harris and J.~F.~Owens,
  Phys.\ Rev.\ D {\bf 65}, 094032 (2002)
  doi:10.1103/PhysRevD.65.094032
  [hep-ph/0102128].

\bibitem{Aivazis:1993pi} 
  M.~A.~G.~Aivazis, J.~C.~Collins, F.~I.~Olness and W.~K.~Tung,
  Phys.\ Rev.\ D {\bf 50}, 3102 (1994)
  doi:10.1103/PhysRevD.50.3102
  [hep-ph/9312319].
  
  \bibitem{Collins:1998rz} 
  J.~C.~Collins,
  Phys.\ Rev.\ D {\bf 58}, 094002 (1998)
  doi:10.1103/PhysRevD.58.094002
  [hep-ph/9806259].

\bibitem{Collins:2011zzd} 
  J.~Collins,
  Camb.\ Monogr.\ Part.\ Phys.\ Nucl.\ Phys.\ Cosmol.\  {\bf 32}, 1 (2011).

\bibitem{Dawson:2014pea} 
  S.~Dawson, A.~Ismail and I.~Low,
  Phys.\ Rev.\ D {\bf 90}, no. 1, 014005 (2014)
  doi:10.1103/PhysRevD.90.014005
  [arXiv:1405.6211 [hep-ph]].	
  
  \bibitem{Han:2014nja} 
  T.~Han, J.~Sayre and S.~Westhoff,
  JHEP {\bf 1504}, 145 (2015)
  doi:10.1007/JHEP04(2015)145
  [arXiv:1411.2588 [hep-ph]].

\bibitem{Pumplin:2001ct} 
  J.~Pumplin, D.~Stump, R.~Brock, D.~Casey, J.~Huston, J.~Kalk, H.~L.~Lai and W.~K.~Tung,
  Phys.\ Rev.\ D {\bf 65}, 014013 (2001)
  doi:10.1103/PhysRevD.65.014013
  [hep-ph/0101032].
  
  \bibitem{Martin:2009iq} 
  A.~D.~Martin, W.~J.~Stirling, R.~S.~Thorne and G.~Watt,
  Eur.\ Phys.\ J.\ C {\bf 63}, 189 (2009)
  doi:10.1140/epjc/s10052-009-1072-5
  [arXiv:0901.0002 [hep-ph]].

\bibitem{Tackmann:2016jyb} 
  F.~J.~Tackmann, W.~J.~Waalewijn and L.~Zeune,
  JHEP {\bf 1607}, 119 (2016)
  doi:10.1007/JHEP07(2016)119
  [arXiv:1603.03052 [hep-ph]].



\bibitem{Fuks:2017vtl} 
  B.~Fuks and R.~Ruiz,
  JHEP {\bf 1705}, 032 (2017)
  doi:10.1007/JHEP05(2017)032
  [arXiv:1701.05263 [hep-ph]].



\bibitem{Arpino:2019fmo} 
  L.~Arpino, A.~Banfi, S.~Jäger and N.~Kauer,
  JHEP {\bf 1908}, 076 (2019)
  doi:10.1007/JHEP08(2019)076
  [arXiv:1905.06646 [hep-ph]].



\bibitem{Becher:2012qa} 
  T.~Becher and M.~Neubert,
  JHEP {\bf 1207}, 108 (2012)
  doi:10.1007/JHEP07(2012)108
  [arXiv:1205.3806 [hep-ph]].



\bibitem{Becher:2013xia} 
  T.~Becher, M.~Neubert and L.~Rothen,
  JHEP {\bf 1310}, 125 (2013)
  doi:10.1007/JHEP10(2013)125
  [arXiv:1307.0025 [hep-ph]].



\bibitem{Cacciari:2008gp} 
  M.~Cacciari, G.~P.~Salam and G.~Soyez,
  JHEP {\bf 0804}, 063 (2008)
  doi:10.1088/1126-6708/2008/04/063
  [arXiv:0802.1189 [hep-ph]].



\bibitem{Michel:2018hui} 
  J.~K.~L.~Michel, P.~Pietrulewicz and F.~J.~Tackmann,
  JHEP {\bf 1904}, 142 (2019)
  doi:10.1007/JHEP04(2019)142
  [arXiv:1810.12911 [hep-ph]].



\bibitem{Aaboud:2018xdt} 
  M.~Aaboud {\it et al.} [ATLAS Collaboration],
  Phys.\ Rev.\ D {\bf 98}, 052005 (2018)
  doi:10.1103/PhysRevD.98.052005
  [arXiv:1802.04146 [hep-ex]].



\bibitem{Dasgupta:2007wa} 
  M.~Dasgupta, L.~Magnea and G.~P.~Salam,
  JHEP {\bf 0802}, 055 (2008)
  doi:10.1088/1126-6708/2008/02/055
  [arXiv:0712.3014 [hep-ph]].



\bibitem{Dasgupta:2014yra} 
  M.~Dasgupta, F.~Dreyer, G.~P.~Salam and G.~Soyez,
  JHEP {\bf 1504}, 039 (2015)
  doi:10.1007/JHEP04(2015)039
  [arXiv:1411.5182 [hep-ph]].



\bibitem{Banfi:2015pju} 
  A.~Banfi, F.~Caola, F.~A.~Dreyer, P.~F.~Monni, G.~P.~Salam, G.~Zanderighi and F.~Dulat,
  JHEP {\bf 1604}, 049 (2016)
  doi:10.1007/JHEP04(2016)049
  [arXiv:1511.02886 [hep-ph]].



\bibitem{Dasgupta:2016bnd} 
  M.~Dasgupta, F.~A.~Dreyer, G.~P.~Salam and G.~Soyez,
  JHEP {\bf 1606}, 057 (2016)
  doi:10.1007/JHEP06(2016)057
  [arXiv:1602.01110 [hep-ph]].



\bibitem{Martin:2014nqa} 
  A.~D.~Martin and M.~G.~Ryskin,
  Eur.\ Phys.\ J.\ C {\bf 74}, 3040 (2014)
  doi:10.1140/epjc/s10052-014-3040-y
  [arXiv:1406.2118 [hep-ph]].



\bibitem{Alva:2014gxa} 
  D.~Alva, T.~Han and R.~Ruiz,
  JHEP {\bf 1502}, 072 (2015)
  doi:10.1007/JHEP02(2015)072
  [arXiv:1411.7305 [hep-ph]].



\bibitem{Schmidt:2015zda} 
  C.~Schmidt, J.~Pumplin, D.~Stump and C.~P.~Yuan,
  Phys.\ Rev.\ D {\bf 93}, no. 11, 114015 (2016)
  doi:10.1103/PhysRevD.93.114015
  [arXiv:1509.02905 [hep-ph]].



\bibitem{Bertone:2017bme} 
  V.~Bertone {\it et al.} [NNPDF Collaboration],
  SciPost Phys.\  {\bf 5}, no. 1, 008 (2018)
  doi:10.21468/SciPostPhys.5.1.008
  [arXiv:1712.07053 [hep-ph]].



\bibitem{Budnev:1974de} 
  V.~M.~Budnev, I.~F.~Ginzburg, G.~V.~Meledin and V.~G.~Serbo,
  Phys.\ Rept.\  {\bf 15}, 181 (1975).
  doi:10.1016/0370-1573(75)90009-5



\bibitem{Ball:2013hta} 
  R.~D.~Ball {\it et al.} [NNPDF Collaboration],
  Nucl.\ Phys.\ B {\bf 877}, 290 (2013)
  doi:10.1016/j.nuclphysb.2013.10.010
  [arXiv:1308.0598 [hep-ph]].



\bibitem{Martin:2004dh} 
  A.~D.~Martin, R.~G.~Roberts, W.~J.~Stirling and R.~S.~Thorne,
  Eur.\ Phys.\ J.\ C {\bf 39}, 155 (2005)
  doi:10.1140/epjc/s2004-02088-7
  [hep-ph/0411040].



\bibitem{Dittmaier:2017bnh} 
  S.~Dittmaier, A.~Huss and G.~Knippen,
  JHEP {\bf 1709}, 034 (2017)
  doi:10.1007/JHEP09(2017)034
  [arXiv:1705.03722 [hep-ph]].



\bibitem{Dawson:1990zj} 
  S.~Dawson,
  Nucl.\ Phys.\ B {\bf 359}, 283 (1991).
  doi:10.1016/0550-3213(91)90061-2



\bibitem{Anastasiou:2015ema} 
  C.~Anastasiou, C.~Duhr, F.~Dulat, F.~Herzog and B.~Mistlberger,
  Phys.\ Rev.\ Lett.\  {\bf 114}, 212001 (2015)
  doi:10.1103/PhysRevLett.114.212001
  [arXiv:1503.06056 [hep-ph]].



\bibitem{Dawson:1998py} 
  S.~Dawson, S.~Dittmaier and M.~Spira,
  Phys.\ Rev.\ D {\bf 58}, 115012 (1998)
  doi:10.1103/PhysRevD.58.115012
  [hep-ph/9805244].



\bibitem{deFlorian:2013jea} 
  D.~de Florian and J.~Mazzitelli,
  Phys.\ Rev.\ Lett.\  {\bf 111}, 201801 (2013)
  doi:10.1103/PhysRevLett.111.201801
  [arXiv:1309.6594 [hep-ph]].



\bibitem{Bonvini:2014joa} 
  M.~Bonvini and S.~Marzani,
  JHEP {\bf 1409}, 007 (2014)
  doi:10.1007/JHEP09(2014)007
  [arXiv:1405.3654 [hep-ph]].



\bibitem{Becher:2007ty} 
  T.~Becher, M.~Neubert and G.~Xu,
  JHEP {\bf 0807}, 030 (2008)
  doi:10.1088/1126-6708/2008/07/030
  [arXiv:0710.0680 [hep-ph]].



\bibitem{photonNLO}
R.~Ruiz,  {\it et al.} , (In Progress)


\bibitem{Campanario:2013fsa} 
  F.~Campanario, T.~M.~Figy, S.~Plätzer and M.~Sjödahl,
  Phys.\ Rev.\ Lett.\  {\bf 111}, no. 21, 211802 (2013)
  doi:10.1103/PhysRevLett.111.211802
  [arXiv:1308.2932 [hep-ph]].



\bibitem{Campanario:2018ppz} 
  F.~Campanario, T.~M.~Figy, S.~Plätzer, M.~Rauch, P.~Schichtel and M.~Sjödahl,
  Phys.\ Rev.\ D {\bf 98}, no. 3, 033003 (2018)
  doi:10.1103/PhysRevD.98.033003
  [arXiv:1802.09955 [hep-ph]].



\bibitem{Ballestrero:2018anz} 
  A.~Ballestrero {\it et al.},
  Eur.\ Phys.\ J.\ C {\bf 78}, no. 8, 671 (2018)
  doi:10.1140/epjc/s10052-018-6136-y
  [arXiv:1803.07943 [hep-ph]].



\bibitem{Cacciari:2005hq} 
  M.~Cacciari and G.~P.~Salam,
  Phys.\ Lett.\ B {\bf 641}, 57 (2006)
  doi:10.1016/j.physletb.2006.08.037
  [hep-ph/0512210].



\bibitem{Cacciari:2011ma} 
  M.~Cacciari, G.~P.~Salam and G.~Soyez,
  Eur.\ Phys.\ J.\ C {\bf 72}, 1896 (2012)
  doi:10.1140/epjc/s10052-012-1896-2
  [arXiv:1111.6097 [hep-ph]].



\bibitem{Rubin:2010xp} 
  M.~Rubin, G.~P.~Salam and S.~Sapeta,
  JHEP {\bf 1009}, 084 (2010)
  doi:10.1007/JHEP09(2010)084
  [arXiv:1006.2144 [hep-ph]].



\bibitem{Han:1991ia} 
  T.~Han and S.~Willenbrock,
  Phys.\ Lett.\ B {\bf 273}, 167 (1991).
  doi:10.1016/0370-2693(91)90572-8



\bibitem{Bolzoni:2010xr} 
  P.~Bolzoni, F.~Maltoni, S.~O.~Moch and M.~Zaro,
  Phys.\ Rev.\ Lett.\  {\bf 105}, 011801 (2010)
  doi:10.1103/PhysRevLett.105.011801
  [arXiv:1003.4451 [hep-ph]].



\bibitem{Dreyer:2016oyx} 
  F.~A.~Dreyer and A.~Karlberg,
  Phys.\ Rev.\ Lett.\  {\bf 117}, no. 7, 072001 (2016)
  doi:10.1103/PhysRevLett.117.072001
  [arXiv:1606.00840 [hep-ph]].



\bibitem{Degrande:2015xnm} 
  C.~Degrande, K.~Hartling, H.~E.~Logan, A.~D.~Peterson and M.~Zaro,
  Phys.\ Rev.\ D {\bf 93}, no. 3, 035004 (2016)
  doi:10.1103/PhysRevD.93.035004
  [arXiv:1512.01243 [hep-ph]].



    
  
  
  
  
\end{thebibliography}
\end{document}